\documentclass[twocolumn]{aastex701}
\usepackage{graphicx}
\usepackage{hyperref}
\usepackage{booktabs}
\usepackage{makecell}
\usepackage{amsmath}

\begin{document}
\title{Evidence for a Similar Physical Origin Between Prompt Emission and X-Ray Flares of Gamma-Ray Bursts}

\author[0009-0007-4057-8623]{Wen-Yuan Yu}
\affiliation{Guangxi Key Laboratory for Relativistic 
Astrophysics, School of Physical Science and Technology, Guangxi University, Nanning 530004, People's Republic of China}
\email{wyu@gxu.edu.cn}

\author[0009-0008-6247-0645]{Wen-Long Zhang}
\affiliation{Purple Mountain Observatory, Chinese Academy of Sciences, Nanjing 210023, China}
\affiliation{School of Astronomy and Space Sciences, University of Science and Technology of China, Hefei 230026, China}
\email{wlzhang@pmo.ac.cn}

\author[0000-0001-6396-9386]{Hou-Jun L\"{u}}
\altaffiliation{Corresponding author: lhj@gxu.edu.cn (LHJ)}
\affiliation{Guangxi Key Laboratory for Relativistic 
Astrophysics, School of Physical Science and Technology, Guangxi University, Nanning 530004, People's Republic of China}
\email{lhj@gxu.edu.cn}

\begin{abstract}
Statistical properties of X-ray flares are highly similar to those of the prompt emission in gamma-ray bursts (GRBs), suggesting that they may share a common physical origin (e.g., central engine activity). However, comparable statistical studies of afterglow X-ray flares and prompt gamma-ray pulses that are not observed simultaneously remain incomplete. In this paper, by systematically searching for gamma-ray emission simultaneous with X-ray flares, we identify 38 GRBs containing 50 pulse--flare pairs, 19 of which have measured redshifts. We perform joint temporal and spectral analyses of the gamma-ray pulses and X-ray flares over their overlapping time intervals, and find that the spectrum of each pair can be well described by an absorbed power-law (PL), PL plus blackbody (PL+BB), cutoff power-law (CPL), and Band function. More importantly, by applying self-organized criticality (SOC) analysis to both the prompt gamma-ray pulses and the simultaneously observed X-ray flares, we find that the temporal parameters (e.g., waiting time, rise time, peak time,
decay time, and duration) of the two components follow similar power-law distributions, with the largest absolute difference of $\Delta\alpha\sim0.36$ for the waiting time and much smaller differences ($\Delta\alpha\lesssim0.12$) for the other four parameters. For the isotropic energy derived from the subsample with measured redshifts, the difference between the two components is $\Delta\alpha\sim0.54$. These results suggest that the X-ray flares and the prompt emission are likely to share the same physical origin.

\end{abstract}


\section{Introduction}
Gamma-ray burst (GRB) is defined as a burst with a sudden release of gamma-ray emission that lasts from milliseconds to thousands of seconds, and it is believed to be from a catastrophic event, such as the collapsar of massive stars or the merger of double compact 
stars \citep{1989Natur.340..126E, 1993ApJ...405..273W, 2018NatCo...9..447Z, 2025ApJ...988L..46L}. In general, it can be divided into emission phases, e.g., the prompt emission phase and the afterglow phase \citep{2015PhR...561....1K}. Quantitatively, the duration of a burst is usually defined by the so-called $T_{90}$, which is the time interval between the epochs with 5\% and 95\% of the total ﬂuence collected by the detector \citep{1993ApJ...413L.101K}. However, the $T_{90}$ is energy and detector-sensitivity dependent \citep{2013ApJ...763...15Q}, and it cannot provide a reliable measure of the intrinsic timescale of central engine activity \citep{2009ApJ...703.1696Z, 2010ApJ...725.1965L, 2014ApJ...787...66Z,  2017ApJ...837...50G}.

Observational data indicate that a small fraction of GRB prompt emission is composed of two or more emission episodes with a quiescent interval \citep{1995ApJ...452..145K, 2012ApJ...748..132Z, 2014ApJ...789..145H, 2018ApJ...862..155L}. The spectral analysis of those two-episode emissions follows similar evolutionary behaviors and suggests that they are from the same physical origin. Moreover, with the Swift X-Ray Telescope (XRT) observations, an increasing number of GRBs have been found to exhibit X-ray flares \citep{2005Sci...309.1833B, 2006Natur.444.1010Z, 2011MNRAS.417.2144M} and internal plateau emission \citep{2007ApJ...670..565L, 2007ApJ...665..599T, 2014ApJ...785...74L} in their X-ray afterglows, indicating that the intrinsic timescale of central engine activity can extend to several thousand seconds \citep{2018pgrb.book.....Z,2022ApJ...931L..23L}.

One basic question is whether the physical process of X-ray flares (or internal plateau) is similar to that of the prompt emission powered by the activity of the central engine. Previous studies have shown that the X-ray flares and the prompt emission share similar temporal properties, including peak times, widths, rise and decay times, spectral lags, spectral evolution, and profile asymmetries \citep{2007ApJ...671.1903C, 2010A&A...524A..64C, 2007ApJ...667.1024K, 2010MNRAS.406.2149M, 2013ApJ...767L..28S, 2015ApJ...801...57G, 2018A&A...615A..80P, 2023ApJ...948...23Z, 2025ApJ...989..113M}. These similarities suggest that both X-ray flares and the prompt emission are from the same physical origin. On the other hand, a small fraction of X-ray flares can be simultaneously detected by the Burst Alert Telescope (BAT) and the XRT on board the Swift \citep{2014ApJ...789..145H, 2025ApJ...995..201Y}. Joint temporal and spectral analyses have determined that such flares originate from the activity of the central engine \citep{2014JApA...35..423P, 2025ApJ...995..201Y}. 

Self-organized criticality (SOC) describes how complex systems spontaneously evolve to a critical state without external fine-tuning of parameters \citep{1987PhRvL..59..381B}, and such scale-free behavior naturally gives rise to power-law distributions in temporal and energetic observables. SOC-like phenomena have been discussed in a wide range of astrophysical systems, including solar flares, stellar flares, pulsar glitches, soft gamma-ray repeaters, black hole systems, blazars, X-ray bursts, and fast radio bursts (e.g., \citealt{1991ApJ...380L..89L, 2008ApJ...672.1103M, 2013NatPh...9..465W, 2014ApJ...782...54A, 2015ApJ...810...19L,  2017NewAR..79....1L, 2017JCAP...03..023W, 2017MNRAS.471.2517W, 2020MNRAS.491.1498C, 2021ApJ...912...94A, 2021ApJ...920..153W, 2022ApJ...934...33A, 2022RAA....22f5012Z,2023RAA....23k5013Z}). Moreover, the SOC theory has also been applied in GRBs, such as the frequency distributions of energy and waiting time for GRB X-ray flares \citep{2013NatPh...9..465W,2016ApJS..224...20Y}, optical flares \citep{2017ApJ...844...79Y} and the gamma-ray pulses of prompt emission \citep{2021FrPhy..1614501L,2023ApJ...955L..34L}, even in the TeV photons of GRB \citep{2025A&A...693A.290Z}. These studies have shown that X-ray flares and prompt gamma-ray pulses exhibit similar power-law distributions in energy and waiting time, suggesting that they may share a common physical origin. However, such statistical results for both the X-ray flares of afterglow and gamma-ray pulses of prompt emission, which are not simultaneously detected are still incomplete. A convincing statistical comparison of the physical origins of afterglow X-ray flares and prompt gamma-ray pulses requires simultaneous observations of both components in the same GRB.

In this paper, by adopting the SOC theory, we investigate whether the statistical distributions of waiting time, rising time, peak time, decay time, duration, and the isotropic energy of X-ray flares of afterglow are consistent with that of gamma-ray pulses of prompt emission observed simultaneously with X-ray flares in the same GRBs, and then address the physical origin between X-ray flares of afterglow and gamma-ray pulses of prompt emission through a systematic analysis of the Swift BAT and XRT data. This paper is organized as follows. The data reduction and sample selection criteria procedures are described in Section~2. In Section~3, we derive the temporal parameters and joint spectral fitting for both prompt pulses and X-ray flares. The distributions of the temporal parameters by invoking the SOC method for both prompt pulses and X-ray flares are presented in Section~4. Conclusions and discussion are drawn in Section~5.

\section{Data Reduction and Sample Selection} 
\subsection{Light-curve Extraction}
We downloaded data of GRBs that triggered Swift/BAT from the Swift Archive\footnote{\url{https://www.swift.ac.uk/swift_portal/}}, and analyzed them using the standard HEASOFT v6.34 software package. The BAT light curves in the 15--350 keV energy band of GRBs were extracted with a fixed 1 s timebin by using \texttt{batbinevt} and employing the standard mask-weighting technique. The background was extracted using time intervals before and after each burst, and then modeled as Poisson noise, which is the standard background model for prompt emission in BAT events \citep{2007A&A...469..379E, 2009MNRAS.397.1177E}. The method of Bayesian blocks (\citealt{1998ApJ...504..405S}) is performed to identify possible signals in the light curves, and the false-positive probability threshold is set to $p_0=0.05$ by using the \texttt{astropy.stats.bayesian\_blocks} implementation.

The rapid slewing capacity of XRT makes it possible to catch 
the early X-ray emission of GRBs \citep{2005Sci...309.1833B}. We directly downloaded the X-ray light curves in the 0.3--10 keV energy band for all GRBs triggered by Swift/BAT from the Swift/XRT repository\footnote{\url{https://www.swift.ac.uk/xrt_curves/}} whenever XRT follow-up observations were available. We then plotted the BAT and XRT light curves together for each GRB whose prompt emission overlapped with the early X-ray emission.

\subsection{Sample Selection Criteria}
Our entire sample includes more than 1721 GRBs observed between 2005 January and 2025 January. We focus on the GRB events that exhibit both gamma-ray pulses of the prompt emission and X-ray flares of the afterglow observed simultaneously in the BAT and XRT bands. 

Here, an X-ray flare is identified by the following criteria \citep{2025ApJ...989..113M}:
(1) It exhibits a relatively complete structure, consisting of rise, peak, and decay phases;
(2) The flare flux exceeds twice the underlying afterglow background level;
(3) The underlying afterglow emission has sufficient observational data to be fitted with a power-law function. Building on this flare identification, two additional criteria are adopted for our sample selection: (i) The prompt emission phase must contain multiple distinct pulses identifiable in the BAT light curve; (ii) At least one X-ray flare must be present in the XRT light curve and overlap in time with one of the prompt pulses.
There are 38 GRBs with 50 pulse--flare pairs that satisfy our criteria, 19 of which have a 
measured redshift (see Table~\ref{tab:sample} for a summary).

In order to verify the representativeness of our sample, we crossmatched it against two reference X-ray flare catalogs from \cite{2016ApJS..224...20Y} and \cite{2025ApJ...989..113M}. All 16 out of our sample occurring before 2015 are present in the sample from \cite{2016ApJS..224...20Y}, confirming that our sample is drawn from the same underlying flare population. We also compared with the more recent compilation of \cite{2025ApJ...989..113M}, whose flare identification criteria we adopted. 32 out of 38 GRBs in our sample ($\sim$ 84\%) are found in that compilation, and the 6 GRBs not present there (e.g., GRBs 240912A, 250114A, and 250823A) are bursts from 2024--2025 that extend beyond its temporal coverage. For overlapping GRBs, \cite{2025ApJ...989..113M} generally lists more individual flare entries per GRB than our sample. The reason is that we additionally require that each flare has a temporally coincident prompt gamma-ray pulse.

\subsection{Morphological Classification}

\begin{figure}[htbp]
\centering
\begin{minipage}[b]{0.32\textwidth}
  \centering
  \includegraphics[width=\linewidth]{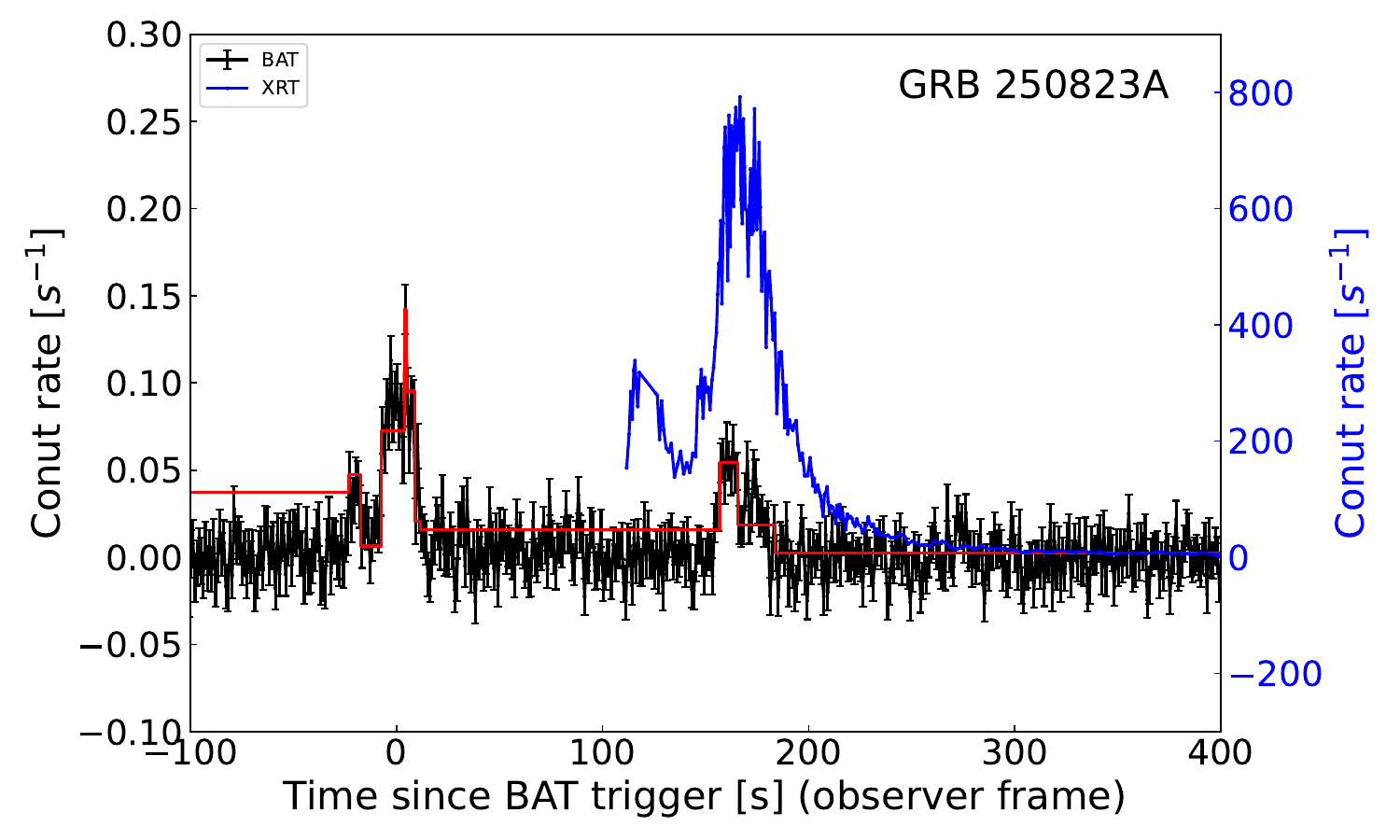}
\end{minipage}
\hfill
\begin{minipage}[b]{0.32\textwidth}
  \centering
  \includegraphics[width=\linewidth]{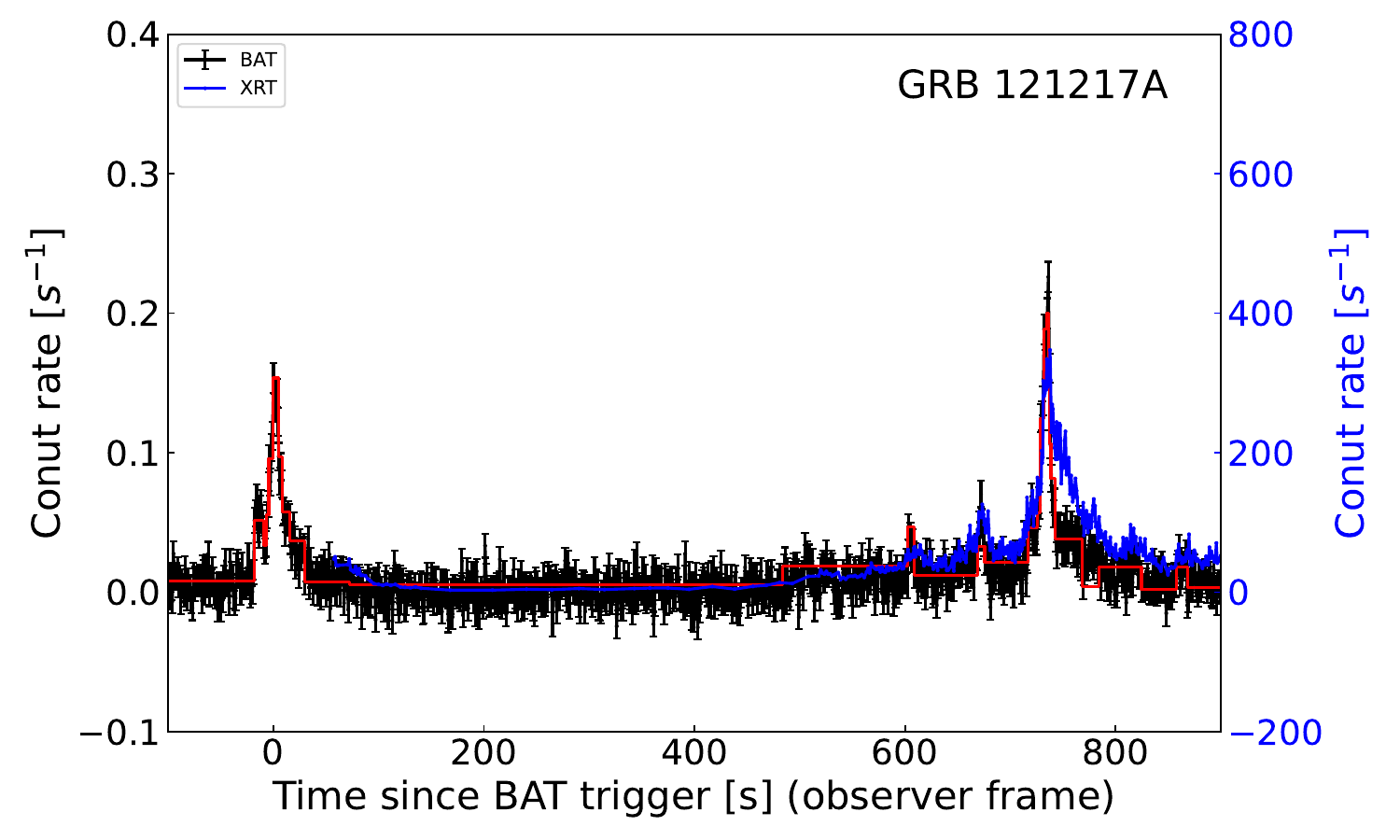}
\end{minipage}
\hfill
\begin{minipage}[b]{0.32\textwidth}
  \centering
  \includegraphics[width=\linewidth]{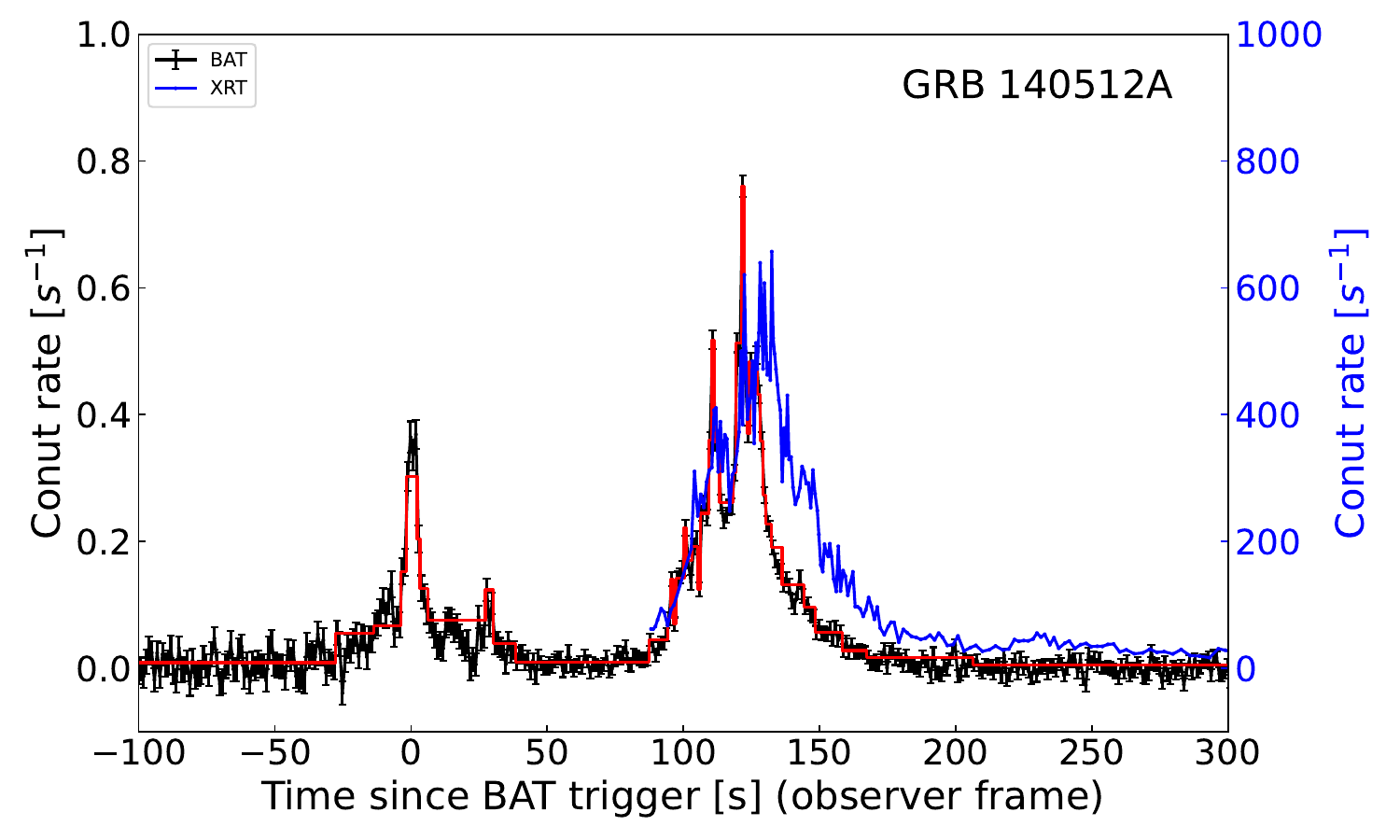}
\end{minipage}
\caption{Examples of BAT (black) and XRT (blue) light curves for Type I (left), Type II (middle), and Type III (right). The red solid lines are the Bayesian block analysis. The complete figure set (38 images) is available in the online journal.}
\label{example GRB}
\end{figure}

Phenomenologically, the morphological light curves can be broadly classified into three categories according to whether X-ray flares are present during the quiescent intervals of the prompt emission. Type I includes bursts in which at least one X-ray flare is observed during a quiescent interval of the prompt emission. This behavior suggests that the central engine remains active during the quiescent phase, with the emission appearing mainly in the X-ray band and being too weak to be detected by current gamma-ray instruments. Type II includes bursts that exhibit X-ray emission during the quiescent interval, but without distinct X-ray flares. In this case, the central engine may have already ceased its activity, and the observed X-ray emission is more likely to be the tail of the prompt emission or the onset of the afterglow. Type III includes bursts in which X-ray emission is either absent during the quiescent interval or too short-lived to place meaningful constraints on the emission properties during that phase. In such cases, it is difficult to determine whether the central engine is still active. In our sample, 12, 12, and 14 out of 38 GRBs belong to Types I, II, and III, respectively. Examples of the light curves for the three types are shown in Figure~\ref{example GRB}.\footnote{The remaining cases are available in the figure set or at the link: \url{https://astro.gxu.edu.cn/info/1067/2547.htm}.}

\section{Temporal Analysis and Joint Spectral Fitting} 
\subsection{Light-curve Fitting}
\begin{figure}[htbp]
\centering
\begin{minipage}[b]{0.45\textwidth} 
  \centering
  \includegraphics[width=\linewidth]{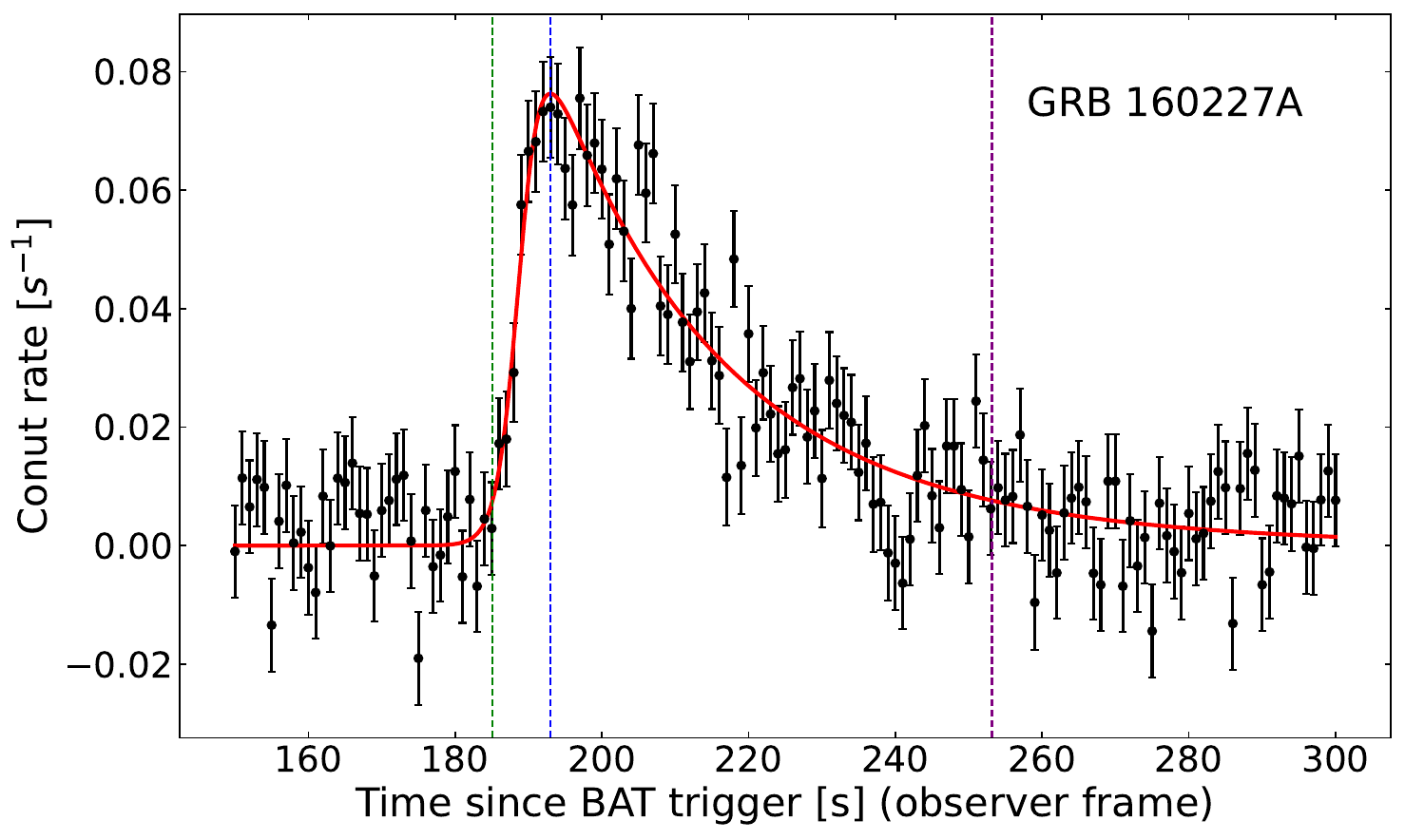} 
\end{minipage}
\hfill
\begin{minipage}[b]{0.45\textwidth}
  \centering
  \includegraphics[width=\linewidth]{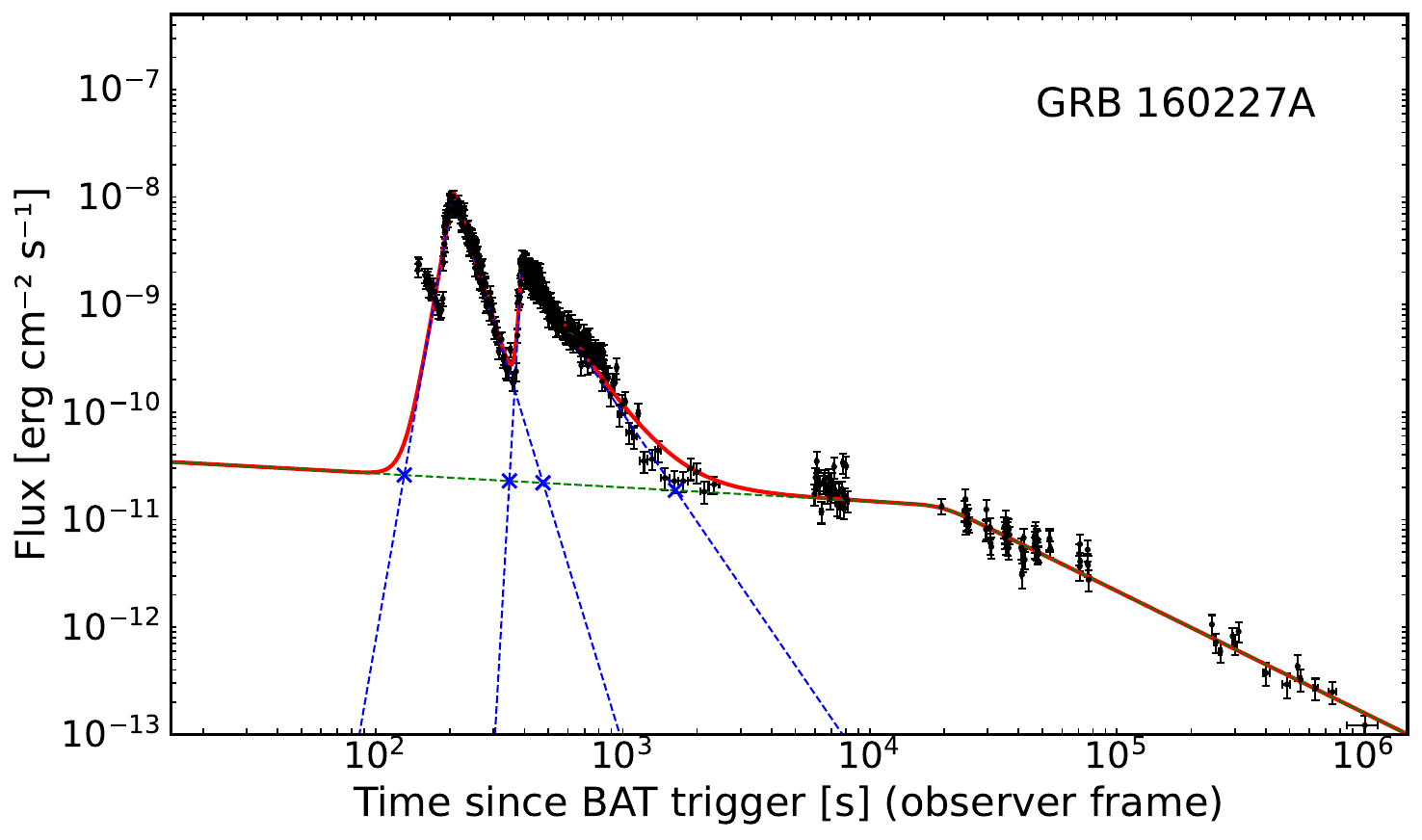}
\end{minipage}
\caption{Left: an example of pulse fitting by adopting Kocevski power-law function (red solid line). The blue, green, and purple vertical dashed lines are the times of pulse peak, start, and end, respectively. Right: an example of X-ray light curve fitting by adopting afterglow component (green dashed line) and two flare components (blue dashed lines). The intersection points between the flare and afterglow components define the start and end times of flare.}
\label{Lightcurve fitting}
\end{figure}

For the prompt pulses observed by BAT in our sample, we adopt the Kocevski power-law (KPL) function, which has been widely used to characterize the temporal profiles of GRB pulses to fit the pulses \citep{2003ApJ...596..389K, 2021ApJS..252...16L}. The function is expressed as
\begin{equation}
\begin{split}
f(t) = f_m \left( \frac{t + t_0}{t_m + t_0} \right)^r \left[ \frac{d}{d + r} + \frac{r}{d + r} \left( \frac{t + t_0}{t_m + t_0} \right)^{(r+1)} \right]^{-\left( \frac{r+d}{r+1} \right)},
\end{split}
\end{equation}
where $f_m$ is the peak flux, $t_m$ is the peak time, $t_0$ is a time offset, and $r$ and $d$ are the rising and decay slopes, respectively. Moreover, this empirical function can also provide a good description of both fast rise--exponential decay (FRED) and symmetric pulse shapes. An example of a pulse light curve is shown in Figure~\ref{Lightcurve fitting}. 

For the X-ray flares observed by XRT of our sample, we adopt the fitting procedure described in \cite{2025ApJ...989..113M} to fit. The light curve is modeled as the sum of an afterglow component and a flare component. The afterglow component is described by either a single power law or a broken power law, while each individual flare is modeled with a smoothly broken power law \citep{2007ApJ...671.1903C, 2010MNRAS.406.2113C, 2016ApJS..224...20Y}. The fitting function can be expressed as
\begin{equation}
F(t) = A \left( \frac{t}{t_b} \right)^{-\alpha_1} \left\{ \frac{1}{2} \left[ 1 + \left( \frac{t}{t_b} \right)^{1/\omega} \right] \right\}^{(\alpha_1 - \alpha_2)\omega},
\end{equation}
where $t_{\rm b}$ is the break time, $A$ is the model amplitude at the break time $t_{\rm b}$, $\alpha_1$ and $\alpha_2$ are the rising and decay indexes before and after $t_{\rm b}$, respectively, and $\omega$ represents the smoothness of the peak. An example of an X-ray flare light curve is shown in Figure~\ref{Lightcurve fitting}.

The parameters of the fitting model and their uncertainties are estimated with a Markov chain Monte Carlo (MCMC) approach implemented in the \texttt{emcee} package \citep{2013PASP..125..306F}. By adopting the best-fit parameters from the MCMC fitting, we reconstruct the light-curve profiles for both prompt pulses and X-ray flares, from which the following temporal parameters are then inferred. The detailed fitting parameters vary among bursts depending on the number of pulses and flares, and are not reported here for brevity. For prompt pulses, we define the peak time $t_{\rm p,peak}$ (subscript ``p'' denotes pulse) as the time from trigger to the peak of pulse, the rising time $t_{\rm p,rise}$ as the time interval from the point where the count rate reaches 10\% of the peak value to the peak, and the decay time $t_{\rm p,decay}$ as the time interval from the peak to the point where the count rate drops to 10\% of the peak value on the decay side. The waiting time $t_{\rm p,wait}$ of a given pulse is defined as the time interval between its start time and that of the immediately preceding pulse in the same GRB, i.e., $t_{\rm p,wait,i} = t_{\rm p,start,i} - t_{\rm p,start,i-1}$. The duration is defined as $t_{\rm p,duration} = t_{\rm p,rise} + t_{\rm p,decay}$. With analogous definitions, for X-ray flares (subscript ``f'' denotes flare), $t_{\rm f,peak}$, $t_{\rm f,rise}$, $t_{\rm f,decay}$, $t_{\rm f,wait}$, and $t_{\rm f,duration}$ represent the peak, rising, decay, waiting, and duration times, respectively. For the first flare appearing in the X-ray afterglow, the waiting time is taken as the start time of that flare itself, i.e., $t_{\rm f,wait} = t_{\rm f,start}$. The flare onset and end times are determined by the intersection points between the fitted flare profile and the underlying afterglow continuum. The fitting results are shown in Table~\ref{tab:sample}.

For the prompt pulses, it is found that the peak times $t_{\rm p,peak}$ extend from $\sim 59$ s to $\sim 864$ s, whereas the rising and decay times range from a few seconds to more than $10^2$ s. The X-ray flares cover an even wider range, with peak times exceeding $10^3$ s in some events and decay times approaching the same scale. 

\subsection{Joint Spectral Fitting}
The motivation for the joint spectral fitting is to test whether the prompt pulse and the contemporaneous X-ray flare can be described by a common spectral model. Previous studies have shown that the joint spectra of X-ray flares and prompt pulses can be well fitted by an absorbed single power-law (PL) model or a Band-function model \citep{2014JApA...35..423P}, suggesting that they may share a common physical origin.

For each pulse--flare pair, in this section, we also perform joint spectral fitting to the BAT and XRT data extracted from the overlapping time interval defined in Section~3. The BAT spectra are extracted using the standard HEASOFT v6.34 software package. We use \texttt{batbinevt} to generate the detector-plane spectra for the selected interval, \texttt{batphasyserr} to apply the systematic uncertainties, and \texttt{batdrmgen} to generate the corresponding response matrices. For the XRT data, we directly use the processed spectral products provided by the Swift/XRT spectral repository\footnote{\url{https://www.swift.ac.uk/xrt_spectra}}, including the source and background spectra, the response matrix files, and the ancillary response files for the same time intervals.

\begin{figure}[htbp]
\centering
\begin{minipage}[b]{0.3\textwidth}
  \centering
  \includegraphics[width=\linewidth]{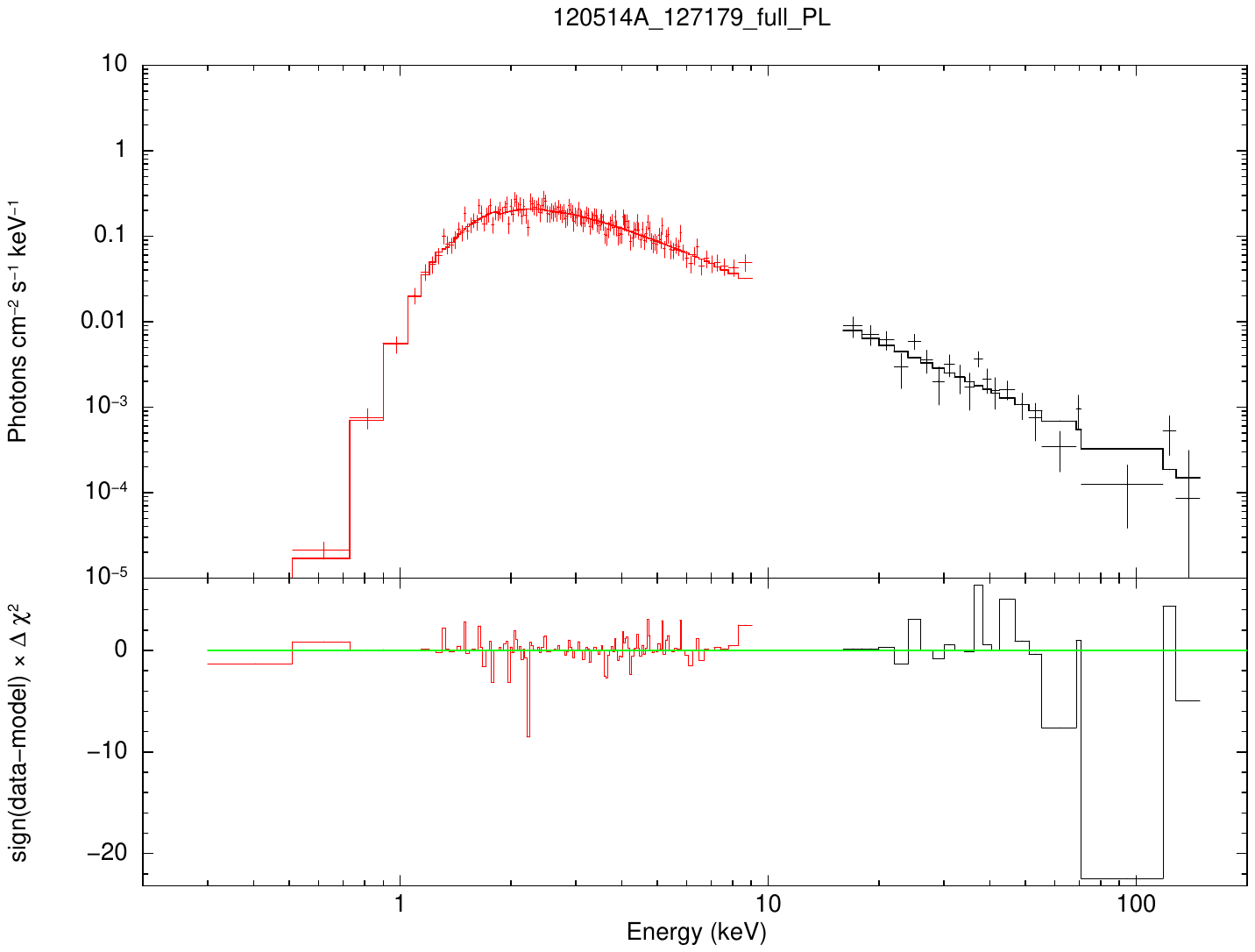}
\end{minipage}
\hfill
\begin{minipage}[b]{0.3\textwidth}
  \centering
  \includegraphics[width=\linewidth]{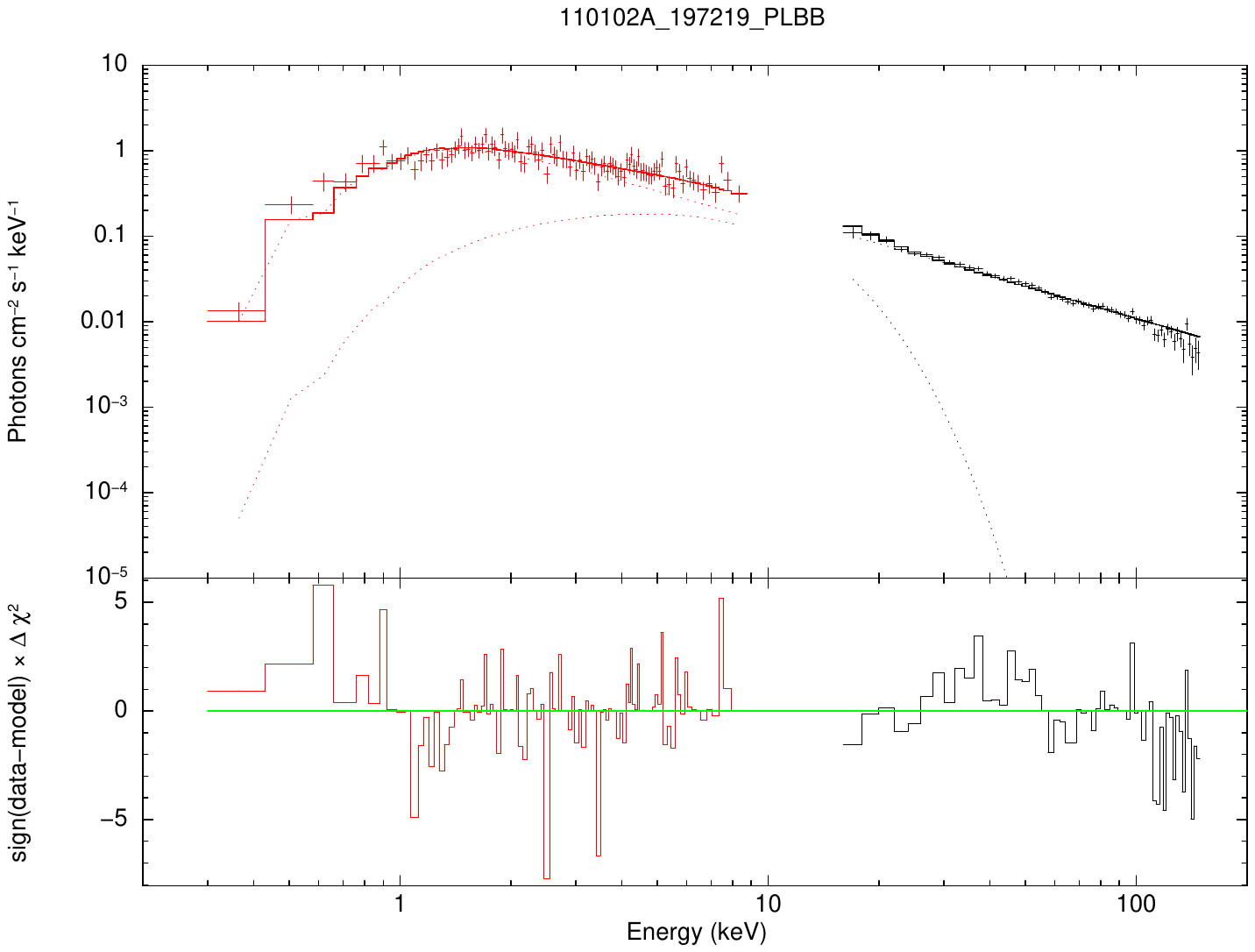}
\end{minipage}
\hfill
\begin{minipage}[b]{0.3\textwidth}
  \centering
  \includegraphics[width=\linewidth]{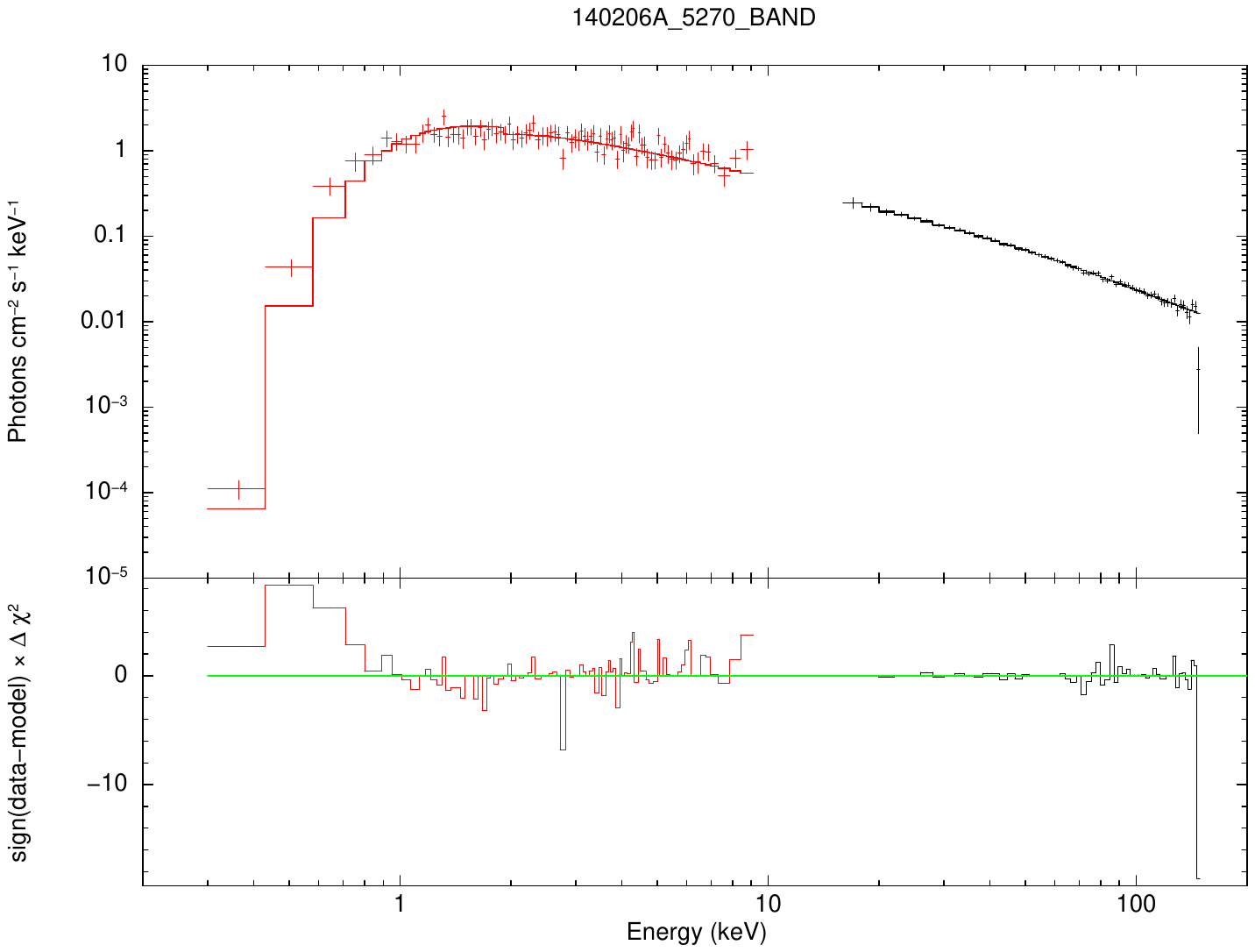}
\end{minipage}
\caption{Examples of joint spectral fitting with absorbed PL (left), PL+BB (middle), and Band function (right) models for representative pulse--flare pairs. The complete figure set (50 images) is available in the online journal.}
\label{Spectral fitting}
\end{figure}

The strategy and procedure of our spectral fits are described as follows.
\begin{enumerate}
    \item We first fit each joint BAT+XRT spectrum in the full 0.3--150\,keV band with an absorbed PL model. For pairs where the fit yields a reduced $\chi^2<1.2$, the absorbed PL model is adopted as the preferred model. Pairs with a reduced $\chi^2\geq1.2$ are passed to the next step.
    \item For the pairs with a reduced $\chi^2\geq1.2$, we adopt an absorbed PL+BB model. If this model provides a satisfactory fit with a reduced $\chi^2<1.2$, we adopt it as the preferred model. The remaining pairs are passed to the next step.
    \item For the pairs that still yield a reduced $\chi^2\geq1.2$, we adopt an absorbed cutoff power-law (CPL) model. If this model provides a satisfactory fit with a reduced $\chi^2<1.2$, we adopt it as the preferred model. The remaining pairs are passed to the next step.
    \item The final pairs are fitted with the absorbed Band function. For pairs where the Band fit yields a reduced $\chi^2<1.2$ with well-constrained parameters, we adopt it as the preferred model. For those where the reduced $\chi^2$ remains $\geq1.2$, we compare the fitting results among the PL, PL+BB, CPL, and Band models, and adopt the one that offers the best combination of fit quality and well-constrained parameters.
\end{enumerate}
Thus, across the 50 pairs, the best-fit models are: absorbed PL for 21 pairs, absorbed PL+BB for 20 pairs, absorbed CPL for 5 pairs, and absorbed Band function for 4 pairs. The best-fit model for each pair is listed in Table~\ref{tab:spectral_fit}. Representative spectral fits for different models are shown in Figure~\ref{Spectral fitting}.\footnote{Or at the link: \url{https://astro.gxu.edu.cn/info/1067/2551.htm}.}

\section{SOC Analysis}
The joint spectral fitting between gamma-ray pulses of BAT observed simultaneously with X-ray flares of XRT can provide indirect evidence or clues as to whether they are from the same origin. SOC, on the other hand, is another independent method to provide a natural framework for describing systems that exhibit scale-invariant, avalanche-like behavior under continuous driving. Within this scenario, a slowly driven system evolves toward a critical state in which small perturbations can trigger avalanches over a broad range of scales, naturally leading to power-law distributions in temporal and energetic observables \citep{1987PhRvL..59..381B}. \cite{2013NatPh...9..465W} found that the distributions of the durations, energies, and waiting times for X-ray ﬂares and solar ﬂares have similar statistical distributions, and suggest that they are from a similar physical mechanism (e.g., driven by magnetic reconnection).

In this section, we invoke the SOC method to examine the frequency distributions of the temporal parameters (e.g., waiting time $t_{\rm wait}$, rising time $t_{\rm rise}$, peak time $t_{\rm peak}$, decay time $t_{\rm decay}$, and duration $t_{\rm duration}$) for both X-ray flares and the simultaneously observed gamma-ray pulses. In general, the differential distribution can be described by a threshold power-law distribution \citep[also called a generalized Pareto type II distribution;][]{2015ApJ...814...19A} 
\begin{equation}
\frac{dN}{dx}\propto (x+x_{0})^{-\alpha_{x}},
\label{eq_diff}
\end{equation}
where $x_0$ is a constant introduced to account for threshold effects, such as incomplete sampling below $x_0$ and possible background contamination, and $\alpha_x$ is the power-law slope of the distribution. The cumulative distribution function corresponding to Equation~(\ref{eq_diff}) can therefore be written as ($\alpha_x \neq 1$)
\begin{equation}
N_{\mathrm{cum}} (>x) = A+B\times(x+x_{0})^{1-\alpha_x},
\label{eq_cum}
\end{equation}
where $N_{\mathrm{cum}}(>x)$ is the cumulative number of events with parameter values greater than $x$, and $A$ and $B$ are constants. We fit the cumulative distributions of the five temporal parameters by using Equation~(\ref{eq_cum}).

The uncertainty of the cumulative distribution in a given bin, $i$, was estimated as $\sigma_{\mathrm{cum}, i}=\sqrt{N_{\mathrm{cum},i}}$, where $N_{\mathrm{cum},i}$ is the number of events in the $i$th bin. The fits were performed using a MCMC method, which is equivalent to a maximum likelihood estimation under uniform priors. The goodness of fit was assessed with the reduced chi-square statistic for the cumulative distribution function \citep{2015ApJ...814...19A}:
\begin{equation}
\chi^2_{\mathrm{cum,\nu}} = \frac{1}{n_x - n_{\mathrm{par}}} \sum_{i=1}^{n_x} \frac{\left[N_{\mathrm{cum,th}}(x_i) - N_{\mathrm{cum,obs}}(x_i)\right]^2}{\sigma_{\mathrm{cum},i}^2},
\end{equation}
where $n_x$ is the number of logarithmic bins, $n_{\mathrm{par}}$ the number of free parameters, $N_{\mathrm{cum,obs}}(x_i)$ the observed values, and $N_{\mathrm{cum,th}}(x_i)$ the corresponding theoretical values of the cumulative distribution. Uncertainties in the fitted power-law indices were estimated from the marginal posterior distributions derived from the MCMC.

\begin{figure}[htbp]
\centering
\begin{minipage}[b]{0.3\textwidth}
  \centering
  \includegraphics[width=\linewidth]{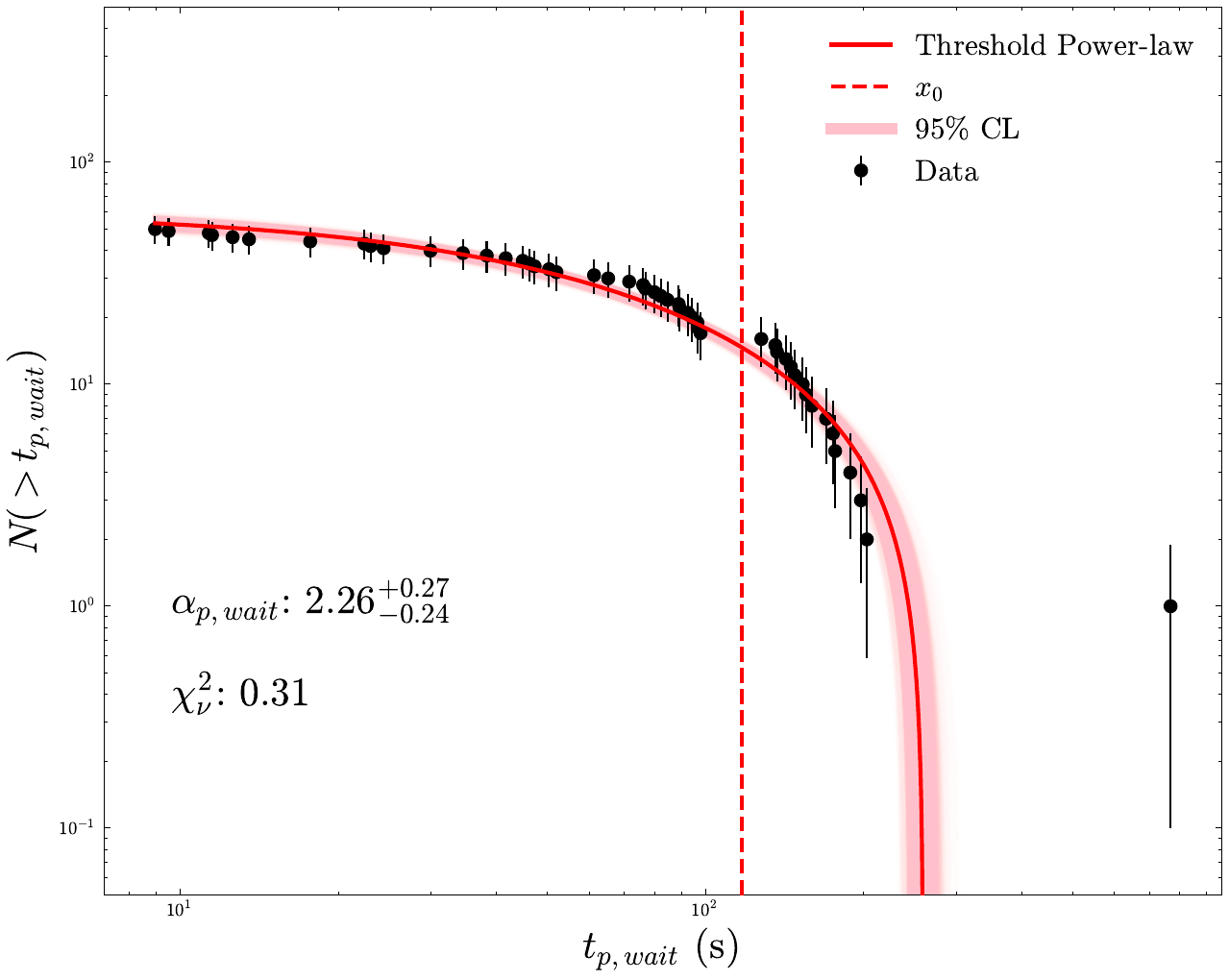}
\end{minipage}
\hfill
\begin{minipage}[b]{0.3\textwidth}
  \centering
  \includegraphics[width=\linewidth]{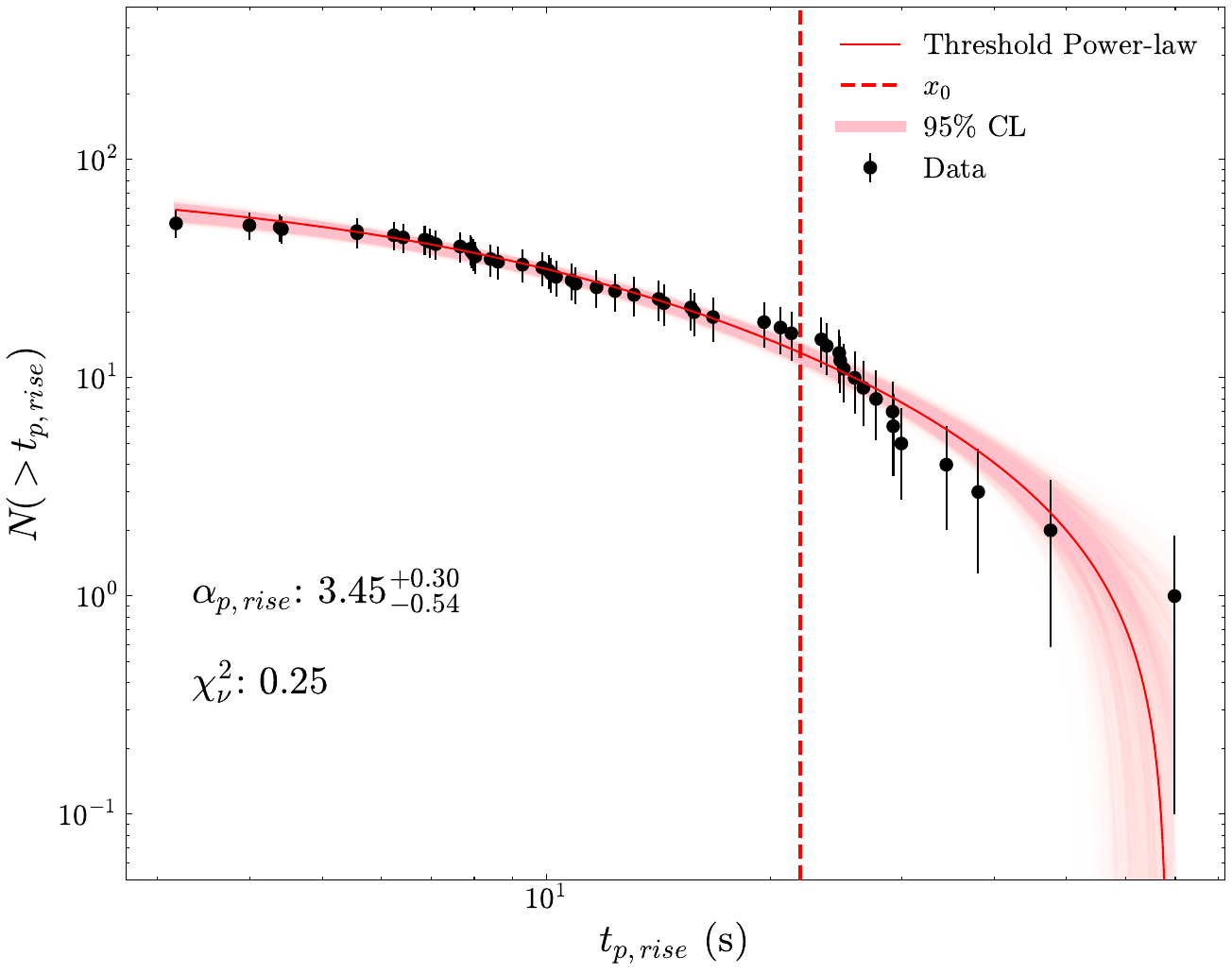}
\end{minipage}
\hfill
\begin{minipage}[b]{0.3\textwidth}
  \centering
  \includegraphics[width=\linewidth]{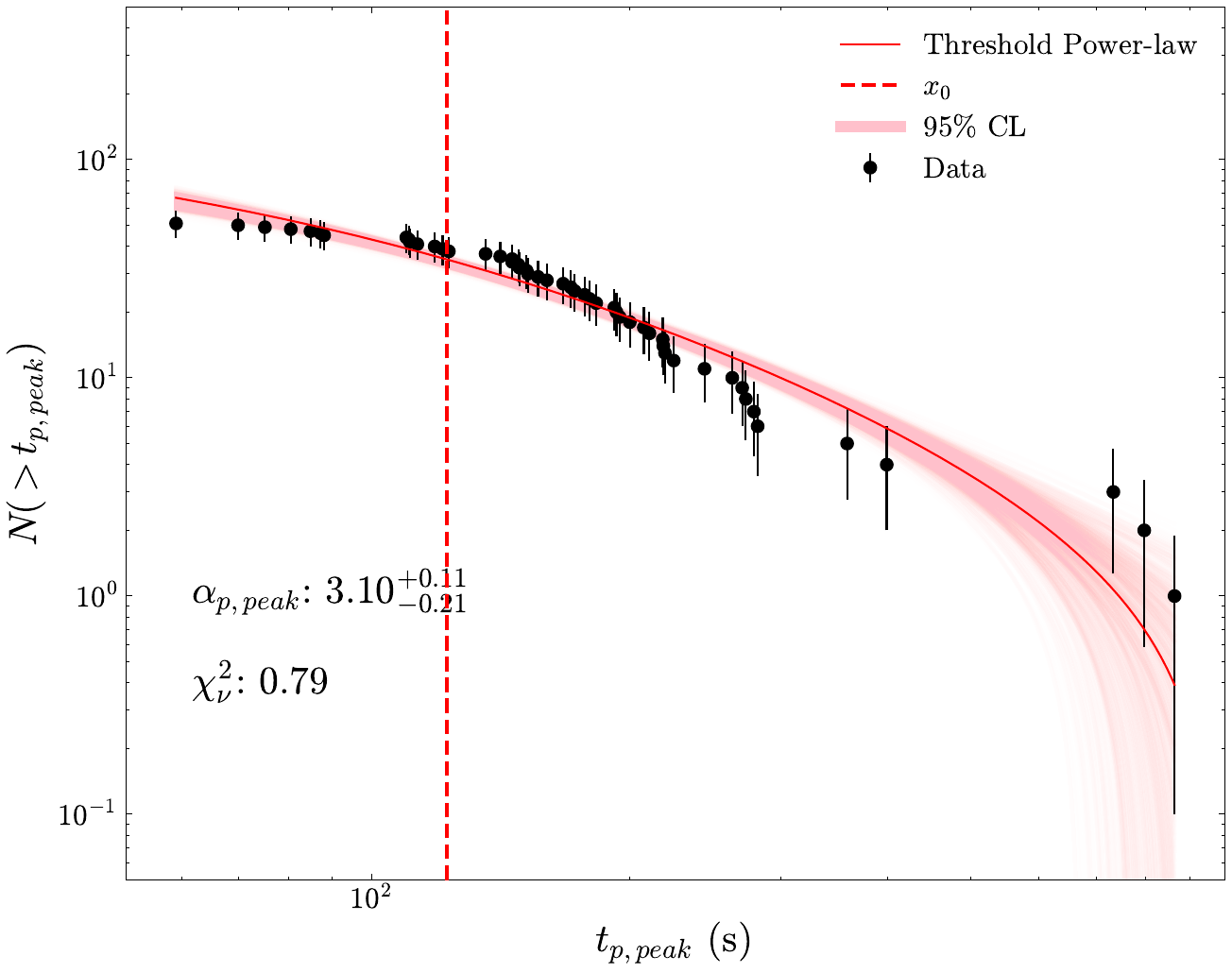}
\end{minipage}

\vspace{0.5em}

\begin{minipage}[b]{0.3\textwidth}
  \centering
  \includegraphics[width=\linewidth]{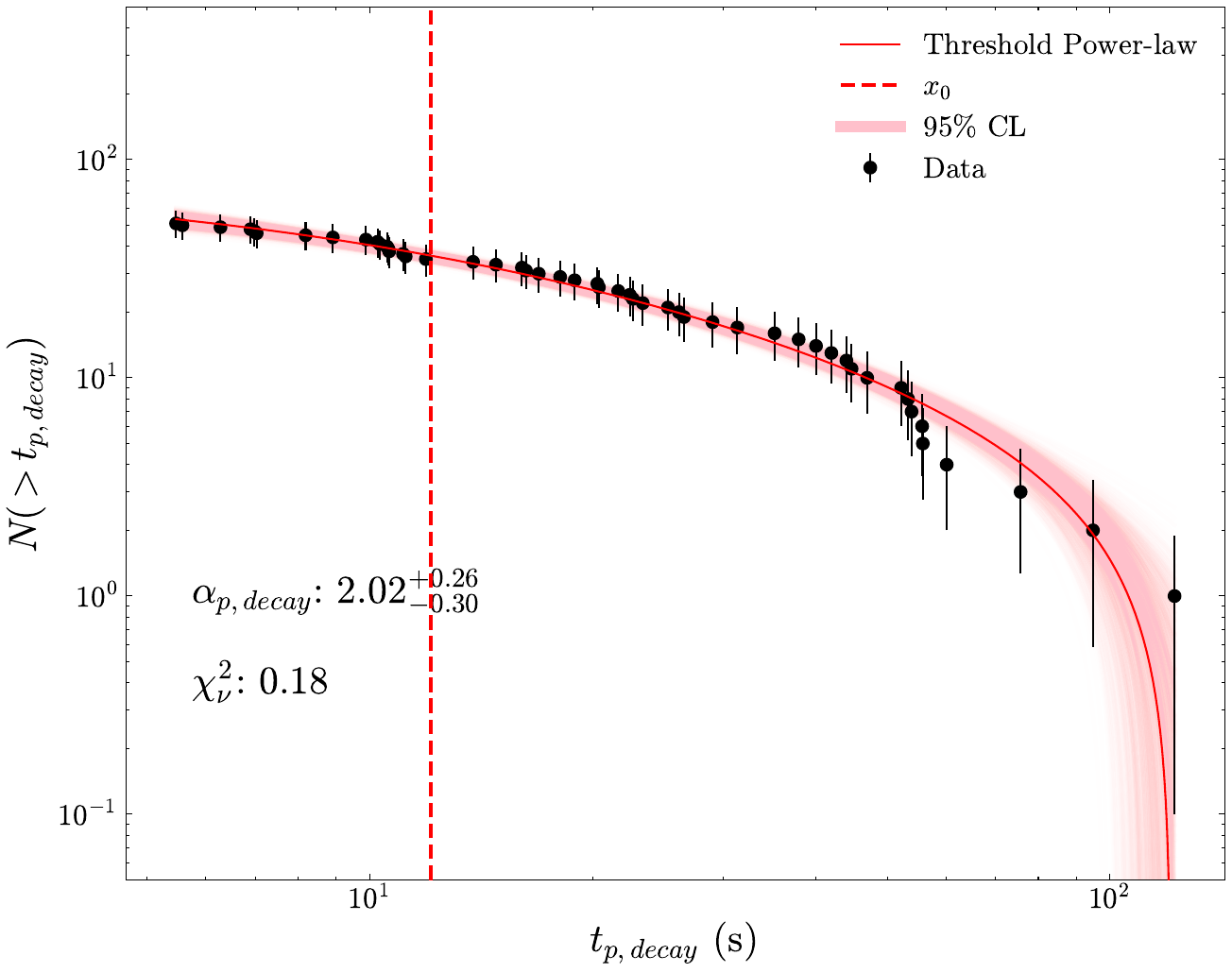}
\end{minipage}
\hfill
\begin{minipage}[b]{0.3\textwidth}
  \centering
  \includegraphics[width=\linewidth]{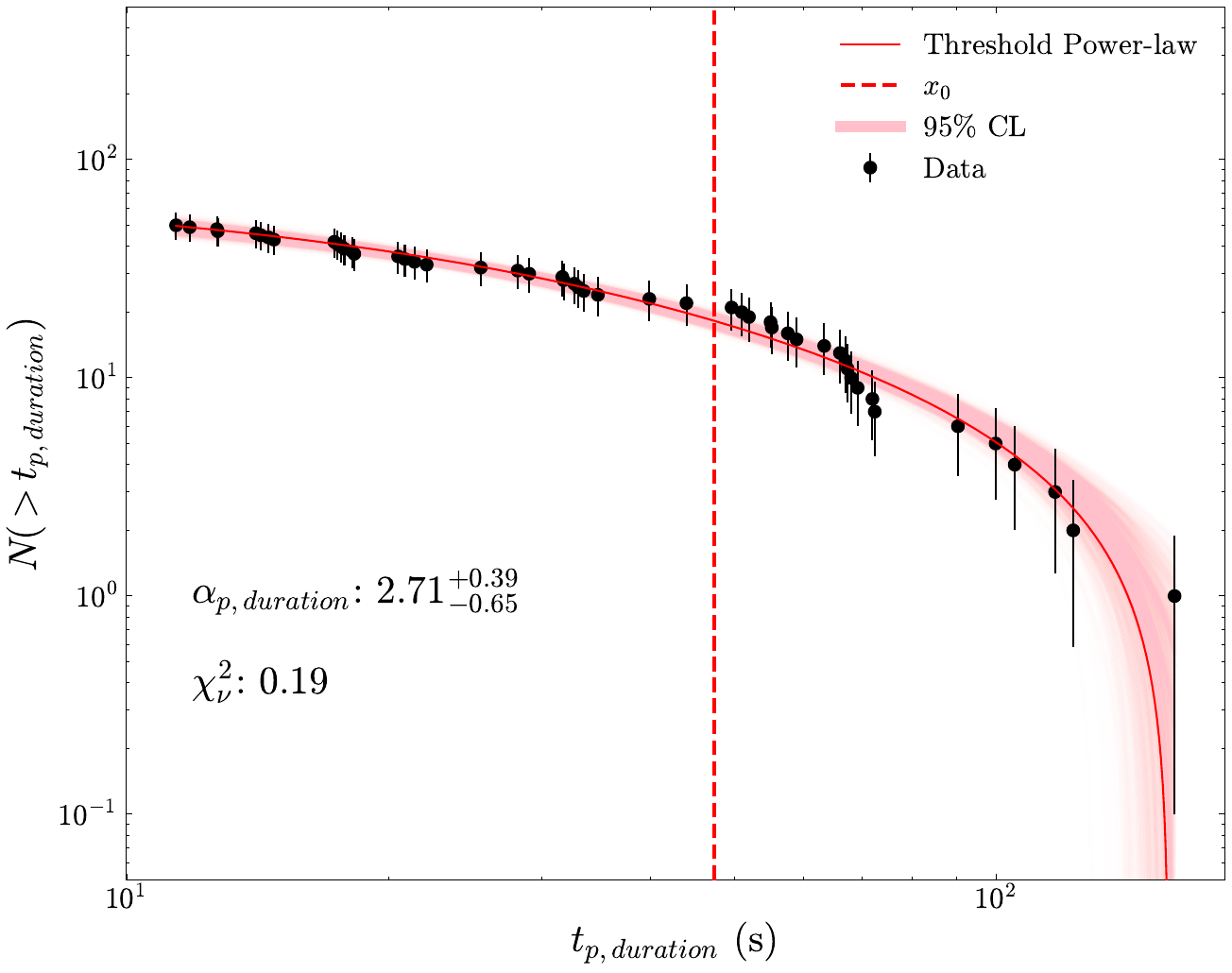}
\end{minipage}
\hfill
\begin{minipage}[b]{0.3\textwidth}
  \centering
  \includegraphics[width=\linewidth]{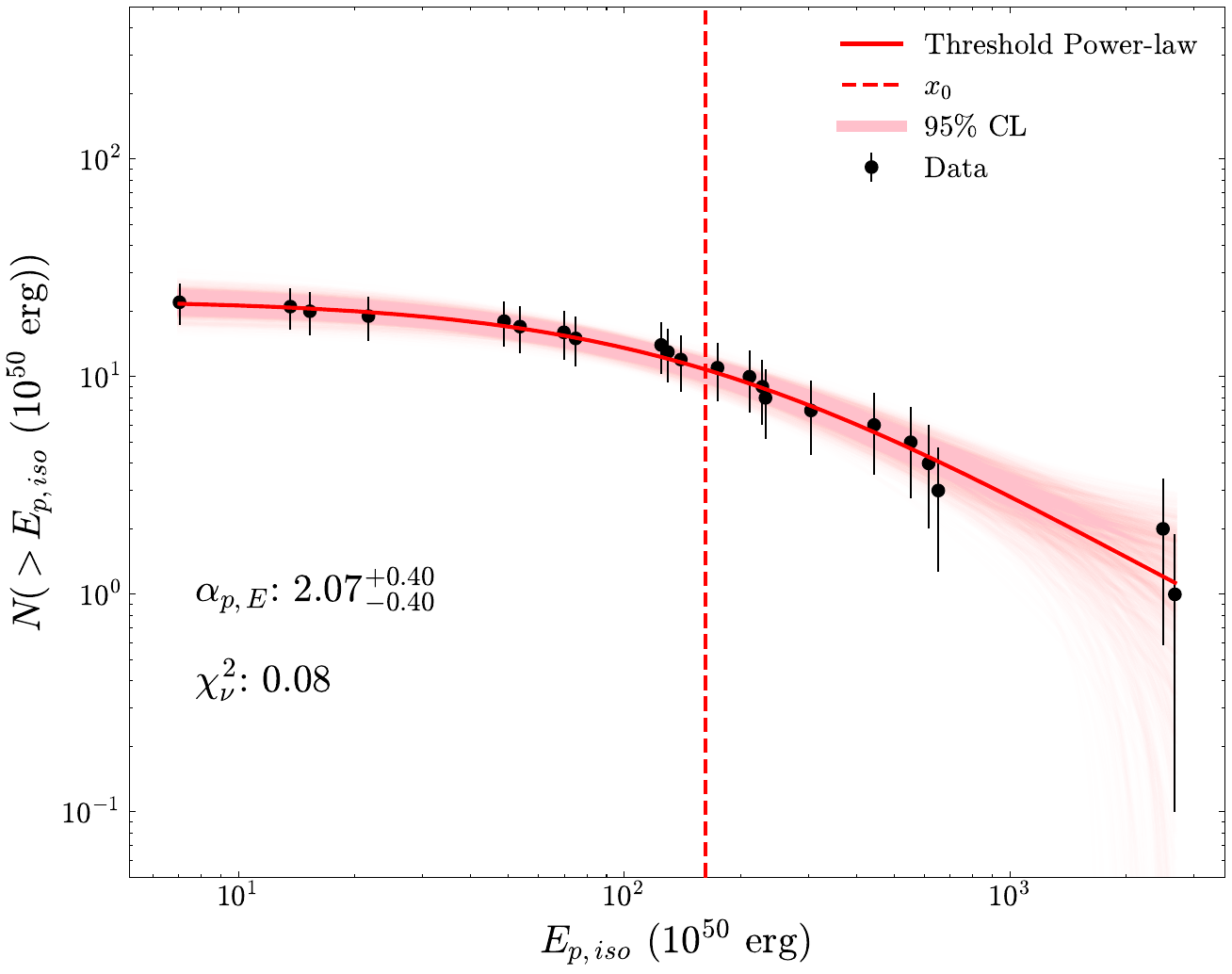}
\end{minipage}
\caption{Cumulative distributions of temporal parameters and isotropic energy for prompt $\gamma$-ray pulses. The black points show the observed distributions, and the red solid lines represent the best-fit power-law models. The power-law indices $\alpha$ are indicated in each panel.}
\label{SOC distributions BAT}
\end{figure}

\begin{figure}[htbp]
\centering
\begin{minipage}[b]{0.3\textwidth}
  \centering
  \includegraphics[width=\linewidth]{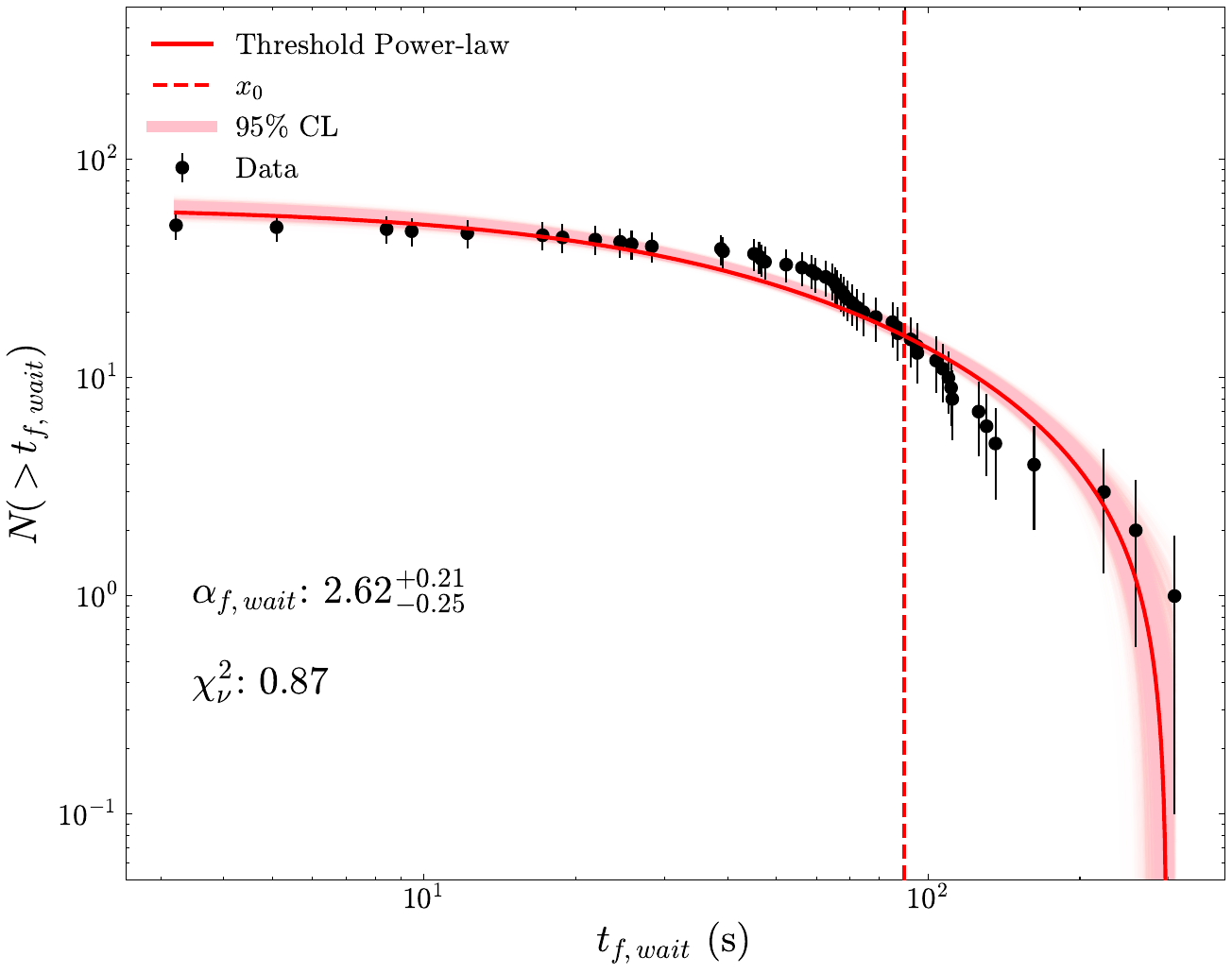}
\end{minipage}
\hfill
\begin{minipage}[b]{0.3\textwidth}
  \centering
  \includegraphics[width=\linewidth]{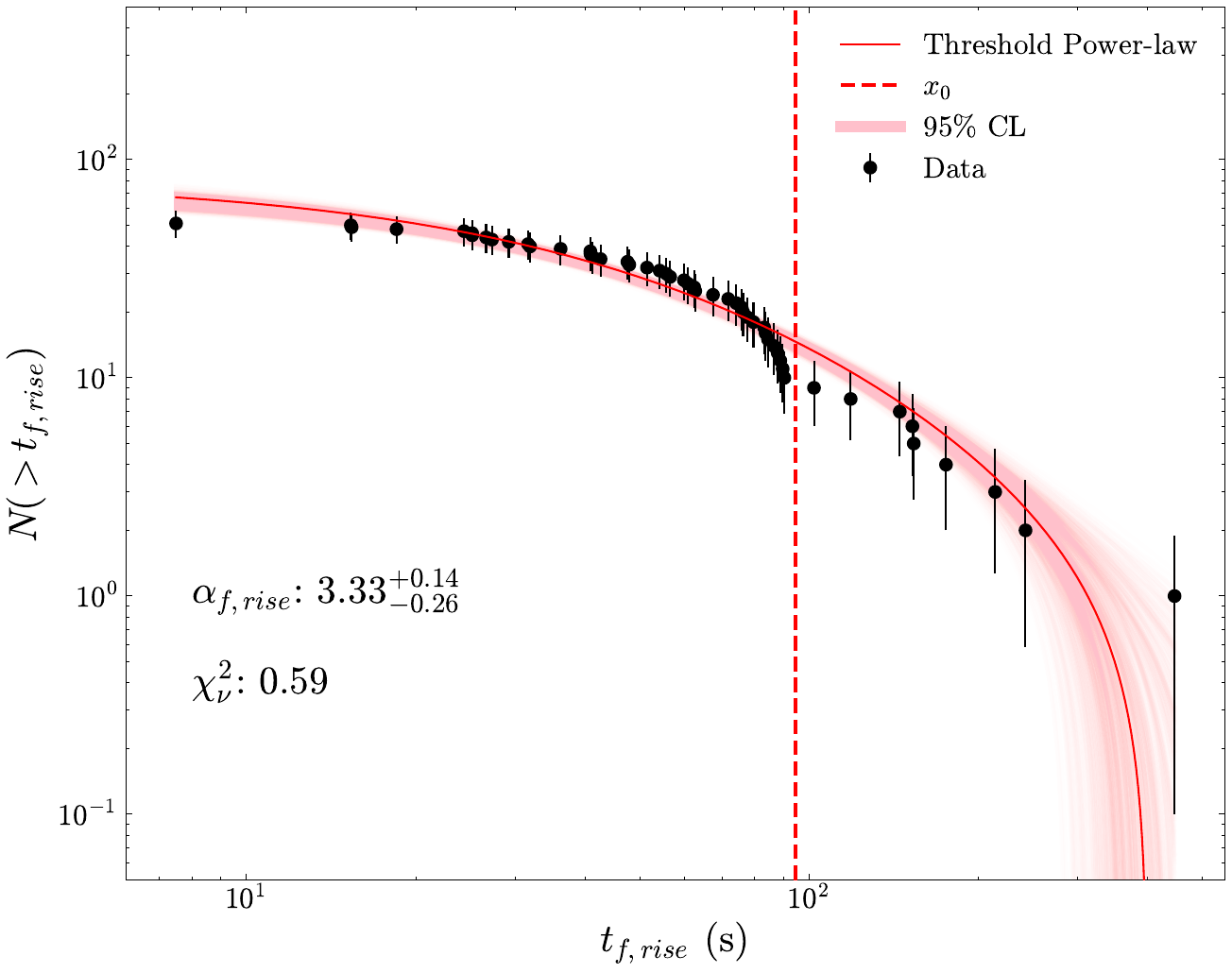}
\end{minipage}
\hfill
\begin{minipage}[b]{0.3\textwidth}
  \centering
  \includegraphics[width=\linewidth]{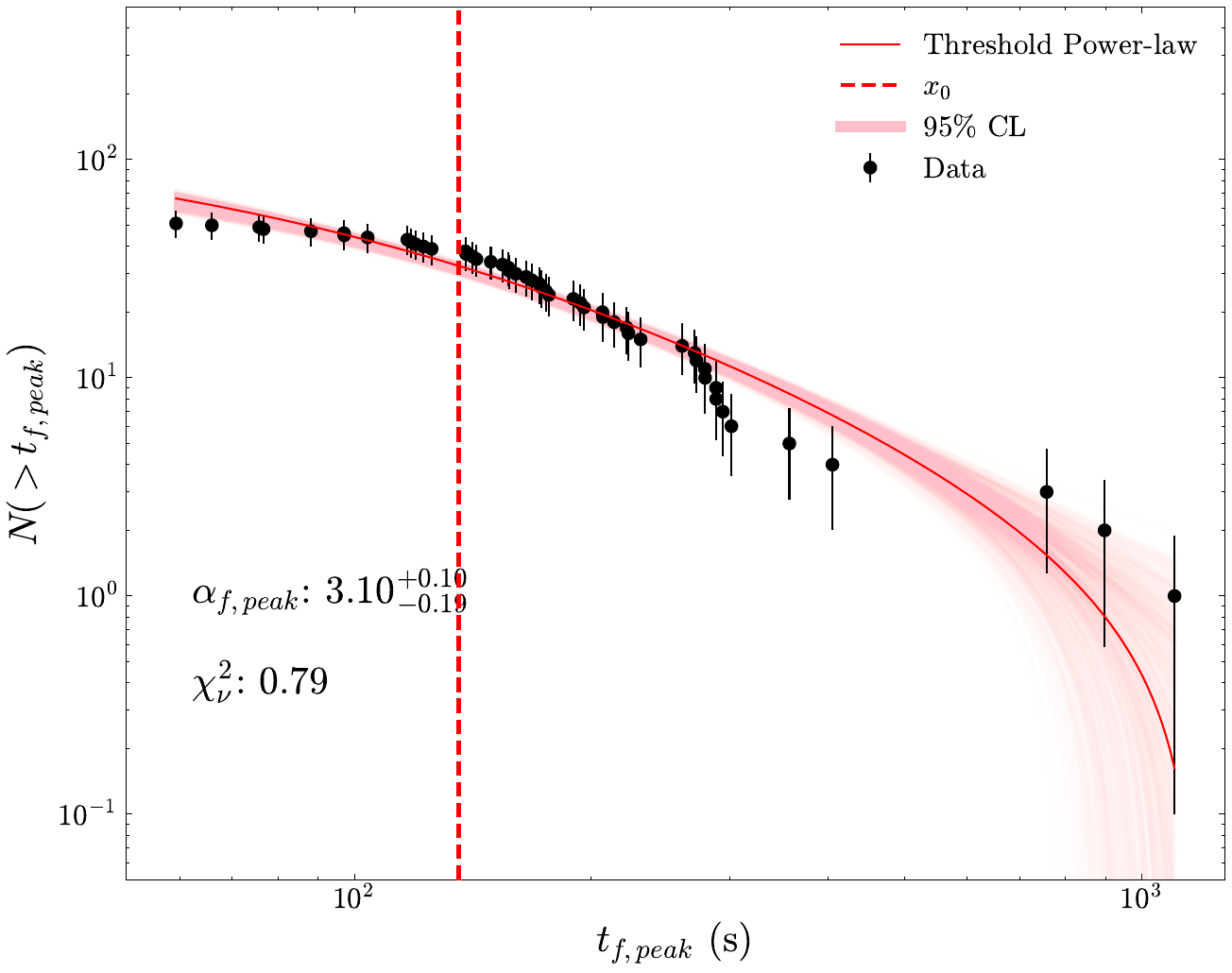}
\end{minipage}

\vspace{0.5em}

\begin{minipage}[b]{0.3\textwidth}
  \centering
  \includegraphics[width=\linewidth]{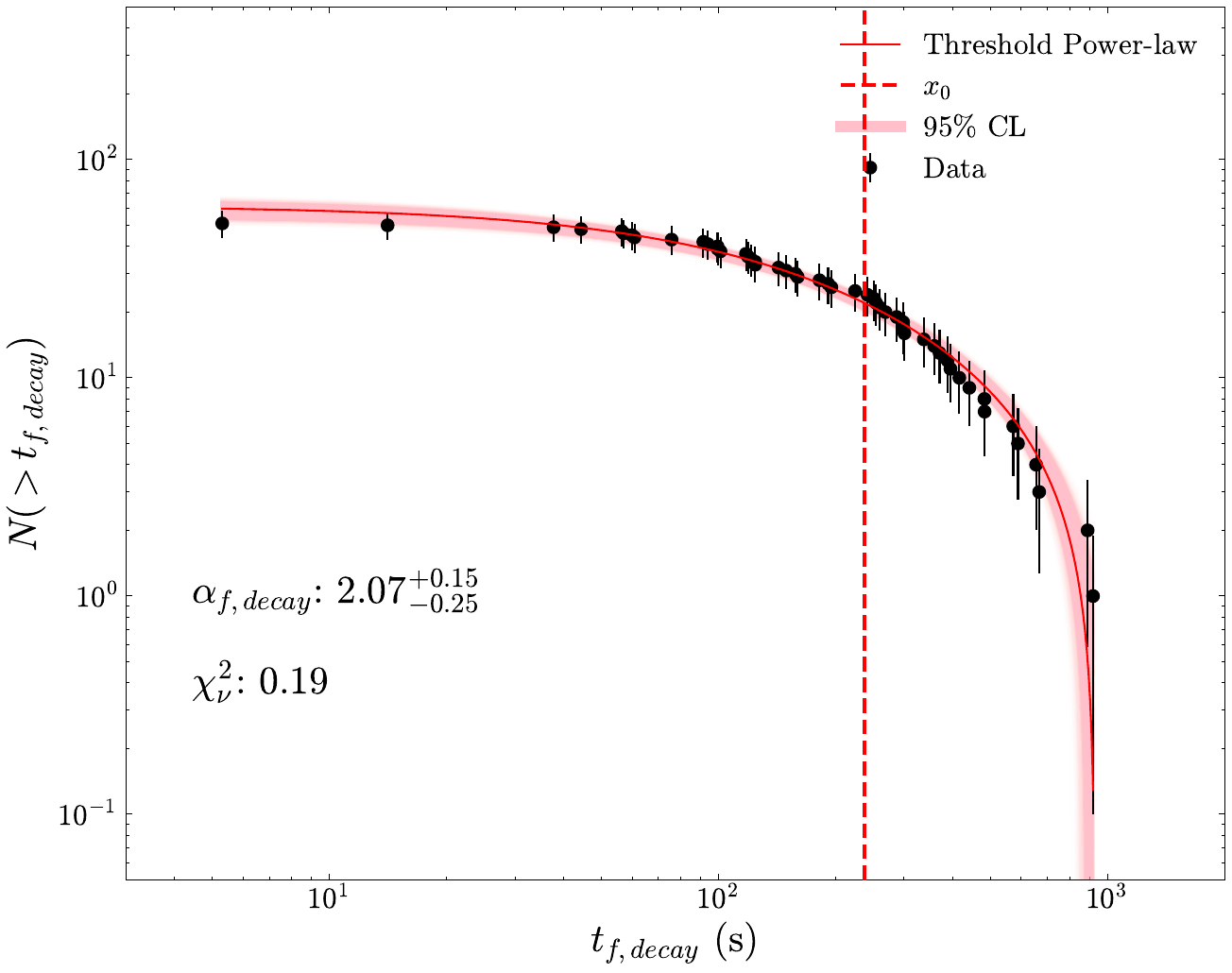}
\end{minipage}
\hfill
\begin{minipage}[b]{0.3\textwidth}
  \centering
  \includegraphics[width=\linewidth]{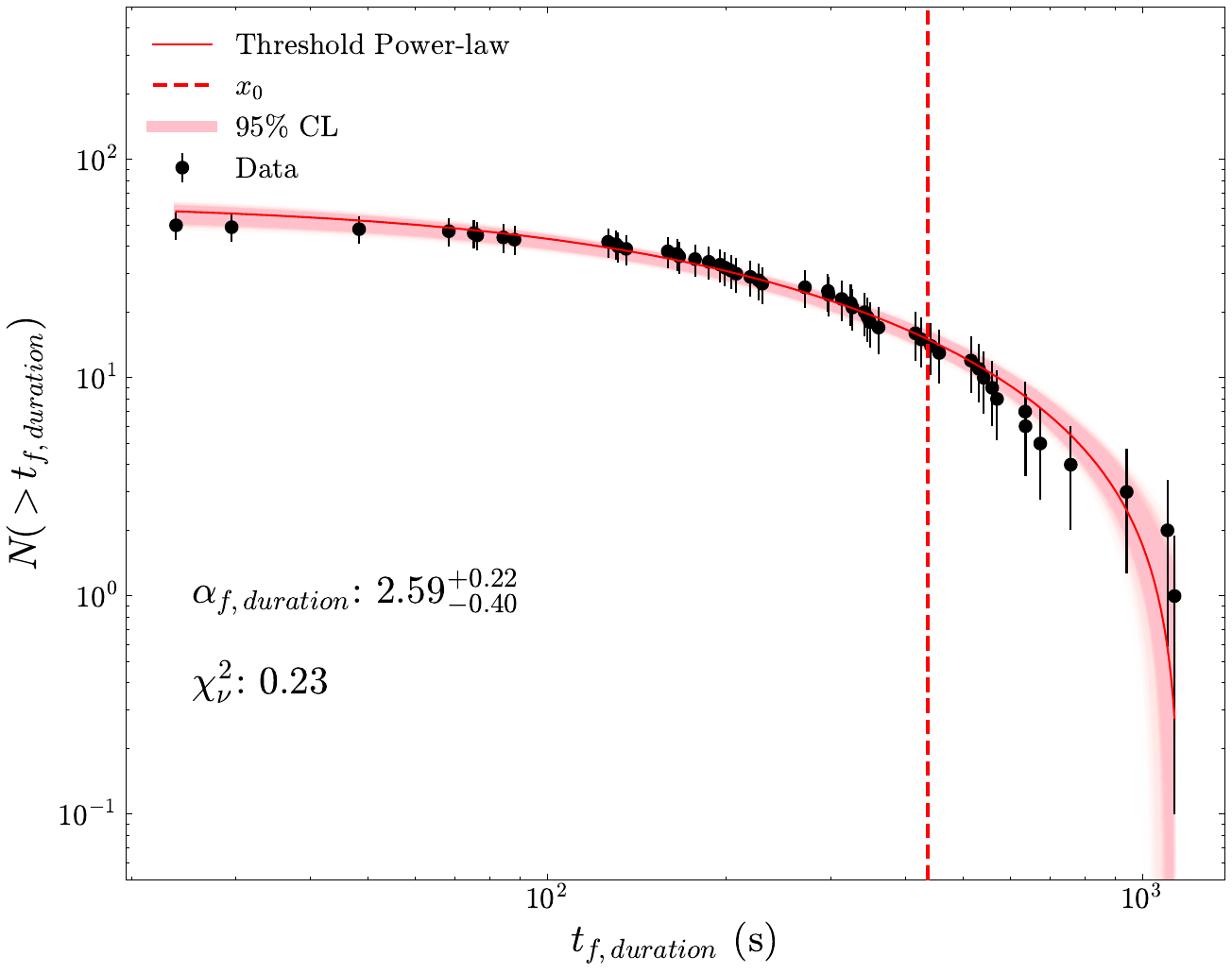}
\end{minipage}
\hfill
\begin{minipage}[b]{0.3\textwidth}
  \centering
  \includegraphics[width=\linewidth]{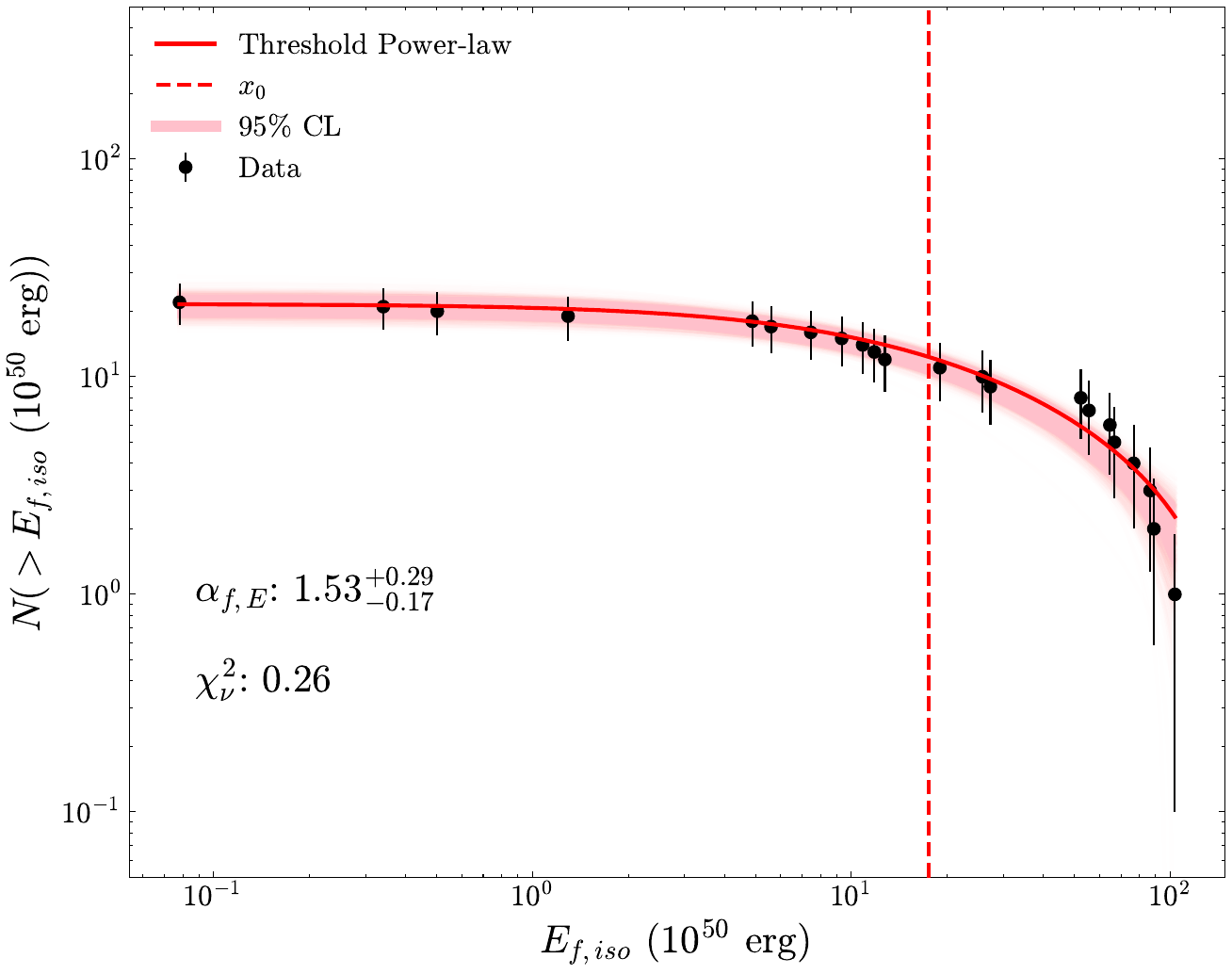}
\end{minipage}
\caption{Cumulative distributions of temporal parameters and isotropic energy for X-ray flares. The black points show the observed distributions, and the red solid lines represent the best-fit power-law models. The power-law indices $\alpha$ are indicated in each panel.}
\label{SOC distributions XRT}
\end{figure}

\begin{figure}[htbp]
  \centering
  \includegraphics[width=0.5\linewidth]{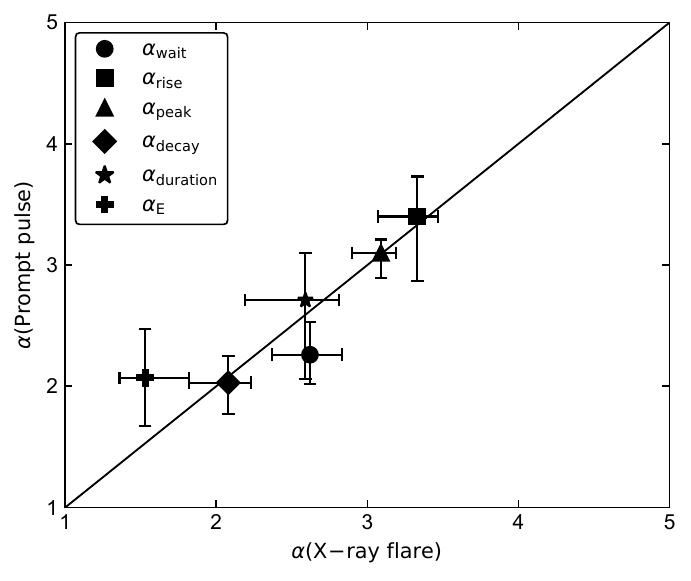}
  \caption{Comparisons of the power-law indices derived from the SOC analysis for X-ray flares and prompt $\gamma$-ray pulses. The solid line presents the case where the power-law index of a given temporal parameter is identical for X-ray flares and prompt pulses.}
  \label{fig:flare_pulse_compare}
\end{figure}

For gamma-ray pulses in the prompt emission, the cumulative distributions of waiting time, rising time, peak time, decay time, and duration are all well fitted by a power-law function, with best-fit indices of $\alpha_{\rm p,wait}=2.26^{+0.27}_{-0.24}$ (after excluding a pulse with an anomalously long waiting time of ${\sim}700$\,s), $\alpha_{\rm p,rise}=3.40^{+0.33}_{-0.53}$, $\alpha_{\rm p,peak}=3.10^{+0.11}_{-0.21}$, $\alpha_{\rm p,decay}=2.03^{+0.22}_{-0.26}$, and $\alpha_{\rm p,duration}=2.71^{+0.39}_{-0.65}$, respectively. Similarly, for X-ray flares in the afterglow phase, the corresponding distributions also follow power-law behaviors, with $\alpha_{\rm f,wait}=2.62^{+0.21}_{-0.25}$, $\alpha_{\rm f,rise}=3.33^{+0.14}_{-0.26}$, $\alpha_{\rm f,peak}=3.09^{+0.10}_{-0.19}$, $\alpha_{\rm f,decay}=2.08^{+0.15}_{-0.26}$, and $\alpha_{\rm f,duration}=2.59^{+0.22}_{-0.40}$, respectively. The cumulative distributions of the five temporal parameters are shown in Figure~\ref{SOC distributions BAT} for prompt gamma-ray pulses and Figure~\ref{SOC distributions XRT} for X-ray flares. Moreover, we also examine the cumulative distribution of the isotropic energy $E_{\rm iso}$ for both prompt gamma-ray pulses and X-ray flares in the subsample with measured redshifts. The energy distribution of the prompt pulses follows a power-law function with an index of $\alpha_{\rm p,E}=2.07^{+0.40}_{-0.40}$, and that of the X-ray flares yields $\alpha_{\rm f,E}=1.53^{+0.29}_{-0.17}$, as shown in Figures~\ref{SOC distributions BAT} and~\ref{SOC distributions XRT}, respectively. We further compare the inferred power-law indices for both gamma-ray pulses and X-ray flares to test whether the two components exhibit consistent SOC behavior. It is found that the inferred indices of rising time, peak time, and decay time for X-ray flares are roughly consistent with that of prompt pulses. The power-law index of waiting time between X-ray flares and prompt pulses is slightly different, but is still in the reasonable range. Except for the waiting time, which shows a moderate offset of $\Delta\alpha\sim0.36$, the other four parameters exhibit much smaller differences ($\Delta\alpha\lesssim0.12$) between the two components (see Figure~\ref{fig:flare_pulse_compare}).

\section{Conclusions and Discussion} 
The prompt emission and X-ray flares are thought to arise from a common physical origin, based on their similar temporal properties and analogous spectral evolutionary behaviors \citep{2023ApJ...948...23Z, 2023ApJS..269....2S, 2025ApJ...989..113M, 2025ApJ...995..201Y}. Alternatively, the SOC framework has been applied to the frequency distributions of energy and waiting time for GRB X-ray flares \citep{2013NatPh...9..465W} and prompt gamma-ray pulses \citep{2021FrPhy..1614501L}, respectively. However, SOC studies focusing on prompt pulses and X-ray flares observed simultaneously in the same GRBs are still lacking. Motivated by this issue, we systematically searched the Swift/BAT and XRT archive data and constructed a sample of 38 GRBs containing 50 simultaneous pulse--flare pairs that satisfy our selection criteria. We find that the morphological light curves can be broadly classified into three categories, according to whether X-ray flares are present during the quiescent intervals of the prompt emission.

First, we invoked an empirical function to fit the BAT and XRT data of each GRB in our sample. Based on the reconstructed light curves derived from these empirical fits, we extracted five temporal parameters for each prompt pulse and X-ray flare, namely the waiting time, rising time, peak time, decay time, and duration. Second, we performed joint spectral fits to the overlapping time intervals between X-ray flares and prompt pulses, and found that the spectra can be well described by an absorbed PL, PL+BB, CPL, or Band-function model. This indicates that the X-ray flares and the simultaneously observed gamma-ray pulses are likely to share a similar physical origin.

Then, we investigate the frequency distributions of five temporal parameters, namely the waiting time, rising time, peak time, decay time, and duration, for both X-ray flares and the simultaneously observed gamma-ray pulses by using the SOC framework. We find that all five parameters follow power-law distributions in both components. The isotropic energy $E_{\rm iso}$, computed for the subsample with measured redshifts, also follows a power-law distribution in both components. More importantly, the corresponding power-law indices for X-ray flares are broadly consistent with those for gamma-ray pulses. These results further support the idea that X-ray flares and prompt gamma-ray pulses share the same physical origin. In this sense, the SOC analysis provides independent evidence that X-ray flares and the simultaneously observed gamma-ray pulses share a common origin.

Among the five temporal parameters, the waiting-time indices exhibit the largest offset between the two components ($\alpha_{\rm p,wait}=2.26^{+0.27}_{-0.24}$ vs. $\alpha_{\rm f,wait}=2.62^{+0.21}_{-0.25}$, $\Delta\alpha\sim0.36$), while the other four parameters all show much smaller differences ($\Delta\alpha\lesssim0.12$). There are two possible ways to result in such a larger offset in the waiting-time indices. One is possible from the different definitions of waiting time for both pulses and flares, i.e., for the first flare in each GRB, the waiting time is taken as its start time, and more than half of the flares in our sample are such first flares, which tends to bias the flare waiting-time distribution. The other one is possible from a single long-waiting-time pulse outlier (${\sim}700$\,s) for the pulse sample (see Figure~\ref{SOC distributions BAT}) and the limited sample size. Regarding the isotropic energy, the cumulative $E_{\rm iso}$ distribution computed for the subsample with measured redshifts (19 GRBs, 22 pairs) yields $\alpha_{\rm p,E}=2.07^{+0.40}_{-0.40}$ and $\alpha_{\rm f,E}=1.53^{+0.29}_{-0.17}$ for prompt pulses and X-ray flares, respectively, corresponding to a difference of $\Delta\alpha\sim0.54$. Such a difference may mainly arise from the relatively large uncertainties of the inferred $E_{\rm p,iso}$, which rely on the fluence derived from the spectral fitting of the prompt pulses, as well as the very limited number of $E_{\rm iso}$ measurements in the redshift subsample (22 pulse--flare pairs). Therefore, given the large uncertainties and the limited sample size, this difference does not affect our conclusion that the X-ray flares and the prompt gamma-ray pulses share a common physical origin, which is primarily supported by the consistent temporal power-law indices.

To further explore the physical implications, we compute the rest-frame power-law indices for the subsample with measured redshifts (19 GRBs including 22 pairs). By correcting the redshift effects, the indices of the five temporal parameters for prompt pulses are $\alpha_{\rm p,wait,rest}=1.83^{+0.48}_{-0.40}$, $\alpha_{\rm p,rise,rest}=2.01^{+0.67}_{-0.53}$, $\alpha_{\rm p,peak,rest}=2.58^{+0.73}_{-0.73}$, $\alpha_{\rm p,decay,rest}=2.09^{+0.64}_{-0.57}$, and $\alpha_{\rm p,duration,rest}=2.18^{+0.73}_{-0.64}$; those of X-ray flares are $\alpha_{\rm f,wait,rest}=2.51^{+0.48}_{-0.51}$, $\alpha_{\rm f,rise,rest}=1.54^{+0.32}_{-0.16}$, $\alpha_{\rm f,peak,rest}=2.50^{+0.71}_{-0.68}$, $\alpha_{\rm f,decay,rest}=1.80^{+0.51}_{-0.42}$, and $\alpha_{\rm f,duration,rest}=1.90^{+0.55}_{-0.46}$. For a fractal-diffusive SOC model, \citet{2012A&A...539A...2A, 2014ApJ...782...54A} predicted power-law distributions for the duration and peak flux with indices $\alpha_T = (1+S)\beta/2$ and $\alpha_P = 2-1/S$, where $S=1,2,3$ are the Euclidean space dimensions, and $\beta$ is the diffusive spreading exponent. By adopting the classical diffusion $\beta=1$, the Euclidean dimension $S$ can be estimated from the temporal index via $S = 2\alpha-1$. If this is the case, the inferred $S$ values for individual parameters carry substantial uncertainties due to the limited redshift subsample (e.g., $S = 2.66^{+0.96}_{-0.80}$ for the pulse waiting time and $S = 2.08^{+0.64}_{-0.32}$ for the flare rise time), and all of them are broadly consistent with $S\sim3$ within their $1\sigma$ ranges. To obtain more robust estimates, we compute the mean rest-frame indices. The mean indices are $\langle\alpha_{\rm p,rest}\rangle=2.14$ for pulses and $\langle\alpha_{\rm f,rest}\rangle=2.05$ for flares, corresponding to $S\sim3.3$ and $S\sim3.1$, respectively. Both of them are consistent with $S=3$, the Euclidean dimension of a three-dimensional system, supporting a common physical origin and similar spatial configuration for both prompt pulses and X-ray flares.

We note that the power-law indices of those distributions of flares obtained in this work are different from those reported earlier in studies focusing on prompt-only or flare-only samples \citep{2013NatPh...9..465W, 2016ApJS..224...20Y, 2021FrPhy..1614501L, 2023ApJ...955L..34L}. For example, the cumulative waiting-time indices, rise-time indices, peak-time indices, decay-time indices, and duration for prompt pulses and X-ray flares in this work are larger than those of indices from \cite{2013NatPh...9..465W}, \cite{2016ApJS..224...20Y}, and \cite{2021FrPhy..1614501L}. Such discrepancies may be caused by fundamental differences in sample selection and the statistical analysis method of SOC. Our sample uniquely requires both multiple prompt pulses and a temporally overlapping X-ray flare from the same GRB, yielding 50 pulse--flare pairs from 38 GRBs. The number of both prompt pulses and X-ray flares is much less than that reported by \cite{2021FrPhy..1614501L} (e.g., 454 prompt-only pulses) and \cite{2016ApJS..224...20Y} (e.g., 468 only X-ray flares). This means that our sample is a subsample of previous studies, which can result in statistical differences. The other one is possible due to the adopted method. Previous studies with large samples directly fit the differential distribution, whereas we adopt the cumulative distribution due to our limited sample size.

One needs to note that the X-ray flares and the simultaneously observed gamma-ray pulses indeed share a common origin (e.g., the central engine reactivity). However, the physical origin of those X-ray flares that are not observed simultaneously with gamma-ray pulses remains under debate. For example, \cite{2006ApJ...636L..29P} and \cite{2017MNRAS.464.4399D} proposed that fragmentation in the accretion disk can reproduce the empirical statistical properties of the X-ray flares, which are not from the central engine reactivity. \cite{2017MNRAS.470.4925G, 2018MNRAS.478.4323G} proposed that the magnetar central engine, with the delayed onset of a propeller regime, could explain the observed X-ray flares. \cite{2007ApJ...670.1247W} proposed the jet bow shock and the relativistic jet itself to explain the early prompt emission and later X-ray flares. 

On the other hand, the origin of the X-ray flare, observed during the quiescent intervals of prompt emission, also remains uncertain. There are two possible ways to interpret such behavior. One possibility is that these X-ray flares still reflect continuous central engine activity, while the corresponding gamma-ray emission during the quiescent interval is too weak to be detected by current gamma-ray instruments, for example owing to the tip-of-the-iceberg effect \citep{2014MNRAS.442.1922L}.  The alternative is that X-ray flares should be from a different physical process which is unrelated to the central engine reactivity.

\section{Acknowledgments}
We acknowledge the use of the public data from the Swift/BAT and XRT Science Data Center, and the comments from the anonymous referee. This work is supported by the Natural Science Foundation of China (grant No. 12494574), the Natural Science Foundation of Guangxi (grant Nos. 2025GXNSFDA02850010, 2023GXNSFDA026007, and Guike LT2600640026), the Bagui Scholars Program (LHJ), and the Guangxi Talent Program (Highland of Innovation Talents).


\bibliographystyle{aasjournal}
\bibliography{references}

@ARTICLE{2025ApJ...988L..46L,
       author = {{Liu}, Xiao-Xuan and {L{\"u}}, Hou-Jun and {Chen}, Qiu-Hong and {Du}, Zhao-Wei and {Liang}, En-Wei},
        title = "{Neutron Star─White Dwarf Merger as One Possible Optional Source of Kilonova-like Emission: Implications for GRB 211211A}",
      journal = {\apjl},
         year = 2025,
        month = aug,
       volume = {988},
       number = {2},
          eid = {L46},
        pages = {L46},
          doi = {10.3847/2041-8213/adec83},
archivePrefix = {arXiv},
       eprint = {2507.04318},
 primaryClass = {astro-ph.HE},
       adsurl = {https://ui.adsabs.harvard.edu/abs/2025ApJ...988L..46L}
}

@ARTICLE{2025ApJ...989..113M,
       author = {{Ma}, Yinuo and {Gao}, He},
        title = "{Comparative Statistical Analysis of Prompt and Afterglow X-Ray Flares in Gamma-Ray Bursts: Insights into Extended Central Engine Activity}",
      journal = {\apj},
         year = 2025,
        month = aug,
       volume = {989},
       number = {1},
          eid = {113},
        pages = {113},
          doi = {10.3847/1538-4357/ade4cb},
archivePrefix = {arXiv},
       eprint = {2510.18317},
 primaryClass = {astro-ph.HE},
       adsurl = {https://ui.adsabs.harvard.edu/abs/2025ApJ...989..113M}
}

@ARTICLE{2025A&A...693A.290Z,
       author = {{Zhang}, Wen-Long and {Yi}, Shuang-Xi and {Zou}, Yuan-Chuan and {Wang}, Fa-Yin and {Li}, Cheng-Kui and {Xie}, Sheng-Lun},
        title = "{Self-organized critical characteristics of teraelectronvolt photons from GRB 221009A}",
      journal = {\aap},
         year = 2025,
        month = jan,
       volume = {693},
          eid = {A290},
        pages = {A290},
          doi = {10.1051/0004-6361/202453174},
archivePrefix = {arXiv},
       eprint = {2412.16052},
 primaryClass = {astro-ph.HE},
       adsurl = {https://ui.adsabs.harvard.edu/abs/2025A&A...693A.290Z}
}

@ARTICLE{2023ApJS..269....2S,
       author = {{Saji}, Joseph and {Iyyani}, Shabnam and {Mazde}, Kratika},
        title = "{Statistical Analysis of Long GRBs' Prompt Emission and X-Ray Flares: Multivariate Clustering and Correlations}",
      journal = {\apjs},
         year = 2023,
        month = nov,
       volume = {269},
       number = {1},
          eid = {2},
        pages = {2},
          doi = {10.3847/1538-4365/acf4ef},
archivePrefix = {arXiv},
       eprint = {2309.07224},
 primaryClass = {astro-ph.HE},
       adsurl = {https://ui.adsabs.harvard.edu/abs/2023ApJS..269....2S}
}

@ARTICLE{2023RAA....23k5013Z,
       author = {{Zhang}, Wen-Long and {Li}, Xiu-Juan and {Yang}, Yu-Peng and {Yi}, Shuang-Xi and {Li}, Cheng-Kui and {Tang}, Qing-Wen and {Qin}, Ying and {Wang}, Fa-Yin},
        title = "{Statistical Properties of X-Ray Bursts from SGR J1935+2154 Detected by Insight-HXMT}",
      journal = {Research in Astronomy and Astrophysics},
         year = 2023,
        month = nov,
       volume = {23},
       number = {11},
          eid = {115013},
        pages = {115013},
          doi = {10.1088/1674-4527/acf979},
archivePrefix = {arXiv},
       eprint = {2310.06299},
 primaryClass = {astro-ph.HE},
       adsurl = {https://ui.adsabs.harvard.edu/abs/2023RAA....23k5013Z}
}

@ARTICLE{2023ApJ...955L..34L,
       author = {{Li}, Xiu-Juan and {Yang}, Yu-Peng},
        title = "{Signatures of the Self-organized Criticality Phenomenon in Precursors of Gamma-Ray Bursts}",
      journal = {\apjl},
         year = 2023,
        month = oct,
       volume = {955},
       number = {2},
          eid = {L34},
        pages = {L34},
          doi = {10.3847/2041-8213/acf12c},
archivePrefix = {arXiv},
       eprint = {2308.14281},
 primaryClass = {astro-ph.GA},
       adsurl = {https://ui.adsabs.harvard.edu/abs/2023ApJ...955L..34L}
}

@ARTICLE{2023ApJ...948...23Z,
       author = {{Zhu}, Dan and {Peng}, Zhao-Yang and {Chen}, Jia-Ming and {Li}, Ting},
        title = "{A Study of the Overall Evolution Behaviors of Pulses and Flares within Gamma-Ray Bursts}",
      journal = {\apj},
         year = 2023,
        month = may,
       volume = {948},
       number = {1},
          eid = {23},
        pages = {23},
          doi = {10.3847/1538-4357/acbe9f},
       adsurl = {https://ui.adsabs.harvard.edu/abs/2023ApJ...948...23Z}
}

@ARTICLE{2022ApJ...934...33A,
       author = {{Aschwanden}, Markus J.},
        title = "{The Fractality and Size Distributions of Astrophysical Self-Organized Criticality Systems}",
      journal = {\apj},
         year = 2022,
        month = jul,
       volume = {934},
       number = {1},
          eid = {33},
        pages = {33},
          doi = {10.3847/1538-4357/ac6bf2},
       adsurl = {https://ui.adsabs.harvard.edu/abs/2022ApJ...934...33A}
}

@ARTICLE{2022RAA....22f5012Z,
       author = {{Zhang}, Wen-Long and {Yi}, Shuang-Xi and {Yang}, Yu-Peng and {Qin}, Ying},
        title = "{Statistical Properties of X-Ray Flares from the Supergiant Fast X-Ray Transients}",
      journal = {Research in Astronomy and Astrophysics},
         year = 2022,
        month = jun,
       volume = {22},
       number = {6},
          eid = {065012},
        pages = {065012},
          doi = {10.1088/1674-4527/ac6aac},
       adsurl = {https://ui.adsabs.harvard.edu/abs/2022RAA....22f5012Z}
}

@ARTICLE{2021ApJ...920..153W,
       author = {{Wei}, Jun-Jie and {Wu}, Xue-Feng and {Dai}, Zi-Gao and {Wang}, Fa-Yin and {Wang}, Pei and {Li}, Di and {Zhang}, Bing},
        title = "{Similar Scale-invariant Behaviors between Soft Gamma-Ray Repeaters and an Extreme Epoch from FRB 121102}",
      journal = {\apj},
         year = 2021,
        month = oct,
       volume = {920},
       number = {2},
          eid = {153},
        pages = {153},
          doi = {10.3847/1538-4357/ac2604},
archivePrefix = {arXiv},
       eprint = {2107.12605},
 primaryClass = {astro-ph.HE},
       adsurl = {https://ui.adsabs.harvard.edu/abs/2021ApJ...920..153W}
}

@ARTICLE{2021ApJ...912...94A,
       author = {{Aschwanden}, Markus J. and {Dudok de Wit}, Thierry},
        title = "{Correlation of the Sunspot Number and the Waiting-time Distribution of Solar Flares, Coronal Mass Ejections, and Solar Wind Switchback Events Observed with the Parker Solar Probe}",
      journal = {\apj},
         year = 2021,
        month = may,
       volume = {912},
       number = {2},
          eid = {94},
        pages = {94},
          doi = {10.3847/1538-4357/abef69},
archivePrefix = {arXiv},
       eprint = {2102.02305},
 primaryClass = {astro-ph.SR},
       adsurl = {https://ui.adsabs.harvard.edu/abs/2021ApJ...912...94A}
}

@ARTICLE{2021ApJS..252...16L,
       author = {{Li}, X.~J. and {Zhang}, Z.~B. and {Zhang}, X.~L. and {Zhen}, H.~Y.},
        title = "{Temporal Properties of Precursors, Main Peaks, and Extended Emissions of Short GRBs in the Third Swift/BAT GRB Catalog}",
      journal = {\apjs},
         year = 2021,
        month = feb,
       volume = {252},
       number = {2},
          eid = {16},
        pages = {16},
          doi = {10.3847/1538-4365/abd3fd},
archivePrefix = {arXiv},
       eprint = {2010.10929},
 primaryClass = {astro-ph.HE},
       adsurl = {https://ui.adsabs.harvard.edu/abs/2021ApJS..252...16L}
}

@ARTICLE{2021FrPhy..1614501L,
       author = {{Lyu}, Fen and {Li}, Ya-Ping and {Hou}, Shu-Jin and {Wei}, Jun-Jie and {Geng}, Jin-Jun and {Wu}, Xue-Feng},
        title = "{Self-organized criticality in multi-pulse gamma-ray bursts}",
      journal = {Frontiers of Physics},
         year = 2021,
        month = feb,
       volume = {16},
       number = {1},
          eid = {14501},
        pages = {14501},
          doi = {10.1007/s11467-020-0989-x},
archivePrefix = {arXiv},
       eprint = {2009.06180},
 primaryClass = {astro-ph.HE},
       adsurl = {https://ui.adsabs.harvard.edu/abs/2021FrPhy..1614501L}
}

@ARTICLE{2020MNRAS.491.1498C,
       author = {{Cheng}, Yingjie and {Zhang}, G.~Q. and {Wang}, F.~Y.},
        title = "{Statistical properties of magnetar bursts and FRB 121102}",
      journal = {\mnras},
         year = 2020,
        month = jan,
       volume = {491},
       number = {1},
        pages = {1498-1505},
          doi = {10.1093/mnras/stz3085},
archivePrefix = {arXiv},
       eprint = {1910.14201},
 primaryClass = {astro-ph.HE},
       adsurl = {https://ui.adsabs.harvard.edu/abs/2020MNRAS.491.1498C}
}

@ARTICLE{2018ApJ...862..155L,
       author = {{Lan}, Lin and {L{\"u}}, Hou-Jun and {Zhong}, Shu-Qing and {Zhang}, Hai-Ming and {Rice}, Jared and {Cheng}, Ji-Gui and {Du}, Shen-Shi and {Li}, Long and {Lin}, Jie and {Lu}, Rui-Jing and {Liang}, En-Wei},
        title = "{Characteristics of Two-episode Emission Patterns in Fermi Long Gamma-Ray Bursts}",
      journal = {\apj},
         year = 2018,
        month = aug,
       volume = {862},
       number = {2},
          eid = {155},
        pages = {155},
          doi = {10.3847/1538-4357/aacda6},
archivePrefix = {arXiv},
       eprint = {1806.06690},
 primaryClass = {astro-ph.HE},
       adsurl = {https://ui.adsabs.harvard.edu/abs/2018ApJ...862..155L}
}

@ARTICLE{2018A&A...615A..80P,
       author = {{Pescalli}, A. and {Ronchi}, M. and {Ghirlanda}, G. and {Ghisellini}, G.},
        title = "{From the earliest pulses to the latest flares in long gamma-ray bursts}",
      journal = {\aap},
         year = 2018,
        month = jul,
       volume = {615},
          eid = {A80},
        pages = {A80},
          doi = {10.1051/0004-6361/201732270},
archivePrefix = {arXiv},
       eprint = {1801.08146},
 primaryClass = {astro-ph.HE},
       adsurl = {https://ui.adsabs.harvard.edu/abs/2018A&A...615A..80P}
}

@ARTICLE{2018NatCo...9..447Z,
       author = {{Zhang}, B.-B. and {Zhang}, B. and {Sun}, H. and {Lei}, W.-H. and {Gao}, H. and {Li}, Y. and {Shao}, L. and {Zhao}, Y. and {Hu}, Y.-D. and {L{\"u}}, H.-J. and {Wu}, X.-F. and {Fan}, X.-L. and {Wang}, G. and {Castro-Tirado}, A.~J. and {Zhang}, S. and {Yu}, B.-Y. and {Cao}, Y.-Y. and {Liang}, E.-W.},
        title = "{A peculiar low-luminosity short gamma-ray burst from a double neutron star merger progenitor}",
      journal = {Nature Communications},
         year = 2018,
        month = jan,
       volume = {9},
          eid = {447},
        pages = {447},
          doi = {10.1038/s41467-018-02847-3},
archivePrefix = {arXiv},
       eprint = {1710.05851},
 primaryClass = {astro-ph.HE},
       adsurl = {https://ui.adsabs.harvard.edu/abs/2018NatCo...9..447Z}
}

@ARTICLE{2017MNRAS.471.2517W,
       author = {{Wang}, J.~S. and {Wang}, F.~Y. and {Dai}, Z.~G.},
        title = "{Self-organized criticality in type I X-ray bursts}",
      journal = {\mnras},
         year = 2017,
        month = nov,
       volume = {471},
       number = {3},
        pages = {2517-2522},
          doi = {10.1093/mnras/stx1728},
archivePrefix = {arXiv},
       eprint = {1707.02173},
 primaryClass = {astro-ph.HE},
       adsurl = {https://ui.adsabs.harvard.edu/abs/2017MNRAS.471.2517W}
}

@ARTICLE{2017NewAR..79....1L,
       author = {{Liu}, Tong and {Gu}, Wei-Min and {Zhang}, Bing},
        title = "{Neutrino-dominated accretion flows as the central engine of gamma-ray bursts}",
      journal = {\nar},
         year = 2017,
        month = nov,
       volume = {79},
        pages = {1-25},
          doi = {10.1016/j.newar.2017.07.001},
archivePrefix = {arXiv},
       eprint = {1705.05516},
 primaryClass = {astro-ph.HE},
       adsurl = {https://ui.adsabs.harvard.edu/abs/2017NewAR..79....1L}
}

@ARTICLE{2017ApJ...844...79Y,
       author = {{Yi}, Shuang-Xi and {Yu}, Hai and {Wang}, F.~Y. and {Dai}, Zi-Gao},
        title = "{Statistical Distributions of Optical Flares from Gamma-Ray Bursts}",
      journal = {\apj},
         year = 2017,
        month = jul,
       volume = {844},
       number = {1},
          eid = {79},
        pages = {79},
          doi = {10.3847/1538-4357/aa7b7b},
archivePrefix = {arXiv},
       eprint = {1706.08716},
 primaryClass = {astro-ph.HE},
       adsurl = {https://ui.adsabs.harvard.edu/abs/2017ApJ...844...79Y}
}

@ARTICLE{2017ApJ...837...50G,
       author = {{Gao}, He and {Zhang}, Bing and {L{\"u}}, Hou-Jun and {Li}, Ye},
        title = "{Searching for Magnetar-powered Merger-novae from Short GRBS}",
      journal = {\apj},
         year = 2017,
        month = mar,
       volume = {837},
       number = {1},
          eid = {50},
        pages = {50},
          doi = {10.3847/1538-4357/aa5be3},
archivePrefix = {arXiv},
       eprint = {1608.03375},
 primaryClass = {astro-ph.HE},
       adsurl = {https://ui.adsabs.harvard.edu/abs/2017ApJ...837...50G}
}

@ARTICLE{2017JCAP...03..023W,
       author = {{Wang}, F.~Y. and {Yu}, H.},
        title = "{SGR-like behaviour of the repeating FRB 121102}",
      journal = {\jcap},
         year = 2017,
        month = mar,
       volume = {2017},
       number = {3},
          eid = {023},
        pages = {023},
          doi = {10.1088/1475-7516/2017/03/023},
archivePrefix = {arXiv},
       eprint = {1604.08676},
 primaryClass = {astro-ph.HE},
       adsurl = {https://ui.adsabs.harvard.edu/abs/2017JCAP...03..023W}
}

@ARTICLE{2016ApJS..224...20Y,
       author = {{Yi}, Shuang-Xi and {Xi}, Shao-Qiang and {Yu}, Hai and {Wang}, F.~Y. and {Mu}, Hui-Jun and {L{\"u}}, Lian-Zhong and {Liang}, En-Wei},
        title = "{Comprehensive Study of the X-Ray Flares from Gamma-ray Bursts Observed by Swift}",
      journal = {\apjs},
         year = 2016,
        month = jun,
       volume = {224},
       number = {2},
          eid = {20},
        pages = {20},
          doi = {10.3847/0067-0049/224/2/20},
archivePrefix = {arXiv},
       eprint = {1603.07429},
 primaryClass = {astro-ph.HE},
       adsurl = {https://ui.adsabs.harvard.edu/abs/2016ApJS..224...20Y}
}

@ARTICLE{2015ApJ...810...19L,
       author = {{Li}, Ya-Ping and {Yuan}, Feng and {Yuan}, Qiang and {Wang}, Q. Daniel and {Chen}, P.~F. and {Neilsen}, Joseph and {Fang}, Taotao and {Zhang}, Shuo and {Dexter}, Jason},
        title = "{Statistics of X-Ray Flares of Sagittarius A*: Evidence for Solar-like Self-organized Criticality Phenomena}",
      journal = {\apj},
         year = 2015,
        month = sep,
       volume = {810},
       number = {1},
          eid = {19},
        pages = {19},
          doi = {10.1088/0004-637X/810/1/19},
archivePrefix = {arXiv},
       eprint = {1506.02946},
 primaryClass = {astro-ph.HE},
       adsurl = {https://ui.adsabs.harvard.edu/abs/2015ApJ...810...19L}
}

@ARTICLE{2015ApJ...801...57G,
       author = {{Guidorzi}, C. and {Dichiara}, S. and {Frontera}, F. and {Margutti}, R. and {Baldeschi}, A. and {Amati}, L.},
        title = "{A Common Stochastic Process Rules Gamma-ray Burst Prompt Emission and X-ray Flares}",
      journal = {\apj},
         year = 2015,
        month = mar,
       volume = {801},
       number = {1},
          eid = {57},
        pages = {57},
          doi = {10.1088/0004-637X/801/1/57},
archivePrefix = {arXiv},
       eprint = {1501.02706},
 primaryClass = {astro-ph.HE},
       adsurl = {https://ui.adsabs.harvard.edu/abs/2015ApJ...801...57G}
}

@ARTICLE{2015PhR...561....1K,
       author = {{Kumar}, Pawan and {Zhang}, Bing},
        title = "{The physics of gamma-ray bursts \& relativistic jets}",
      journal = {\physrep},
         year = 2015,
        month = feb,
       volume = {561},
        pages = {1-109},
          doi = {10.1016/j.physrep.2014.09.008},
archivePrefix = {arXiv},
       eprint = {1410.0679},
 primaryClass = {astro-ph.HE},
       adsurl = {https://ui.adsabs.harvard.edu/abs/2015PhR...561....1K}
}

@ARTICLE{2014JApA...35..423P,
       author = {{Peng}, Fang-Kun and {Hu}, You-Dong and {Wang}, Xiang-Gao and {Lu}, Rui-Jing and {Liang}, En-Wei},
        title = "{Joint Spectral Analysis for Early Bright X-ray Flares of {\ensuremath{\gamma}}-Ray Bursts with Swift BAT and XRT Data}",
      journal = {Journal of Astrophysics and Astronomy},
         year = 2014,
        month = sep,
       volume = {35},
       number = {3},
        pages = {423-427},
          doi = {10.1007/s12036-014-9246-4},
       adsurl = {https://ui.adsabs.harvard.edu/abs/2014JApA...35..423P}
}

@ARTICLE{2022ApJ...931L..23L,
       author = {{L{\"u}}, Hou-Jun and {Yuan}, Hao-Yu and {Yi}, Ting-Feng and {Wang}, Xiang-Gao and {Hu}, You-Dong and {Yuan}, Yong and {Rice}, Jared and {Wang}, Jian-Guo and {Cao}, Jia-Xin and {Kong}, De-Feng and {Fernandez-Garc{\'\i}a}, Emilio and {Castro-Tirado}, Alberto J. and {Lian}, Ji-Shun and {Gan}, Wen-Pei and {Wang}, Shan-Qin and {Xin}, Li-Ping and {Caballero-Garc{\'\i}a}, M.~D. and {Fan}, Yu-Feng and {Liang}, En-Wei},
        title = "{GRB 211227A as a Peculiar Long Gamma-Ray Burst from a Compact Star Merger}",
      journal = {\apjl},
         year = 2022,
        month = jun,
       volume = {931},
       number = {2},
          eid = {L23},
        pages = {L23},
          doi = {10.3847/2041-8213/ac6e3a},
archivePrefix = {arXiv},
       eprint = {2201.06395},
 primaryClass = {astro-ph.HE},
       adsurl = {https://ui.adsabs.harvard.edu/abs/2022ApJ...931L..23L}
}

@ARTICLE{2014ApJ...785...74L,
       author = {{L{\"u}}, Hou-Jun and {Zhang}, Bing},
        title = "{A Test of the Millisecond Magnetar Central Engine Model of Gamma-Ray Bursts with Swift Data}",
      journal = {\apj},
         year = 2014,
        month = apr,
       volume = {785},
       number = {1},
          eid = {74},
        pages = {74},
          doi = {10.1088/0004-637X/785/1/74},
archivePrefix = {arXiv},
       eprint = {1401.1562},
 primaryClass = {astro-ph.HE},
       adsurl = {https://ui.adsabs.harvard.edu/abs/2014ApJ...785...74L}
}

@ARTICLE{2014ApJ...789..145H,
       author = {{Hu}, You-Dong and {Liang}, En-Wei and {Xi}, Shao-Qiang and {Peng}, Fang-Kun and {Lu}, Rui-Jing and {L{\"u}}, Lian-Zhong and {Zhang}, Bing},
        title = "{Internal Energy Dissipation of Gamma-Ray Bursts Observed with Swift: Precursors, Prompt Gamma-Rays, Extended Emission, and Late X-Ray Flares}",
      journal = {\apj},
         year = 2014,
        month = jul,
       volume = {789},
       number = {2},
          eid = {145},
        pages = {145},
          doi = {10.1088/0004-637X/789/2/145},
archivePrefix = {arXiv},
       eprint = {1405.5949},
 primaryClass = {astro-ph.HE},
       adsurl = {https://ui.adsabs.harvard.edu/abs/2014ApJ...789..145H}
}

@ARTICLE{2014ApJ...787...66Z,
       author = {{Zhang}, Bin-Bin and {Zhang}, Bing and {Murase}, Kohta and {Connaughton}, Valerie and {Briggs}, Michael S.},
        title = "{How Long does a Burst Burst?}",
      journal = {\apj},
         year = 2014,
        month = may,
       volume = {787},
       number = {1},
          eid = {66},
        pages = {66},
          doi = {10.1088/0004-637X/787/1/66},
archivePrefix = {arXiv},
       eprint = {1310.2540},
 primaryClass = {astro-ph.HE},
       adsurl = {https://ui.adsabs.harvard.edu/abs/2014ApJ...787...66Z}
}

@ARTICLE{2014ApJ...782...54A,
       author = {{Aschwanden}, Markus J.},
        title = "{A Macroscopic Description of a Generalized Self-organized Criticality System: Astrophysical Applications}",
      journal = {\apj},
         year = 2014,
        month = feb,
       volume = {782},
       number = {1},
          eid = {54},
        pages = {54},
          doi = {10.1088/0004-637X/782/1/54},
archivePrefix = {arXiv},
       eprint = {1310.4191},
 primaryClass = {astro-ph.SR},
       adsurl = {https://ui.adsabs.harvard.edu/abs/2014ApJ...782...54A}
}

@ARTICLE{2013NatPh...9..465W,
       author = {{Wang}, F.~Y. and {Dai}, Z.~G.},
        title = "{Self-organized criticality in X-ray flares of gamma-ray-burst afterglows}",
      journal = {Nature Physics},
         year = 2013,
        month = aug,
       volume = {9},
       number = {8},
        pages = {465-467},
          doi = {10.1038/nphys2670},
archivePrefix = {arXiv},
       eprint = {1308.1253},
 primaryClass = {astro-ph.HE},
       adsurl = {https://ui.adsabs.harvard.edu/abs/2013NatPh...9..465W}
}

@ARTICLE{2013ApJ...767L..28S,
       author = {{Sonbas}, E. and {MacLachlan}, G.~A. and {Shenoy}, A. and {Dhuga}, K.~S. and {Parke}, W.~C.},
        title = "{A New Correlation between GRB X-Ray Flares and the Prompt Emission}",
      journal = {\apjl},
         year = 2013,
        month = apr,
       volume = {767},
       number = {2},
          eid = {L28},
        pages = {L28},
          doi = {10.1088/2041-8205/767/2/L28},
archivePrefix = {arXiv},
       eprint = {1210.6850},
 primaryClass = {astro-ph.HE},
       adsurl = {https://ui.adsabs.harvard.edu/abs/2013ApJ...767L..28S}
}

@ARTICLE{2013PASP..125..306F,
       author = {{Foreman-Mackey}, Daniel and {Hogg}, David W. and {Lang}, Dustin and {Goodman}, Jonathan},
        title = "{emcee: The MCMC Hammer}",
      journal = {\pasp},
         year = 2013,
        month = mar,
       volume = {125},
       number = {925},
        pages = {306},
          doi = {10.1086/670067},
archivePrefix = {arXiv},
       eprint = {1202.3665},
 primaryClass = {astro-ph.IM},
       adsurl = {https://ui.adsabs.harvard.edu/abs/2013PASP..125..306F}
}

@ARTICLE{2013ApJ...763...15Q,
       author = {{Qin}, Ying and {Liang}, En-Wei and {Liang}, Yun-Feng and {Yi}, Shuang-Xi and {Lin}, Lin and {Zhang}, Bin-Bin and {Zhang}, Jin and {L{\"u}}, Hou-Jun and {Lu}, Rui-Jing and {L{\"u}}, Lian-Zhong and {Zhang}, Bing},
        title = "{A Comprehensive Analysis of Fermi Gamma-Ray Burst Data. III. Energy-dependent T $_{90}$ Distributions of GBM GRBs and Instrumental Selection Effect on Duration Classification}",
      journal = {\apj},
         year = 2013,
        month = jan,
       volume = {763},
       number = {1},
          eid = {15},
        pages = {15},
          doi = {10.1088/0004-637X/763/1/15},
archivePrefix = {arXiv},
       eprint = {1205.1188},
 primaryClass = {astro-ph.HE},
       adsurl = {https://ui.adsabs.harvard.edu/abs/2013ApJ...763...15Q}
}

@ARTICLE{2012ApJ...748..132Z,
       author = {{Zhang}, Bin-Bin and {Burrows}, David N. and {Zhang}, Bing and {M{\'e}sz{\'a}ros}, Peter and {Wang}, Xiang-Yu and {Stratta}, Giulia and {D'Elia}, Valerio and {Frederiks}, Dmitry and {Golenetskii}, Sergey and {Cummings}, Jay R. and {Norris}, Jay P. and {Falcone}, Abraham D. and {Barthelmy}, Scott D. and {Gehrels}, Neil},
        title = "{Unusual Central Engine Activity in the Double Burst GRB 110709B}",
      journal = {\apj},
         year = 2012,
        month = apr,
       volume = {748},
       number = {2},
          eid = {132},
        pages = {132},
          doi = {10.1088/0004-637X/748/2/132},
archivePrefix = {arXiv},
       eprint = {1111.2922},
 primaryClass = {astro-ph.HE},
       adsurl = {https://ui.adsabs.harvard.edu/abs/2012ApJ...748..132Z}
}

@ARTICLE{2012A&A...539A...2A,
       author = {{Aschwanden}, M.~J.},
        title = "{A statistical fractal-diffusive avalanche model of a slowly-driven self-organized criticality system}",
      journal = {\aap},
         year = 2012,
        month = mar,
       volume = {539},
          eid = {A2},
        pages = {A2},
          doi = {10.1051/0004-6361/201118237},
archivePrefix = {arXiv},
       eprint = {1112.4859},
 primaryClass = {astro-ph.SR},
       adsurl = {https://ui.adsabs.harvard.edu/abs/2012A&A...539A...2A}
}

@ARTICLE{2011MNRAS.417.2144M,
       author = {{Margutti}, R. and {Chincarini}, G. and {Granot}, J. and {Guidorzi}, C. and {Berger}, E. and {Bernardini}, M.~G. and {Gehrels}, N. and {Soderberg}, A.~M. and {Stamatikos}, M. and {Zaninoni}, E.},
        title = "{X-ray flare candidates in short gamma-ray bursts}",
      journal = {\mnras},
         year = 2011,
        month = nov,
       volume = {417},
       number = {3},
        pages = {2144-2160},
          doi = {10.1111/j.1365-2966.2011.19397.x},
archivePrefix = {arXiv},
       eprint = {1107.1740},
 primaryClass = {astro-ph.HE},
       adsurl = {https://ui.adsabs.harvard.edu/abs/2011MNRAS.417.2144M}
}

@ARTICLE{2010A&A...524A..64C,
       author = {{Cusumano}, G. and {La Parola}, V. and {Segreto}, A. and {Ferrigno}, C. and {Maselli}, A. and {Sbarufatti}, B. and {Romano}, P. and {Chincarini}, G. and {Giommi}, P. and {Masetti}, N. and {Moretti}, A. and {Parisi}, P. and {Tagliaferri}, G.},
        title = "{The Palermo Swift-BAT hard X-ray catalogue. III. Results after 54 months of sky survey}",
      journal = {\aap},
         year = 2010,
        month = dec,
       volume = {524},
          eid = {A64},
        pages = {A64},
          doi = {10.1051/0004-6361/201015249},
archivePrefix = {arXiv},
       eprint = {1009.0522},
 primaryClass = {astro-ph.HE},
       adsurl = {https://ui.adsabs.harvard.edu/abs/2010A&A...524A..64C}
}

@ARTICLE{2010ApJ...725.1965L,
       author = {{L{\"u}}, Hou-Jun and {Liang}, En-Wei and {Zhang}, Bin-Bin and {Zhang}, Bing},
        title = "{A New Classification Method for Gamma-ray Bursts}",
      journal = {\apj},
         year = 2010,
        month = dec,
       volume = {725},
       number = {2},
        pages = {1965-1970},
          doi = {10.1088/0004-637X/725/2/1965},
archivePrefix = {arXiv},
       eprint = {1001.0598},
 primaryClass = {astro-ph.HE},
       adsurl = {https://ui.adsabs.harvard.edu/abs/2010ApJ...725.1965L}
}

@ARTICLE{2010MNRAS.406.2149M,
       author = {{Margutti}, R. and {Guidorzi}, C. and {Chincarini}, G. and {Bernardini}, M.~G. and {Genet}, F. and {Mao}, J. and {Pasotti}, F.},
        title = "{Lag-luminosity relation in {\ensuremath{\gamma}}-ray burst X-ray flares: a direct link to the prompt emission}",
      journal = {\mnras},
         year = 2010,
        month = aug,
       volume = {406},
       number = {4},
        pages = {2149-2167},
          doi = {10.1111/j.1365-2966.2010.16824.x},
archivePrefix = {arXiv},
       eprint = {1004.1568},
 primaryClass = {astro-ph.HE},
       adsurl = {https://ui.adsabs.harvard.edu/abs/2010MNRAS.406.2149M}
}

@ARTICLE{2010MNRAS.406.2113C,
       author = {{Chincarini}, G. and {Mao}, J. and {Margutti}, R. and {Bernardini}, M.~G. and {Guidorzi}, C. and {Pasotti}, F. and {Giannios}, D. and {Della Valle}, M. and {Moretti}, A. and {Romano}, P. and {D'Avanzo}, P. and {Cusumano}, G. and {Giommi}, P.},
        title = "{Unveiling the origin of X-ray flares in gamma-ray bursts}",
      journal = {\mnras},
         year = 2010,
        month = aug,
       volume = {406},
       number = {4},
        pages = {2113-2148},
          doi = {10.1111/j.1365-2966.2010.17037.x},
archivePrefix = {arXiv},
       eprint = {1004.0901},
 primaryClass = {astro-ph.HE},
       adsurl = {https://ui.adsabs.harvard.edu/abs/2010MNRAS.406.2113C}
}

@ARTICLE{2009ApJ...703.1696Z,
       author = {{Zhang}, Bing and {Zhang}, Bin-Bin and {Virgili}, Francisco J. and {Liang}, En-Wei and {Kann}, D. Alexander and {Wu}, Xue-Feng and {Proga}, Daniel and {Lv}, Hou-Jun and {Toma}, Kenji and {M{\'e}sz{\'a}ros}, Peter and {Burrows}, David N. and {Roming}, Peter W.~A. and {Gehrels}, Neil},
        title = "{Discerning the Physical Origins of Cosmological Gamma-ray Bursts Based on Multiple Observational Criteria: The Cases of z = 6.7 GRB 080913, z = 8.2 GRB 090423, and Some Short/Hard GRBs}",
      journal = {\apj},
         year = 2009,
        month = oct,
       volume = {703},
       number = {2},
        pages = {1696-1724},
          doi = {10.1088/0004-637X/703/2/1696},
archivePrefix = {arXiv},
       eprint = {0902.2419},
 primaryClass = {astro-ph.HE},
       adsurl = {https://ui.adsabs.harvard.edu/abs/2009ApJ...703.1696Z}
}

@ARTICLE{2009MNRAS.397.1177E,
       author = {{Evans}, P.~A. and {Beardmore}, A.~P. and {Page}, K.~L. and {Osborne}, J.~P. and {O'Brien}, P.~T. and {Willingale}, R. and {Starling}, R.~L.~C. and {Burrows}, D.~N. and {Godet}, O. and {Vetere}, L. and {Racusin}, J. and {Goad}, M.~R. and {Wiersema}, K. and {Angelini}, L. and {Capalbi}, M. and {Chincarini}, G. and {Gehrels}, N. and {Kennea}, J.~A. and {Margutti}, R. and {Morris}, D.~C. and {Mountford}, C.~J. and {Pagani}, C. and {Perri}, M. and {Romano}, P. and {Tanvir}, N.},
        title = "{Methods and results of an automatic analysis of a complete sample of Swift-XRT observations of GRBs}",
      journal = {\mnras},
         year = 2009,
        month = aug,
       volume = {397},
       number = {3},
        pages = {1177-1201},
          doi = {10.1111/j.1365-2966.2009.14913.x},
archivePrefix = {arXiv},
       eprint = {0812.3662},
 primaryClass = {astro-ph},
       adsurl = {https://ui.adsabs.harvard.edu/abs/2009MNRAS.397.1177E}
}

@ARTICLE{2008ApJ...672.1103M,
       author = {{Melatos}, A. and {Peralta}, C. and {Wyithe}, J.~S.~B.},
        title = "{Avalanche Dynamics of Radio Pulsar Glitches}",
      journal = {\apj},
         year = 2008,
        month = jan,
       volume = {672},
       number = {2},
        pages = {1103-1118},
          doi = {10.1086/523349},
archivePrefix = {arXiv},
       eprint = {0710.1021},
 primaryClass = {astro-ph},
       adsurl = {https://ui.adsabs.harvard.edu/abs/2008ApJ...672.1103M}
}

@ARTICLE{2007ApJ...671.1903C,
       author = {{Chincarini}, G. and {Moretti}, A. and {Romano}, P. and {Falcone}, A.~D. and {Morris}, D. and {Racusin}, J. and {Campana}, S. and {Covino}, S. and {Guidorzi}, C. and {Tagliaferri}, G. and {Burrows}, D.~N. and {Pagani}, C. and {Stroh}, M. and {Grupe}, D. and {Capalbi}, M. and {Cusumano}, G. and {Gehrels}, N. and {Giommi}, P. and {La Parola}, V. and {Mangano}, V. and {Mineo}, T. and {Nousek}, J.~A. and {O'Brien}, P.~T. and {Page}, K.~L. and {Perri}, M. and {Troja}, E. and {Willingale}, R. and {Zhang}, B.},
        title = "{The First Survey of X-Ray Flares from Gamma-Ray Bursts Observed by Swift: Temporal Properties and Morphology}",
      journal = {\apj},
         year = 2007,
        month = dec,
       volume = {671},
       number = {2},
        pages = {1903-1920},
          doi = {10.1086/521591},
archivePrefix = {arXiv},
       eprint = {astro-ph/0702371},
 primaryClass = {astro-ph},
       adsurl = {https://ui.adsabs.harvard.edu/abs/2007ApJ...671.1903C}
}

@ARTICLE{2007ApJ...670..565L,
       author = {{Liang}, En-Wei and {Zhang}, Bin-Bin and {Zhang}, Bing},
        title = "{A Comprehensive Analysis of Swift XRT Data. II. Diverse Physical Origins of the Shallow Decay Segment}",
      journal = {\apj},
         year = 2007,
        month = nov,
       volume = {670},
       number = {1},
        pages = {565-583},
          doi = {10.1086/521870},
archivePrefix = {arXiv},
       eprint = {0705.1373},
 primaryClass = {astro-ph},
       adsurl = {https://ui.adsabs.harvard.edu/abs/2007ApJ...670..565L}
}

@ARTICLE{2007ApJ...667.1024K,
       author = {{Kocevski}, Daniel and {Butler}, Nathaniel and {Bloom}, Joshua S.},
        title = "{Pulse Width Evolution of Late-Time X-Ray Flares in Gamma-Ray Bursts}",
      journal = {\apj},
         year = 2007,
        month = oct,
       volume = {667},
       number = {2},
        pages = {1024-1032},
          doi = {10.1086/520041},
archivePrefix = {arXiv},
       eprint = {astro-ph/0702452},
 primaryClass = {astro-ph},
       adsurl = {https://ui.adsabs.harvard.edu/abs/2007ApJ...667.1024K}
}

@ARTICLE{2007ApJ...665..599T,
       author = {{Troja}, E. and {Cusumano}, G. and {O'Brien}, P.~T. and {Zhang}, B. and {Sbarufatti}, B. and {Mangano}, V. and {Willingale}, R. and {Chincarini}, G. and {Osborne}, J.~P. and {Marshall}, F.~E. and {Burrows}, D.~N. and {Campana}, S. and {Gehrels}, N. and {Guidorzi}, C. and {Krimm}, H.~A. and {La Parola}, V. and {Liang}, E.~W. and {Mineo}, T. and {Moretti}, A. and {Page}, K.~L. and {Romano}, P. and {Tagliaferri}, G. and {Zhang}, B.~B. and {Page}, M.~J. and {Schady}, P.},
        title = "{Swift Observations of GRB 070110: An Extraordinary X-Ray Afterglow Powered by the Central Engine}",
      journal = {\apj},
         year = 2007,
        month = aug,
       volume = {665},
       number = {1},
        pages = {599-607},
          doi = {10.1086/519450},
archivePrefix = {arXiv},
       eprint = {astro-ph/0702220},
 primaryClass = {astro-ph},
       adsurl = {https://ui.adsabs.harvard.edu/abs/2007ApJ...665..599T}
}

@ARTICLE{2007A&A...469..379E,
       author = {{Evans}, P.~A. and {Beardmore}, A.~P. and {Page}, K.~L. and {Tyler}, L.~G. and {Osborne}, J.~P. and {Goad}, M.~R. and {O'Brien}, P.~T. and {Vetere}, L. and {Racusin}, J. and {Morris}, D. and {Burrows}, D.~N. and {Capalbi}, M. and {Perri}, M. and {Gehrels}, N. and {Romano}, P.},
        title = "{An online repository of Swift/XRT light curves of {\ensuremath{\gamma}}-ray bursts}",
      journal = {\aap},
         year = 2007,
        month = jul,
       volume = {469},
       number = {1},
        pages = {379-385},
          doi = {10.1051/0004-6361:20077530},
archivePrefix = {arXiv},
       eprint = {0704.0128},
 primaryClass = {astro-ph},
       adsurl = {https://ui.adsabs.harvard.edu/abs/2007A&A...469..379E}
}

@ARTICLE{2006Natur.444.1010Z,
       author = {{Zhang}, Bing},
        title = "{Astrophysics: A burst of new ideas}",
      journal = {\nat},
         year = 2006,
        month = dec,
       volume = {444},
       number = {7122},
        pages = {1010-1011},
          doi = {10.1038/4441010a},
archivePrefix = {arXiv},
       eprint = {astro-ph/0612614},
 primaryClass = {astro-ph},
       adsurl = {https://ui.adsabs.harvard.edu/abs/2006Natur.444.1010Z}
}

@ARTICLE{2005Sci...309.1833B,
       author = {{Burrows}, D.~N. and {Romano}, P. and {Falcone}, A. and {Kobayashi}, S. and {Zhang}, B. and {Moretti}, A. and {O'Brien}, P.~T. and {Goad}, M.~R. and {Campana}, S. and {Page}, K.~L. and {Angelini}, L. and {Barthelmy}, S. and {Beardmore}, A.~P. and {Capalbi}, M. and {Chincarini}, G. and {Cummings}, J. and {Cusumano}, G. and {Fox}, D. and {Giommi}, P. and {Hill}, J.~E. and {Kennea}, J.~A. and {Krimm}, H. and {Mangano}, V. and {Marshall}, F. and {M{\'e}sz{\'a}ros}, P. and {Morris}, D.~C. and {Nousek}, J.~A. and {Osborne}, J.~P. and {Pagani}, C. and {Perri}, M. and {Tagliaferri}, G. and {Wells}, A.~A. and {Woosley}, S. and {Gehrels}, N.},
        title = "{Bright X-ray Flares in Gamma-Ray Burst Afterglows}",
      journal = {Science},
         year = 2005,
        month = sep,
       volume = {309},
       number = {5742},
        pages = {1833-1835},
          doi = {10.1126/science.1116168},
archivePrefix = {arXiv},
       eprint = {astro-ph/0506130},
 primaryClass = {astro-ph},
       adsurl = {https://ui.adsabs.harvard.edu/abs/2005Sci...309.1833B}
}

@ARTICLE{2003ApJ...596..389K,
       author = {{Kocevski}, Dan and {Ryde}, Felix and {Liang}, Edison},
        title = "{Search for Relativistic Curvature Effects in Gamma-Ray Burst Pulses}",
      journal = {\apj},
         year = 2003,
        month = oct,
       volume = {596},
       number = {1},
        pages = {389-400},
          doi = {10.1086/377707},
archivePrefix = {arXiv},
       eprint = {astro-ph/0303556},
 primaryClass = {astro-ph},
       adsurl = {https://ui.adsabs.harvard.edu/abs/2003ApJ...596..389K}
}

@ARTICLE{1998ApJ...504..405S,
       author = {{Scargle}, Jeffrey D.},
        title = "{Studies in Astronomical Time Series Analysis. V. Bayesian Blocks, a New Method to Analyze Structure in Photon Counting Data}",
      journal = {\apj},
         year = 1998,
        month = sep,
       volume = {504},
       number = {1},
        pages = {405-418},
          doi = {10.1086/306064},
archivePrefix = {arXiv},
       eprint = {astro-ph/9711233},
 primaryClass = {astro-ph},
       adsurl = {https://ui.adsabs.harvard.edu/abs/1998ApJ...504..405S}
}

@ARTICLE{1995ApJ...452..145K,
       author = {{Koshut}, Thomas M. and {Kouveliotou}, Chryssa and {Paciesas}, William S. and {van Paradijs}, Jan and {Pendleton}, Geoffrey N. and {Briggs}, Michael S. and {Fishman}, Gerald J. and {Meegan}, Charles A.},
        title = "{Gamma-Ray Burst Precursor Activity as Observed with BATSE}",
      journal = {\apj},
         year = 1995,
        month = oct,
       volume = {452},
        pages = {145},
          doi = {10.1086/176286},
       adsurl = {https://ui.adsabs.harvard.edu/abs/1995ApJ...452..145K}
}

@ARTICLE{1993ApJ...413L.101K,
       author = {{Kouveliotou}, Chryssa and {Meegan}, Charles A. and {Fishman}, Gerald J. and {Bhat}, Narayana P. and {Briggs}, Michael S. and {Koshut}, Thomas M. and {Paciesas}, William S. and {Pendleton}, Geoffrey N.},
        title = "{Identification of Two Classes of Gamma-Ray Bursts}",
      journal = {\apjl},
         year = 1993,
        month = aug,
       volume = {413},
        pages = {L101},
          doi = {10.1086/186969},
       adsurl = {https://ui.adsabs.harvard.edu/abs/1993ApJ...413L.101K}
}

@ARTICLE{1993ApJ...405..273W,
        author = {{Woosley}, S.~E.},
         title = "{Gamma-Ray Bursts from Stellar Mass Accretion Disks around Black Holes}",
       journal = {\apj},
          year = 1993,
         month = mar,
        volume = {405},
         pages = {273},
           doi = {10.1086/172359},
        adsurl = {https://ui.adsabs.harvard.edu/abs/1993ApJ...405..273W}
}

@ARTICLE{2010GCN.11230....1T,
       author = {{Tanvir}, N.~R. and {Malesani}, D. and {Fynbo}, J.~P.~U. and {Levan}, A.~J. and {Xu}, D.},
        title = "{GRB 100906A: VLT/X-shooter redshift}",
      journal = {GCN},
         year = 2010,
        month = sep,
       number = {11230},
        pages = {1},
      adsurl = {https://ui.adsabs.harvard.edu/abs/2010GCN.11230....1T}
}

@ARTICLE{2011GCN.12234....1C,
       author = {{Cabrera Lavers}, J. and {de Ugarte Postigo}, A. and {Castro-Tirado}, A.~J. and others},
        title = "{GRB 110801A: Redshift from GTC}",
      journal = {GCN},
         year = 2011,
        month = aug,
       number = {12234},
        pages = {1},
      adsurl = {https://ui.adsabs.harvard.edu/abs/2011GCN.12234....1C}
}

@ARTICLE{2012GCN.14059....1P,
       author = {{Perley}, D.~A. and {Cenko}, S.~B. and {Bloom}, J.~S.},
        title = "{GRB 121211A: Redshift from Keck}",
      journal = {GCN},
         year = 2012,
        month = dec,
       number = {14059},
        pages = {1},
      adsurl = {https://ui.adsabs.harvard.edu/abs/2012GCN.14059....1P}
}

@ARTICLE{2014GCN.15800....1M,
       author = {{Malesani}, D. and {Fynbo}, J.~P.~U. and {Kann}, D.~A. and {Xu}, D.},
        title = "{GRB 140206A: VLT/X-shooter redshift}",
      journal = {GCN},
         year = 2014,
        month = feb,
       number = {15800},
        pages = {1},
      adsurl = {https://ui.adsabs.harvard.edu/abs/2014GCN.15800....1M}
}

@ARTICLE{2014GCN.16194....1K,
       author = {{Kann}, D.~A. and {Malesani}, D. and {Fynbo}, J.~P.~U. and {Xu}, D.},
        title = "{GRB 140430A: VLT/X-shooter redshift}",
      journal = {GCN},
         year = 2014,
        month = apr,
       number = {16194},
        pages = {1},
      adsurl = {https://ui.adsabs.harvard.edu/abs/2014GCN.16194....1K}
}

@ARTICLE{2014GCN.16310....1D,
       author = {{de Ugarte Postigo}, A. and {Gorosabel}, J. and {Xu}, D. and {Malesani}, D.},
        title = "{GRB 140512A: Redshift from NOT}",
      journal = {GCN},
         year = 2014,
        month = may,
       number = {16310},
        pages = {1},
      adsurl = {https://ui.adsabs.harvard.edu/abs/2014GCN.16310....1D}
}

@ARTICLE{2015GCN.18487....1P,
       author = {{Perley}, D.~A. and {Hillenbrand}, L. and {Prochaska}, J.~X.},
        title = "{GRB 151027A: Keck/HIRES redshift}",
      journal = {GCN},
         year = 2015,
        month = oct,
       number = {18487},
        pages = {1},
      adsurl = {https://ui.adsabs.harvard.edu/abs/2015GCN.18487....1P}
}

@ARTICLE{2016GCN.19109....1X,
       author = {{Xu}, D. and {Fynbo}, J.~P.~U. and {Malesani}, D. and {de Ugarte Postigo}, A.},
        title = "{GRB 160227A: NOT redshift}",
      journal = {GCN},
         year = 2016,
        month = feb,
       number = {19109},
        pages = {1},
      adsurl = {https://ui.adsabs.harvard.edu/abs/2016GCN.19109....1X}
}

@ARTICLE{2016GCN.19350....1T,
       author = {{Tanvir}, N.~R. and {Malesani}, D. and {Fynbo}, J.~P.~U. and {Levan}, A.~J.},
        title = "{GRB 160425A VLT/X-shooter redshift}",
      journal = {GCN},
         year = 2016,
        month = apr,
       number = {19350},
        pages = {1},
      adsurl = {https://ui.adsabs.harvard.edu/abs/2016GCN.19350....1T}
}

@ARTICLE{2016GCN.20180....1M,
       author = {{Malesani}, D. and {Kruehler}, T. and {Heintz}, K.~E. and {Fynbo}, J.~P.~U.},
        title = "{GRB 161117A: VLT/X-shooter spectroscopy and redshift}",
      journal = {GCN},
         year = 2016,
        month = nov,
       number = {20180},
        pages = {1},
      adsurl = {https://ui.adsabs.harvard.edu/abs/2016GCN.20180....1M}
}

@ARTICLE{2017GCN.21119....1I,
       author = {{Izzo}, L. and {de Ugarte Postigo}, A. and {Kann}, D.~A. and {Thoene}, C.~C.},
        title = "{GRB 170519A: Redshift from GTC/OSIRIS}",
      journal = {GCN},
         year = 2017,
        month = may,
       number = {21119},
        pages = {1},
      adsurl = {https://ui.adsabs.harvard.edu/abs/2017GCN.21119....1I}
}

@ARTICLE{2017GCN.21177....1D,
       author = {{de Ugarte Postigo}, A. and {Izzo}, L. and {Kann}, D.~A. and {Thoene}, C.~C.},
        title = "{GRB 170531B: Redshift from GTC/OSIRIS}",
      journal = {GCN},
         year = 2017,
        month = jun,
       number = {21177},
        pages = {1},
      adsurl = {https://ui.adsabs.harvard.edu/abs/2017GCN.21177....1D}
}

@ARTICLE{2017GCN.21298....1D,
       author = {{de Ugarte Postigo}, A. and {Kann}, D.~A. and {Izzo}, L. and {Thoene}, C.~C.},
        title = "{GRB 170705A: Redshift from 10.4m GTC/OSIRIS}",
      journal = {GCN},
         year = 2017,
        month = jul,
       number = {21298},
        pages = {1},
      adsurl = {https://ui.adsabs.harvard.edu/abs/2017GCN.21298....1D}
}

@ARTICLE{2018GCN.22555....1D,
       author = {{D'Avanzo}, P. and {Bolmer}, J. and {D'Elia}, V. and {Fynbo}, J.~P.~U.},
        title = "{GRB 180325A: VLT/X-shooter spectroscopic observations}",
      journal = {GCN},
         year = 2018,
        month = mar,
       number = {22555},
        pages = {1},
      adsurl = {https://ui.adsabs.harvard.edu/abs/2018GCN.22555....1D}
}

@ARTICLE{2018GCN.22567....1I,
       author = {{Izzo}, L. and {Heintz}, K.~E. and {Malesani}, D. and {Milvang-Jensen}, B.},
        title = "{GRB 180329B: VLT/X-shooter spectroscopic observations}",
      journal = {GCN},
         year = 2018,
        month = mar,
       number = {22567},
        pages = {1},
      adsurl = {https://ui.adsabs.harvard.edu/abs/2018GCN.22567....1I}
}

@ARTICLE{2018GCN.22845....1R,
       author = {{Rossi}, A. and {Gargiulo}, A. and {Malesani}, D.~B. and {Vergani}, S.~D.},
        title = "{GRB 180624A: LBT/MODS redshift measurement}",
      journal = {GCN},
         year = 2018,
        month = jun,
       number = {22845},
        pages = {1},
      adsurl = {https://ui.adsabs.harvard.edu/abs/2018GCN.22845....1R}
}

@ARTICLE{2024GCN.37469....1D,
       author = {{de Ugarte Postigo}, A. and {Izzo}, L. and {Malesani}, D.~B. and others},
        title = "{GRB 240912A: Redshift from GTC}",
      journal = {GCN},
         year = 2024,
        month = sep,
       number = {37469},
        pages = {1},
      adsurl = {https://ui.adsabs.harvard.edu/abs/2024GCN.37469....1D}
}

@ARTICLE{2025GCN.38934....1M,
       author = {{Malesani}, D.~B. and {Saccardi}, A. and {Vergani}, S.~D. and {D'Avanzo}, P.},
        title = "{GRB 250114A: VLT/X-shooter spectroscopic redshift z = 4.732}",
      journal = {GCN},
         year = 2025,
        month = jan,
       number = {38934},
        pages = {1},
      adsurl = {https://ui.adsabs.harvard.edu/abs/2025GCN.38934....1M}
}

@ARTICLE{1991ApJ...380L..89L,
       author = {{Lu}, Edward T. and {Hamilton}, Russell J.},
        title = "{Avalanches and the Distribution of Solar Flares}",
      journal = {\apjl},
         year = 1991,
        month = oct,
       volume = {380},
        pages = {L89},
          doi = {10.1086/186180},
       adsurl = {https://ui.adsabs.harvard.edu/abs/1991ApJ...380L..89L}
}

@ARTICLE{1989Natur.340..126E,
       author = {{Eichler}, David and {Livio}, Mario and {Piran}, Tsvi and {Schramm}, David N.},
        title = "{Nucleosynthesis, neutrino bursts and {\ensuremath{\gamma}}-rays from coalescing neutron stars}",
      journal = {\nat},
         year = 1989,
        month = jul,
       volume = {340},
       number = {6229},
        pages = {126-128},
          doi = {10.1038/340126a0},
       adsurl = {https://ui.adsabs.harvard.edu/abs/1989Natur.340..126E}
}

@ARTICLE{1987PhRvL..59..381B,
       author = {{Bak}, Per and {Tang}, Chao and {Wiesenfeld}, Kurt},
        title = "{Self-organized criticality: An explanation of the 1/f noise}",
      journal = {\prl},
         year = 1987,
        month = jul,
       volume = {59},
       number = {4},
        pages = {381-384},
          doi = {10.1103/PhysRevLett.59.381},
       adsurl = {https://ui.adsabs.harvard.edu/abs/1987PhRvL..59..381B}
}

@ARTICLE{2025ApJ...995..201Y,
       author = {{Yu}, Wen-Yuan and {L{\"u}}, Hou-Jun and {Tian}, Xiao and {Chen}, Liang-Jun and {Liang}, En-Wei},
        title = "{Probing the Nature of High-redshift Long GRB 250114A and Its Magnetar Central Engine}",
      journal = {\apj},
         year = 2025,
        month = dec,
       volume = {995},
       number = {2},
          eid = {201},
        pages = {201},
          doi = {10.3847/1538-4357/ae2009},
archivePrefix = {arXiv},
       eprint = {2511.21419},
 primaryClass = {astro-ph.HE},
       adsurl = {https://ui.adsabs.harvard.edu/abs/2025ApJ...995..201Y}
}

@BOOK{2018pgrb.book.....Z,
       author = {{Zhang}, Bing},
        title = "{The Physics of Gamma-Ray Bursts}",
         year = 2018,
          doi = {10.1017/9781139226530},
       adsurl = {https://ui.adsabs.harvard.edu/abs/2018pgrb.book.....Z}
}

@ARTICLE{2015ApJ...814...19A,
       author = {{Aschwanden}, Markus J.},
        title = "{Thresholded Power law Size Distributions of Instabilities in Astrophysics}",
      journal = {\apj},
         year = 2015,
        month = nov,
       volume = {814},
       number = {1},
          eid = {19},
        pages = {19},
          doi = {10.1088/0004-637X/814/1/19},
archivePrefix = {arXiv},
       eprint = {1510.01987},
 primaryClass = {astro-ph.SR},
       adsurl = {https://ui.adsabs.harvard.edu/abs/2015ApJ...814...19A}
}

@ARTICLE{2014MNRAS.442.1922L,
       author = {{L{\"u}}, Hou-Jun and {Zhang}, Bing and {Liang}, En-Wei and {Zhang}, Bin-Bin and {Sakamoto}, Takanori},
        title = "{The `amplitude' parameter of gamma-ray bursts and its implications for GRB classification}",
      journal = {\mnras},
         year = 2014,
        month = aug,
       volume = {442},
       number = {3},
        pages = {1922-1929},
          doi = {10.1093/mnras/stu982},
archivePrefix = {arXiv},
       eprint = {1211.1117},
 primaryClass = {astro-ph.HE},
       adsurl = {https://ui.adsabs.harvard.edu/abs/2014MNRAS.442.1922L}
}

@ARTICLE{2006ApJ...636L..29P,
       author = {{Perna}, Rosalba and {Armitage}, Philip J. and {Zhang}, Bing},
        title = "{Flares in Long and Short Gamma-Ray Bursts: A Common Origin in a Hyperaccreting Accretion Disk}",
      journal = {\apjl},
         year = 2006,
        month = jan,
       volume = {636},
       number = {1},
        pages = {L29-L32},
          doi = {10.1086/499775},
archivePrefix = {arXiv},
       eprint = {astro-ph/0511506},
 primaryClass = {astro-ph},
       adsurl = {https://ui.adsabs.harvard.edu/abs/2006ApJ...636L..29P}
}

@ARTICLE{2017MNRAS.464.4399D,
       author = {{Dall'Osso}, Simone and {Perna}, Rosalba and {Tanaka}, Takamitsu L. and {Margutti}, Raffaella},
        title = "{Flares in gamma-ray bursts: disc fragmentation and evolution}",
      journal = {\mnras},
         year = 2017,
        month = feb,
       volume = {464},
       number = {4},
        pages = {4399-4407},
          doi = {10.1093/mnras/stw2695},
archivePrefix = {arXiv},
       eprint = {1610.05302},
 primaryClass = {astro-ph.HE},
       adsurl = {https://ui.adsabs.harvard.edu/abs/2017MNRAS.464.4399D}
}

@ARTICLE{2017MNRAS.470.4925G,
       author = {{Gibson}, S.~L. and {Wynn}, G.~A. and {Gompertz}, B.~P. and {O'Brien}, P.~T.},
        title = "{Fallback accretion on to a newborn magnetar: short GRBs with extended emission}",
      journal = {\mnras},
         year = 2017,
        month = oct,
       volume = {470},
       number = {4},
        pages = {4925-4940},
          doi = {10.1093/mnras/stx1531},
archivePrefix = {arXiv},
       eprint = {1706.04802},
 primaryClass = {astro-ph.HE},
       adsurl = {https://ui.adsabs.harvard.edu/abs/2017MNRAS.470.4925G}
}

@ARTICLE{2018MNRAS.478.4323G,
       author = {{Gibson}, S.~L. and {Wynn}, G.~A. and {Gompertz}, B.~P. and {O'Brien}, P.~T.},
        title = "{Fallback accretion on to a newborn magnetar: long GRBs with giant X-ray flares}",
      journal = {\mnras},
         year = 2018,
        month = aug,
       volume = {478},
       number = {4},
        pages = {4323-4335},
          doi = {10.1093/mnras/sty1363},
archivePrefix = {arXiv},
       eprint = {1805.09022},
 primaryClass = {astro-ph.HE},
       adsurl = {https://ui.adsabs.harvard.edu/abs/2018MNRAS.478.4323G}
}

@ARTICLE{2007ApJ...670.1247W,
       author = {{Wang}, Xiang-Yu and {M{\'e}sz{\'a}ros}, Peter},
        title = "{GRB Precursors in the Fallback Collapsar Scenario}",
      journal = {\apj},
         year = 2007,
        month = dec,
       volume = {670},
       number = {2},
        pages = {1247-1253},
          doi = {10.1086/522820},
archivePrefix = {arXiv},
       eprint = {astro-ph/0702441},
 primaryClass = {astro-ph},
        adsurl = {https://ui.adsabs.harvard.edu/abs/2007ApJ...670.1247W}
}

@ARTICLE{2010GCN.11091....1F,
       author = {{Filgas}, R. and {Schady}, P. and {Greiner}, J.},
        title = "{GRB 100814A: GROND detection of the Optical/NIR afterglow.}",
      journal = {GCN},
         year = 2010,
        month = jan,
       volume = {11091},
        pages = {1},
       adsurl = {https://ui.adsabs.harvard.edu/abs/2010GCN.11091....1F}
}




\begin{deluxetable*}{lcccccccccccccc}
\tabletypesize{\tiny}
\tablecaption{Results of the Temporal Analysis of 38 GRBs in Our Sample.\label{tab:sample}}
\tablehead{\colhead{GRB} & \colhead{$t_{\rm p,wait}$ (s)} & \colhead{$t_{\rm p,rise}$ (s)} & \colhead{$t_{\rm p,peak}$ (s)} & \colhead{$t_{\rm p,decay}$ (s)} & \colhead{$t_{\rm p,duration}$ (s)} & \colhead{$\log E_{\rm p,iso}$ (erg)} & \colhead{$t_{\rm f,wait}$ (s)} & \colhead{$t_{\rm f,rise}$ (s)} & \colhead{$t_{\rm f,peak}$ (s)} & \colhead{$t_{\rm f,decay}$ (s)} & \colhead{$t_{\rm f,duration}$ (s)} & \colhead{$\log E_{\rm f,iso}$ (erg)} & \colhead{Type} & \colhead{Redshift $z$}}
\startdata
100814A & $41.63^{+0.98}_{-0.68}$ & $4.38^{+0.26}_{-0.16}$ & $145.80^{+0.12}_{-0.11}$ & $13.79^{+0.55}_{-0.75}$ & $18.17^{+0.61}_{-0.77}$ & $51.87\pm0.23$ & $58.69^{+1.48}_{-1.52}$ & $90.35^{+1.52}_{-1.48}$ & $149.04^{+0.74}_{-0.70}$ & $667.63^{+25.42}_{-23.84}$ & $757.98^{+25.47}_{-23.89}$ & $48.89\pm0.05$ & III & 1.440\tablenotemark{a} \\
100906A & $12.57^{+0.59}_{-0.59}$ & $6.25^{+0.01}_{-0.01}$ & $105.43^{+0.10}_{-0.11}$ & $6.48^{+0.02}_{-0.02}$ & $12.73^{+0.02}_{-0.02}$ & $51.84\pm0.27$ & $39.18^{+1.49}_{-1.39}$ & $83.18^{+1.39}_{-1.49}$ & $122.36^{+0.48}_{-0.49}$ & $93.92^{+2.09}_{-2.01}$ & $177.10^{+2.51}_{-2.50}$ & $49.53\pm0.04$ & I & 1.727\tablenotemark{b} \\
110102A & $71.52^{+0.15}_{-0.16}$ & $10.30^{+0.03}_{-0.05}$ & $207.74^{+0.07}_{-0.06}$ & $10.59^{+0.05}_{-0.07}$ & $20.89^{+0.06}_{-0.09}$ & \nodata & $65.94^{+18.69}_{-18.77}$ & $24.35^{+7.18}_{-5.15}$ & $206.54^{+2.34}_{-1.51}$ & $195.07^{+15.63}_{-16.98}$ & $219.42^{+17.20}_{-17.74}$ & \nodata & III & \nodata \\
110102A & $50.36^{+0.66}_{-0.38}$ & $15.64^{+0.33}_{-0.66}$ & $263.42^{+0.21}_{-0.18}$ & $16.05^{+0.35}_{-0.66}$ & $31.69^{+0.48}_{-0.93}$ & \nodata & $12.18^{+17.91}_{-17.70}$ & $77.54^{+18.64}_{-13.54}$ & $271.91^{+8.72}_{-2.94}$ & $117.67^{+12.14}_{-20.00}$ & $195.21^{+22.24}_{-24.15}$ & \nodata & III & \nodata \\
110801A & $9.50^{+2.14}_{-2.76}$ & $13.12^{+3.10}_{-2.31}$ & $358.62^{+0.85}_{-0.79}$ & $53.93^{+4.89}_{-4.76}$ & $67.05^{+5.79}_{-5.29}$ & $52.36\pm0.26$ & $111.62^{+2.35}_{-2.36}$ & $42.58^{+1.04}_{-1.06}$ & $356.82^{+0.46}_{-0.46}$ & $254.38^{+4.73}_{-4.16}$ & $296.96^{+4.84}_{-4.29}$ & $51.82\pm0.01$ & II & 1.858\tablenotemark{c} \\
110820A & $174.72^{+8.40}_{-8.36}$ & $69.83^{+8.90}_{-8.55}$ & $244.50^{+2.64}_{-2.95}$ & $46.98^{+10.63}_{-6.54}$ & $116.81^{+13.86}_{-10.76}$ & \nodata & $72.32^{+20.74}_{-32.78}$ & $152.61^{+41.03}_{-4.43}$ & $260.78^{+23.29}_{-0.94}$ & $159.82^{+21.14}_{-41.46}$ & $312.43^{+46.16}_{-41.70}$ & \nodata & II & \nodata \\
120514A & $84.62^{+2.66}_{-3.20}$ & $24.82^{+4.17}_{-4.78}$ & $152.13^{+1.88}_{-2.45}$ & $26.15^{+9.26}_{-5.57}$ & $50.97^{+10.16}_{-7.34}$ & \nodata & $52.27^{+7.18}_{-9.92}$ & $47.89^{+4.03}_{-3.88}$ & $141.25^{+0.61}_{-0.52}$ & $181.84^{+14.14}_{-11.42}$ & $229.73^{+14.70}_{-12.06}$ & \nodata & I & \nodata \\
121123A & $155.26^{+1.02}_{-0.98}$ & $38.04^{+0.78}_{-0.74}$ & $225.07^{+0.35}_{-0.35}$ & $122.22^{+0.52}_{-0.63}$ & $160.26^{+0.94}_{-0.97}$ & \nodata & $87.09^{+0.46}_{-0.50}$ & $214.15^{+0.59}_{-0.60}$ & $301.24^{+0.81}_{-0.82}$ & $918.44^{+7.91}_{-7.16}$ & $1132.59^{+7.93}_{-7.19}$ & \nodata & II & \nodata \\
121211A & $22.41^{+0.27}_{-0.29}$ & $7.30^{+0.01}_{-0.02}$ & $145.25^{+0.13}_{-0.13}$ & $7.47^{+0.01}_{-0.02}$ & $14.77^{+0.01}_{-0.03}$ & $51.19\pm0.44$ & $38.81^{+4.92}_{-5.83}$ & $75.87^{+3.01}_{-3.36}$ & $168.01^{+1.04}_{-1.02}$ & $483.33^{+47.61}_{-43.54}$ & $559.20^{+47.71}_{-43.67}$ & $50.75\pm0.01$ & I & 1.023\tablenotemark{d} \\
121217A & $82.15^{+1.41}_{-1.57}$ & $31.49^{+0.69}_{-1.11}$ & $670.90^{+1.32}_{-1.27}$ & $31.90^{+0.87}_{-1.22}$ & $63.39^{+1.11}_{-1.65}$ & \nodata & $111.02^{+30.18}_{-21.74}$ & $446.21^{+13.52}_{-6.40}$ & $757.25^{+4.02}_{-1.98}$ & $655.62^{+30.80}_{-18.38}$ & $1101.83^{+33.64}_{-19.46}$ & \nodata & II & \nodata \\
130609B & $38.31^{+0.28}_{-0.30}$ & $8.42^{+0.31}_{-0.35}$ & $172.47^{+0.13}_{-0.13}$ & $8.91^{+0.23}_{-0.27}$ & $17.33^{+0.39}_{-0.44}$ & \nodata & $59.79^{+2.96}_{-2.55}$ & $25.62^{+2.37}_{-3.03}$ & $85.42^{+0.76}_{-0.99}$ & $22.66^{+3.73}_{-3.37}$ & $48.28^{+4.42}_{-4.53}$ & \nodata & I & \nodata \\
130609B & $13.51^{+0.44}_{-0.39}$ & $10.22^{+0.22}_{-0.33}$ & $187.81^{+0.40}_{-0.42}$ & $10.30^{+0.35}_{-0.54}$ & $20.52^{+0.41}_{-0.63}$ & \nodata & $62.59^{+3.55}_{-3.89}$ & $54.24^{+2.57}_{-2.41}$ & $176.62^{+0.70}_{-0.60}$ & $149.34^{+13.03}_{-13.13}$ & $203.58^{+13.28}_{-13.35}$ & \nodata & I & \nodata \\
130609B & $11.49^{+0.97}_{-0.99}$ & $10.95^{+0.34}_{-0.53}$ & $200.05^{+0.76}_{-0.77}$ & $11.18^{+0.35}_{-0.57}$ & $22.13^{+0.49}_{-0.78}$ & \nodata & $106.98^{+9.90}_{-15.38}$ & $41.16^{+16.30}_{-7.08}$ & $270.52^{+7.03}_{-2.81}$ & $157.72^{+14.29}_{-12.03}$ & $198.88^{+21.68}_{-13.96}$ & \nodata & I & \nodata \\
140108A & $76.79^{+0.12}_{-0.12}$ & $8.03^{+0.15}_{-0.15}$ & $84.82^{+0.07}_{-0.07}$ & $10.24^{+0.25}_{-0.25}$ & $18.27^{+0.29}_{-0.29}$ & \nodata & $56.18^{+1.08}_{-1.36}$ & $31.91^{+1.31}_{-1.08}$ & $88.09^{+0.35}_{-0.38}$ & $44.34^{+2.22}_{-1.97}$ & $76.25^{+2.58}_{-2.25}$ & \nodata & III & \nodata \\
140206A & $52.01^{+0.04}_{-0.04}$ & $7.10^{+0.07}_{-0.07}$ & $59.12^{+0.04}_{-0.04}$ & $10.55^{+0.12}_{-0.12}$ & $17.65^{+0.14}_{-0.14}$ & $53.43\pm0.16$ & $17.18^{+4.54}_{-5.95}$ & $15.33^{+3.09}_{-3.09}$ & $59.38^{+1.32}_{-1.38}$ & $59.91^{+13.23}_{-16.23}$ & $75.24^{+13.59}_{-16.52}$ & $52.01\pm0.07$ & I & 2.730\tablenotemark{e} \\
140430A & $145.29^{+1.05}_{-0.94}$ & $6.42^{+0.62}_{-0.97}$ & $151.60^{+0.87}_{-0.81}$ & $6.28^{+0.67}_{-0.83}$ & $12.70^{+0.91}_{-1.28}$ & $51.34\pm0.28$ & $68.06^{+10.67}_{-8.67}$ & $88.86^{+8.61}_{-10.78}$ & $156.91^{+0.86}_{-1.14}$ & $37.67^{+22.62}_{-13.89}$ & $126.53^{+24.20}_{-17.58}$ & $50.69\pm0.04$ & III & 1.600\tablenotemark{f} \\
140430A & $17.67^{+0.82}_{-0.76}$ & $7.92^{+0.55}_{-0.86}$ & $170.72^{+0.53}_{-0.50}$ & $9.88^{+0.82}_{-1.16}$ & $17.80^{+0.99}_{-1.44}$ & $51.69\pm0.43$ & $74.36^{+9.99}_{-11.84}$ & $29.24^{+5.13}_{-4.85}$ & $171.66^{+0.62}_{-0.51}$ & $101.18^{+9.49}_{-9.45}$ & $130.42^{+10.79}_{-10.62}$ & $51.07\pm0.05$ & III & 1.600\tablenotemark{f} \\
140512A & $94.32^{+0.25}_{-0.28}$ & $26.66^{+0.31}_{-0.32}$ & $120.97^{+0.14}_{-0.15}$ & $25.28^{+0.51}_{-0.51}$ & $51.94^{+0.60}_{-0.60}$ & $52.37\pm0.09$ & $78.70^{+1.91}_{-2.00}$ & $40.86^{+2.00}_{-1.91}$ & $119.56^{+1.07}_{-0.99}$ & $124.31^{+1.97}_{-3.68}$ & $165.17^{+2.81}_{-4.15}$ & $49.70\pm0.01$ & III & 0.725\tablenotemark{g} \\
140709B & $97.83^{+1.74}_{-1.46}$ & $25.09^{+1.24}_{-1.27}$ & $123.01^{+0.56}_{-0.63}$ & $14.81^{+0.77}_{-0.73}$ & $39.90^{+1.46}_{-1.46}$ & \nodata & $46.20^{+1.19}_{-1.20}$ & $71.80^{+1.16}_{-1.31}$ & $118.00^{+0.83}_{-0.99}$ & $224.47^{+6.54}_{-5.06}$ & $296.27^{+6.64}_{-5.23}$ & \nodata & III & \nodata \\
140817A & $169.60^{+1.72}_{-1.84}$ & $10.08^{+1.82}_{-1.82}$ & $179.57^{+0.32}_{-0.58}$ & $18.08^{+1.02}_{-1.35}$ & $28.16^{+2.09}_{-2.27}$ & \nodata & $95.17^{+6.22}_{-4.58}$ & $118.54^{+4.58}_{-6.22}$ & $213.72^{+1.77}_{-1.48}$ & $241.71^{+46.48}_{-40.12}$ & $360.25^{+46.71}_{-40.60}$ & \nodata & II & \nodata \\
140817A & $23.02^{+1.97}_{-1.86}$ & $25.96^{+1.87}_{-1.94}$ & $218.57^{+0.51}_{-0.50}$ & $29.04^{+1.10}_{-1.08}$ & $55.00^{+2.17}_{-2.22}$ & \nodata & $45.14^{+16.08}_{-15.56}$ & $153.44^{+14.27}_{-15.41}$ & $293.75^{+2.43}_{-2.71}$ & $482.85^{+26.70}_{-32.75}$ & $636.29^{+30.27}_{-36.19}$ & \nodata & II & \nodata \\
141031A & $767.74^{+4.94}_{-8.99}$ & $29.19^{+10.04}_{-5.86}$ & $796.84^{+2.83}_{-2.84}$ & $40.07^{+14.99}_{-9.73}$ & $69.26^{+18.04}_{-11.36}$ & \nodata & $308.19^{+108.17}_{-62.16}$ & $242.45^{+7.58}_{-9.04}$ & $897.08^{+1.19}_{-1.22}$ & $298.89^{+18.41}_{-20.98}$ & $541.34^{+19.91}_{-22.84}$ & \nodata & II & \nodata \\
141031A & $79.87^{+1.12}_{-1.50}$ & $16.74^{+2.00}_{-1.05}$ & $864.44^{+0.94}_{-0.82}$ & $55.73^{+6.80}_{-6.49}$ & $72.47^{+7.09}_{-6.57}$ & \nodata & $258.14^{+28.62}_{-28.72}$ & $175.10^{+26.37}_{-25.45}$ & $1100.69^{+2.93}_{-2.44}$ & $394.78^{+13.45}_{-24.71}$ & $569.88^{+29.60}_{-35.47}$ & \nodata & II & \nodata \\
150724A & $142.15^{+2.18}_{-1.37}$ & $5.56^{+1.70}_{-1.43}$ & $148.66^{+0.65}_{-0.93}$ & $11.91^{+2.06}_{-6.60}$ & $17.47^{+2.67}_{-6.75}$ & \nodata & $64.49^{+13.94}_{-14.14}$ & $89.73^{+14.14}_{-13.94}$ & $154.23^{+5.98}_{-2.27}$ & $441.58^{+41.08}_{-195.87}$ & $531.31^{+43.45}_{-196.37}$ & \nodata & II & \nodata \\
150724A & $65.24^{+1.24}_{-0.61}$ & $12.37^{+0.77}_{-3.48}$ & $219.87^{+0.10}_{-0.24}$ & $22.45^{+0.83}_{-1.58}$ & $34.82^{+1.13}_{-3.82}$ & \nodata & $92.30^{+14.74}_{-15.38}$ & $74.20^{+6.49}_{-4.18}$ & $230.99^{+0.74}_{-0.69}$ & $251.46^{+42.76}_{-39.11}$ & $325.66^{+43.25}_{-39.33}$ & \nodata & II & \nodata \\
150724A & $61.32^{+2.96}_{-2.88}$ & $9.29^{+4.33}_{-3.03}$ & $279.19^{+1.90}_{-1.71}$ & $16.25^{+7.45}_{-6.50}$ & $25.54^{+8.62}_{-7.17}$ & \nodata & $46.64^{+14.20}_{-12.58}$ & $84.52^{+11.86}_{-12.63}$ & $287.95^{+2.63}_{-1.98}$ & $588.99^{+27.05}_{-29.81}$ & $673.51^{+29.54}_{-32.38}$ & \nodata & II & \nodata \\
151027A & $96.58^{+0.31}_{-0.28}$ & $14.15^{+0.36}_{-0.35}$ & $110.77^{+0.12}_{-0.12}$ & $18.91^{+0.47}_{-0.45}$ & $33.06^{+0.59}_{-0.57}$ & $52.24\pm0.09$ & $65.45^{+2.42}_{-2.54}$ & $59.91^{+2.53}_{-2.42}$ & $125.37^{+0.72}_{-0.70}$ & $99.47^{+4.84}_{-4.15}$ & $159.38^{+5.46}_{-4.80}$ & $51.10\pm0.02$ & III & 0.810\tablenotemark{h} \\
160227A & $88.86^{+1.06}_{-0.74}$ & $7.90^{+1.09}_{-0.80}$ & $193.00^{+0.60}_{-0.47}$ & $60.16^{+3.52}_{-3.43}$ & $68.06^{+3.68}_{-3.52}$ & $52.65\pm0.08$ & $130.57^{+3.86}_{-3.13}$ & $76.31^{+3.13}_{-3.76}$ & $206.88^{+1.13}_{-0.90}$ & $268.85^{+10.58}_{-15.12}$ & $345.16^{+11.03}_{-15.58}$ & $51.75\pm0.01$ & III & 2.380\tablenotemark{i} \\
160425A & $202.72^{+0.56}_{-0.50}$ & $14.39^{+0.57}_{-0.40}$ & $270.54^{+0.41}_{-0.36}$ & $35.22^{+2.18}_{-2.05}$ & $49.61^{+2.25}_{-2.09}$ & $51.13\pm0.42$ & $223.20^{+1.23}_{-0.87}$ & $55.72^{+0.86}_{-1.21}$ & $278.92^{+0.29}_{-0.29}$ & $75.76^{+1.70}_{-3.00}$ & $131.48^{+1.91}_{-3.23}$ & $51.04\pm0.01$ & III & 0.555\tablenotemark{j} \\
160912A & $96.44^{+0.27}_{-0.38}$ & $5.57^{+0.53}_{-0.36}$ & $148.34^{+0.29}_{-0.25}$ & $53.35^{+2.50}_{-2.38}$ & $58.92^{+2.56}_{-2.41}$ & \nodata & $84.92^{+8.92}_{-11.77}$ & $87.93^{+14.05}_{-7.14}$ & $172.85^{+9.36}_{-4.18}$ & $337.28^{+81.19}_{-153.65}$ & $425.21^{+82.40}_{-153.82}$ & \nodata & III & \nodata \\
161117A & $34.50^{+1.03}_{-1.19}$ & $8.61^{+0.98}_{-1.09}$ & $87.16^{+0.20}_{-0.22}$ & $5.47^{+0.51}_{-0.51}$ & $14.08^{+1.10}_{-1.20}$ & $51.73\pm0.28$ & $47.44^{+2.24}_{-1.41}$ & $18.48^{+1.13}_{-2.12}$ & $65.92^{+1.18}_{-1.45}$ & $5.30^{+2.87}_{-1.37}$ & $23.78^{+3.08}_{-2.52}$ & $50.87\pm0.03$ & II & 1.549\tablenotemark{k} \\
161117A & $11.33^{+0.20}_{-0.23}$ & $20.64^{+0.29}_{-0.27}$ & $110.50^{+0.09}_{-0.10}$ & $23.38^{+0.33}_{-0.32}$ & $44.02^{+0.44}_{-0.42}$ & $52.81\pm0.10$ & $8.42^{+1.92}_{-2.63}$ & $60.91^{+1.39}_{-1.31}$ & $116.77^{+0.55}_{-0.54}$ & $287.31^{+9.01}_{-8.34}$ & $348.22^{+9.12}_{-8.44}$ & $51.95\pm0.01$ & II & 1.549\tablenotemark{k} \\
170208B & $89.26^{+2.51}_{-2.70}$ & $23.81^{+3.31}_{-3.01}$ & $113.03^{+1.18}_{-1.19}$ & $31.37^{+5.79}_{-4.92}$ & $55.18^{+6.67}_{-5.77}$ & \nodata & $21.85^{+12.21}_{-4.66}$ & $40.84^{+1.92}_{-2.17}$ & $103.96^{+0.66}_{-0.73}$ & $300.65^{+11.43}_{-11.12}$ & $341.49^{+11.59}_{-11.33}$ & \nodata & I & \nodata \\
170519A & $176.42^{+6.23}_{-16.51}$ & $15.80^{+7.61}_{-5.13}$ & $194.63^{+2.02}_{-4.91}$ & $16.91^{+8.15}_{-4.98}$ & $32.71^{+11.15}_{-7.15}$ & $50.85\pm0.45$ & $125.90^{+2.73}_{-9.11}$ & $25.17^{+8.64}_{-0.74}$ & $193.73^{+1.70}_{-0.28}$ & $298.40^{+12.55}_{-12.57}$ & $323.57^{+15.24}_{-12.59}$ & $50.97\pm0.05$ & II & 0.818\tablenotemark{l} \\
170531B & $136.72^{+2.08}_{-2.54}$ & $19.61^{+3.37}_{-3.03}$ & $156.32^{+1.29}_{-1.33}$ & $37.96^{+6.26}_{-5.67}$ & $57.57^{+7.11}_{-6.43}$ & $52.32\pm0.26$ & $103.67^{+1.79}_{-2.31}$ & $56.62^{+2.27}_{-1.80}$ & $160.29^{+0.39}_{-0.41}$ & $358.73^{+26.38}_{-20.28}$ & $415.35^{+26.48}_{-20.36}$ & $51.41\pm0.01$ & III & 2.366\tablenotemark{m} \\
170705A & $147.79^{+2.60}_{-2.49}$ & $29.25^{+3.27}_{-3.05}$ & $177.11^{+1.20}_{-1.19}$ & $75.72^{+7.19}_{-6.96}$ & $104.97^{+7.90}_{-7.60}$ & $52.79\pm0.06$ & $67.20^{+1.29}_{-2.57}$ & $67.46^{+0.81}_{-0.46}$ & $189.75^{+0.41}_{-0.38}$ & $388.11^{+11.80}_{-8.32}$ & $455.57^{+11.83}_{-8.33}$ & $51.72\pm0.01$ & II & 2.010\tablenotemark{n} \\
180325A & $76.04^{+0.09}_{-0.10}$ & $4.41^{+0.13}_{-0.12}$ & $80.45^{+0.04}_{-0.04}$ & $6.99^{+0.20}_{-0.20}$ & $11.40^{+0.24}_{-0.23}$ & $53.40\pm0.06$ & $69.17^{+2.98}_{-3.42}$ & $7.49^{+2.85}_{-1.76}$ & $76.67^{+3.31}_{-3.80}$ & $60.82^{+7.22}_{-5.26}$ & $68.31^{+7.76}_{-5.55}$ & $50.11\pm0.12$ & III & 2.248\tablenotemark{o} \\
180329B & $135.74^{+0.46}_{-0.65}$ & $10.14^{+0.80}_{-0.77}$ & $145.82^{+0.38}_{-0.40}$ & $21.66^{+1.80}_{-1.74}$ & $31.80^{+1.97}_{-1.90}$ & $52.15\pm0.14$ & $109.67^{+1.19}_{-1.54}$ & $47.53^{+1.17}_{-1.05}$ & $157.20^{+1.16}_{-1.49}$ & $119.01^{+3.89}_{-3.16}$ & $166.54^{+4.06}_{-3.33}$ & $51.28\pm0.02$ & III & 1.998\tablenotemark{p} \\
180329B & $44.86^{+8.94}_{-14.98}$ & $34.46^{+18.70}_{-16.47}$ & $210.87^{+7.16}_{-8.81}$ & $55.86^{+41.31}_{-22.87}$ & $90.32^{+45.35}_{-28.18}$ & $52.11\pm0.08$ & $28.29^{+2.18}_{-2.32}$ & $83.69^{+2.00}_{-1.51}$ & $221.65^{+0.47}_{-0.48}$ & $142.75^{+4.82}_{-3.45}$ & $226.44^{+5.22}_{-3.77}$ & $51.44\pm0.01$ & III & 1.998\tablenotemark{p} \\
180624A & $159.41^{+1.57}_{-1.87}$ & $23.42^{+2.14}_{-1.95}$ & $182.77^{+0.94}_{-0.95}$ & $44.03^{+5.04}_{-4.74}$ & $67.45^{+5.48}_{-5.13}$ & $52.48\pm0.22$ & $95.02^{+15.91}_{-15.21}$ & $25.18^{+3.08}_{-1.03}$ & $175.07^{+0.55}_{-0.33}$ & $416.22^{+41.56}_{-47.41}$ & $441.40^{+41.67}_{-47.42}$ & $51.89\pm0.03$ & I & 2.855\tablenotemark{q} \\
180809B & $197.56^{+0.23}_{-0.24}$ & $21.34^{+0.30}_{-0.30}$ & $218.89^{+0.13}_{-0.14}$ & $44.75^{+0.56}_{-0.57}$ & $66.09^{+0.64}_{-0.64}$ & \nodata & $135.95^{+3.53}_{-2.88}$ & $27.30^{+1.83}_{-1.71}$ & $222.90^{+1.01}_{-1.04}$ & $57.05^{+2.83}_{-2.57}$ & $84.35^{+3.37}_{-3.09}$ & \nodata & II & \nodata \\
200122A & $47.26^{+2.21}_{-4.52}$ & $17.97^{+1.16}_{-3.07}$ & $155.66^{+1.71}_{-2.79}$ & $11.06^{+4.56}_{-1.74}$ & $29.03^{+4.71}_{-3.53}$ & \nodata & $9.44^{+15.01}_{-12.66}$ & $86.48^{+1.58}_{-1.51}$ & $195.74^{+0.68}_{-0.63}$ & $121.33^{+2.30}_{-2.38}$ & $207.81^{+2.79}_{-2.82}$ & \nodata & I & \nodata \\
200519A & $46.16^{+1.30}_{-0.75}$ & $6.24^{+0.31}_{-0.59}$ & $69.83^{+4.75}_{-0.50}$ & $5.58^{+0.82}_{-0.96}$ & $11.82^{+0.88}_{-1.13}$ & \nodata & $18.81^{+6.15}_{-10.53}$ & $15.38^{+3.29}_{-6.70}$ & $75.71^{+0.97}_{-11.83}$ & $14.10^{+6.13}_{-7.46}$ & $29.48^{+6.96}_{-10.03}$ & \nodata & I & \nodata \\
200519A & $8.95^{+0.79}_{-1.38}$ & $3.18^{+5.08}_{-0.43}$ & $75.05^{+4.99}_{-0.04}$ & $11.09^{+31.90}_{-1.75}$ & $14.27^{+32.30}_{-1.80}$ & \nodata & $5.10^{+20.97}_{-5.93}$ & $31.66^{+4.91}_{-17.94}$ & $97.09^{+2.86}_{-1.40}$ & $56.42^{+6.09}_{-4.32}$ & $88.08^{+7.82}_{-18.45}$ & \nodata & I & \nodata \\
200528A & $29.94^{+0.82}_{-0.72}$ & $10.81^{+0.54}_{-0.78}$ & $87.98^{+0.36}_{-0.37}$ & $10.62^{+1.08}_{-1.82}$ & $21.43^{+1.21}_{-1.98}$ & \nodata & $3.22^{+9.79}_{-13.24}$ & $62.46^{+3.61}_{-4.77}$ & $97.00^{+0.74}_{-0.74}$ & $124.33^{+25.56}_{-21.88}$ & $186.79^{+25.81}_{-22.39}$ & \nodata & III & \nodata \\
200528A & $24.38^{+4.20}_{-5.93}$ & $7.66^{+4.51}_{-3.82}$ & $109.74^{+1.22}_{-3.64}$ & $6.90^{+2.99}_{-2.57}$ & $14.56^{+5.41}_{-4.60}$ & \nodata & $24.48^{+10.71}_{-7.66}$ & $79.62^{+5.96}_{-10.15}$ & $138.65^{+1.13}_{-1.03}$ & $191.40^{+11.04}_{-13.01}$ & $271.02^{+12.55}_{-16.50}$ & \nodata & III & \nodata \\
230802A & $92.54^{+6.27}_{-6.88}$ & $47.59^{+6.69}_{-6.30}$ & $141.23^{+2.26}_{-2.45}$ & $52.21^{+12.71}_{-11.48}$ & $99.80^{+14.36}_{-13.10}$ & \nodata & $86.88^{+1.85}_{-1.86}$ & $51.49^{+1.84}_{-1.76}$ & $138.37^{+0.86}_{-0.72}$ & $889.50^{+74.25}_{-66.57}$ & $940.99^{+74.27}_{-66.59}$ & \nodata & II & \nodata \\
240912A & $188.53^{+4.46}_{-5.05}$ & $27.73^{+4.21}_{-3.89}$ & $282.01^{+1.92}_{-2.13}$ & $94.82^{+13.30}_{-11.43}$ & $122.55^{+13.95}_{-12.07}$ & $52.10\pm0.07$ & $162.21^{+29.69}_{-37.54}$ & $62.73^{+40.89}_{-8.66}$ & $288.13^{+28.58}_{-4.32}$ & $572.68^{+86.18}_{-66.85}$ & $635.41^{+95.39}_{-67.41}$ & $51.81\pm0.19$ & I & 1.234\tablenotemark{r} \\
250114A & $127.67^{+5.23}_{-7.40}$ & $29.99^{+5.50}_{-4.92}$ & $273.22^{+1.83}_{-1.68}$ & $42.03^{+4.90}_{-4.81}$ & $72.02^{+7.37}_{-6.88}$ & $52.74\pm0.10$ & $70.76^{+17.49}_{-28.08}$ & $144.94^{+7.57}_{-4.17}$ & $278.76^{+1.25}_{-1.03}$ & $370.21^{+22.97}_{-14.15}$ & $515.15^{+24.19}_{-14.75}$ & $51.94\pm0.01$ & I & 4.732\tablenotemark{s} \\
250823A & $153.05^{+1.56}_{-2.16}$ & $6.97^{+1.76}_{-1.31}$ & $160.10^{+0.87}_{-0.73}$ & $26.57^{+3.42}_{-3.00}$ & $33.54^{+3.85}_{-3.27}$ & \nodata & $25.80^{+8.56}_{-10.20}$ & $36.17^{+8.51}_{-4.95}$ & $165.27^{+0.57}_{-0.70}$ & $99.51^{+23.62}_{-17.52}$ & $135.68^{+25.11}_{-18.21}$ & \nodata & I & \nodata \\
\enddata
\tablenote{References of the redshift measured: (a) \cite{2010GCN.11091....1F}; (b) \cite{2010GCN.11230....1T}; (c) \cite{2011GCN.12234....1C}; (d) \cite{2012GCN.14059....1P}; (e) \cite{2014GCN.15800....1M}; (f) \cite{2014GCN.16194....1K}; (g) \cite{2014GCN.16310....1D}; (h) \cite{2015GCN.18487....1P}; (i) \cite{2016GCN.19109....1X}; (j) \cite{2016GCN.19350....1T}; (k) \cite{2016GCN.20180....1M}; (l) \cite{2017GCN.21119....1I}; (m) \cite{2017GCN.21177....1D}; (n) \cite{2017GCN.21298....1D}; (o) \cite{2018GCN.22555....1D}; (p) \cite{2018GCN.22567....1I}; (q) \cite{2018GCN.22845....1R}; (r) \cite{2024GCN.37469....1D}; (s) \cite{2025GCN.38934....1M}.}
\end{deluxetable*}


\begin{deluxetable*}{lcccccccccc}
\tabletypesize{\tiny}
\tablecaption{Results of the Spectral Fitting of Our Sample.\label{tab:spectral_fit}}
\tablehead{\colhead{GRB} & \colhead{Interval (s)} & \colhead{Model} & \colhead{$\Gamma$} & \colhead{$\alpha$} & \colhead{$\beta$} & \colhead{$kT$ (keV)} & \colhead{$E_{\rm c}$ (keV)} & \colhead{$E_{\rm cut}$ (keV)} & \colhead{$N_{\rm H}$ ($10^{22}$ cm$^{-2}$)} & \colhead{$\chi^2_r$}}
\startdata
  100814A & 141--160 & Band & \nodata & $-1.10\pm0.07$ & $-2.09\pm0.68$ & \nodata & $72.42\pm17.01$ & \nodata & $0.25\pm0.12$ & $1.16$ \\
  100906A & 110--127 & PL+BB & $2.40\pm0.11$ & \nodata & \nodata & $5.03\pm0.59$ & \nodata & \nodata & $10.57\pm1.36$ & $1.26$ \\
  110102A & 197--219 & PL+BB & $1.25\pm0.03$ & \nodata & \nodata & $2.76\pm0.38$ & \nodata & \nodata & $0.42\pm0.05$ & $1.14$ \\
  110102A & 247--280 & Band & \nodata & $-0.73\pm0.23$ & $-1.72\pm0.04$ & \nodata & $10.35\pm7.56$ & \nodata & $0.16\pm0.05$ & $1.11$ \\
  110801A & 345--413 & PL+BB & $1.82\pm0.09$ & \nodata & \nodata & $3.46\pm0.14$ & \nodata & \nodata & $1.43\pm0.18$ & $1.17$ \\
  110820A & 174--292 & PL+BB & $2.02\pm0.12$ & \nodata & \nodata & $1.91\pm0.15$ & \nodata & \nodata & $0.51\pm0.07$ & $0.99$ \\
  120514A & 127--179 & PL & $1.89\pm0.07$ & \nodata & \nodata & \nodata & \nodata & \nodata & $1.51\pm0.12$ & $0.79$ \\
  121123A & 187--348 & PL+BB & $1.19\pm0.02$ & \nodata & \nodata & $10.44\pm0.29$ & \nodata & \nodata & $0.12\pm0.01$ & $1.20$ \\
  121211A & 160--175 & Band & \nodata & $-0.86\pm0.15$ & $-2.31\pm0.19$ & \nodata & $15.81\pm9.85$ & \nodata & $0.75\pm0.19$ & $1.23$ \\
  121217A & 721--754 & PL+BB & $1.52\pm0.05$ & \nodata & \nodata & $2.23\pm0.27$ & \nodata & \nodata & $0.09\pm0.06$ & $1.09$ \\
  130609B & 164--182 & PL+BB & $1.75\pm0.05$ & \nodata & \nodata & $1.72\pm0.19$ & \nodata & \nodata & $0.32\pm0.03$ & $1.02$ \\
  130609B & 177--199 & PL & $1.71\pm0.05$ & \nodata & \nodata & \nodata & \nodata & \nodata & $0.33\pm0.03$ & $1.15$ \\
  130609B & 189--212 & PL & $1.74\pm0.05$ & \nodata & \nodata & \nodata & \nodata & \nodata & $0.24\pm0.02$ & $0.89$ \\
  140108A & 76--96 & PL & $1.31\pm0.02$ & \nodata & \nodata & \nodata & \nodata & \nodata & $0.90\pm0.10$ & $0.87$ \\
  140206A & 52--70 & CPL & $0.74\pm0.06$ & \nodata & \nodata & \nodata & \nodata & $84.42\pm8.70$ & $6.64\pm1.65$ & $0.91$ \\
  140430A & 145--158 & PL & $1.77\pm0.10$ & \nodata & \nodata & \nodata & \nodata & \nodata & $0.75\pm0.36$ & $1.11$ \\
  140430A & 162--180 & PL+BB & $1.83\pm0.18$ & \nodata & \nodata & $3.03\pm0.32$ & \nodata & \nodata & $1.03\pm0.49$ & $1.11$ \\
  140512A & 94--147 & PL+BB & $1.24\pm0.02$ & \nodata & \nodata & $4.20\pm0.41$ & \nodata & \nodata & $0.54\pm0.11$ & $1.08$ \\
  140709B & 97--138 & PL+BB & $1.17\pm0.06$ & \nodata & \nodata & $6.07\pm1.22$ & \nodata & \nodata & $0.29\pm0.04$ & $1.09$ \\
  140817A & 169--198 & PL & $1.39\pm0.06$ & \nodata & \nodata & \nodata & \nodata & \nodata & $0.33\pm0.04$ & $1.09$ \\
  140817A & 192--248 & PL & $1.58\pm0.04$ & \nodata & \nodata & \nodata & \nodata & \nodata & $0.32\pm0.03$ & $1.01$ \\
  141031A & 767--837 & PL & $1.38\pm0.08$ & \nodata & \nodata & \nodata & \nodata & \nodata & $0.15\pm0.05$ & $0.98$ \\
  141031A & 847--921 & PL & $1.53\pm0.04$ & \nodata & \nodata & \nodata & \nodata & \nodata & $0.31\pm0.03$ & $0.98$ \\
  150724A & 143--161 & PL & $1.59\pm0.10$ & \nodata & \nodata & \nodata & \nodata & \nodata & $0.27\pm0.07$ & $1.11$ \\
  150724A & 207--243 & PL & $1.76\pm0.06$ & \nodata & \nodata & \nodata & \nodata & \nodata & $0.33\pm0.05$ & $1.08$ \\
  150724A & 269--296 & PL & $1.91\pm0.06$ & \nodata & \nodata & \nodata & \nodata & \nodata & $0.09\pm0.03$ & $1.00$ \\
  151027A & 96--130 & PL+BB & $1.48\pm0.03$ & \nodata & \nodata & $2.90\pm0.26$ & \nodata & \nodata & $1.12\pm0.15$ & $1.07$ \\
  160227A & 185--254 & PL & $1.56\pm0.03$ & \nodata & \nodata & \nodata & \nodata & \nodata & $1.08\pm0.13$ & $0.93$ \\
  160425A & 256--306 & Band & \nodata & $-1.46\pm0.13$ & $-2.44\pm0.09$ & \nodata & $21.72\pm16.48$ & \nodata & $1.14\pm0.11$ & $1.21$ \\
  160912A & 142--202 & PL & $1.39\pm0.04$ & \nodata & \nodata & \nodata & \nodata & \nodata & $0.26\pm0.03$ & $0.96$ \\
  161117A & 78--93 & CPL & $0.55\pm0.12$ & \nodata & \nodata & \nodata & \nodata & $26.32\pm2.42$ & $2.01\pm0.85$ & $0.93$ \\
  161117A & 89--134 & CPL & $0.77\pm0.04$ & \nodata & \nodata & \nodata & \nodata & $37.72\pm1.43$ & $2.37\pm0.31$ & $1.10$ \\
  170208B & 89--145 & PL & $2.28\pm0.05$ & \nodata & \nodata & \nodata & \nodata & \nodata & $0.47\pm0.02$ & $1.01$ \\
  170519A & 178--212 & PL+BB & $2.15\pm0.28$ & \nodata & \nodata & $1.78\pm0.09$ & \nodata & \nodata & $0.65\pm0.13$ & $1.00$ \\
  170531B & 136--195 & PL+BB & $1.60\pm0.10$ & \nodata & \nodata & $4.53\pm0.33$ & \nodata & \nodata & $2.31\pm0.37$ & $1.05$ \\
  170705A & 147--253 & PL & $1.35\pm0.02$ & \nodata & \nodata & \nodata & \nodata & \nodata & $1.28\pm0.10$ & $1.13$ \\
  180325A & 76--88 & PL & $1.00\pm0.02$ & \nodata & \nodata & \nodata & \nodata & \nodata & $3.23\pm0.77$ & $0.91$ \\
  180329B & 135--168 & CPL & $1.04\pm0.06$ & \nodata & \nodata & \nodata & \nodata & $74.13\pm11.76$ & $1.00\pm0.25$ & $1.16$ \\
  180329B & 176--267 & PL & $1.64\pm0.03$ & \nodata & \nodata & \nodata & \nodata & \nodata & $0.99\pm0.08$ & $0.92$ \\
  180624A & 159--227 & PL+BB & $1.65\pm0.08$ & \nodata & \nodata & $5.51\pm0.34$ & \nodata & \nodata & $3.23\pm0.44$ & $1.20$ \\
  180809B & 197--264 & PL+BB & $1.29\pm0.02$ & \nodata & \nodata & $2.70\pm0.35$ & \nodata & \nodata & $0.17\pm0.02$ & $0.98$ \\
  200122A & 187--213 & PL+BB & $2.31\pm0.14$ & \nodata & \nodata & $1.80\pm0.62$ & \nodata & \nodata & $1.63\pm0.10$ & $1.24$ \\
  200519A & 63--76 & PL+BB & $1.41\pm0.09$ & \nodata & \nodata & $4.10\pm0.77$ & \nodata & \nodata & $0.30\pm0.07$ & $0.92$ \\
  200519A & 71--87 & PL & $1.83\pm0.07$ & \nodata & \nodata & \nodata & \nodata & \nodata & $0.79\pm0.09$ & $1.08$ \\
  200528A & 102--117 & PL & $2.39\pm0.09$ & \nodata & \nodata & \nodata & \nodata & \nodata & $0.53\pm0.04$ & $1.09$ \\
  200528A & 77--99 & PL+BB & $2.24\pm0.14$ & \nodata & \nodata & $1.77\pm0.26$ & \nodata & \nodata & $0.53\pm0.04$ & $1.03$ \\
  230802A & 93--194 & PL+BB & $1.69\pm0.10$ & \nodata & \nodata & $3.16\pm0.41$ & \nodata & \nodata & $0.52\pm0.05$ & $1.10$ \\
  240912A & 254--377 & PL & $1.83\pm0.03$ & \nodata & \nodata & \nodata & \nodata & \nodata & $0.59\pm0.04$ & $0.95$ \\
  250114A & 243--316 & CPL & $0.90\pm0.03$ & \nodata & \nodata & \nodata & \nodata & $39.22\pm4.26$ & $3.10\pm0.76$ & $1.03$ \\
  250823A & 153--187 & PL+BB & $1.92\pm0.13$ & \nodata & \nodata & $1.83\pm0.36$ & \nodata & \nodata & $0.17\pm0.02$ & $1.08$ \\
\enddata
\end{deluxetable*}

\end{document}